\documentclass[review]{elsarticle}

\usepackage[capposition=top]{floatrow} % [Figure.. comes on top of figure] notes
\usepackage{float} %introduzido por mim
\usepackage{morefloats} %introduzido por mim. LaTeX by default can cope with
\usepackage{algorithm}
\usepackage{algorithmic}
\usepackage{caption} % Table caption length
\usepackage{setspace}
\usepackage{amssymb}
\usepackage{upquote}
\usepackage{epstopdf}
\usepackage{amsmath}
\newcommand{\comment}[1]{}
\usepackage{indentfirst}
\usepackage[flushleft]{threeparttable} % to add notes below table
\usepackage{array, booktabs, siunitx, makecell} %break up table headings into two lines
\usepackage{enumitem} %
\PassOptionsToPackage{hyphens}{url}
\usepackage{hyperref}
\usepackage{subcaption}
\usepackage{pgf}
\pgfkeys{/pgf/number format/.cd ,precision=2,sci generic={exponent={\times 10^{#1}}}}
\newcommand\convert[1]{\pgfmathprintnumber{#1}}

\makeatletter
\def\ps@pprintTitle{%
	\let\@oddhead\@empty
	\let\@evenhead\@empty
	\def\@oddfoot{\reset@font\hfil\thepage\hfil}
	\let\@evenfoot\@oddfoot
}
\makeatother

\begin{document}
	
	\begin{frontmatter}
		
		%\title{Elsevier \LaTeX\ template\tnoteref{mytitlenote}}
		\title{An Agent-Based Model With Realistic Financial Time Series: A Method for Agent-Based Models Validation}
		
		%\author{Lu\'{i}s Gon\c{c}alves de Faria\fnref{myfootnote}}
		\author{Lu\'{i}s Gon\c{c}alves de Faria}
		\address{Coventry University London, University House, 109-117 Middlesex Street, London, E1 7JF}
		
		
		\ead{ad3940@coventry.ac.uk}
		
		
		\begin{abstract}
			This paper proposes a methodology to empirically validate an agent-based model (ABM) that generates artificial financial time series data comparable with real-world financial data. The approach is based on comparing the results of the ABM against the stylised facts -- the statistical properties of the empirical time-series of financial data. 
			
			The stylised facts appear to be universal and are observed across different markets, financial instruments and time periods%\citep{Cont2001}
			, hence they can serve to validate models of financial markets. If a given model does not consistently replicate these stylised facts, then we can reject it as being empirically inadequate.
			
			We discuss each stylised fact, the empirical evidence for it, and introduce appropriate metrics for testing the presence of these in model generated data. Moreover we investigate the ability of our model to correctly reproduce these stylised facts. We validate our model against a comprehensive list of empirical phenomena that qualify as a stylised fact, of both low and high frequency financial data that can be addressed by means of a relatively simple ABM
			of financial markets. This procedure is able to show whether the model, as an abstraction of reality, has a meaningful empirical counterpart and the significance of this analysis for the purposes of ABM validation and their empirical reliability.
		\end{abstract}
		
		
	\end{frontmatter}

	\section{Introduction}
	\label{sec:Introduction}
	
	\subsection{Agent-based Modelling}
	\label{subsec:Agent-based Modelling}
	
	We implement an ABM of a financial market based on the models of
	\cite{Chiarella2002} and \citep{Chiarella2009b}, in which agents can invest in
	both risky and risk-free assets subject to constraints imposed by both their
	preferences and the Basel III financial regulatory framework. 
	The agents in our model are financial institutions
	trading with each other, based on idiosyncratic characteristics and
	the inadvertent shaping of the market landscape.
	ABMs are, by definition, abstractions from reality but we can identify two
	goals the ABM literature in finance has been trying to achieve:
	
	\begin{itemize}
		\item To replicate statistical properties of financial time-series, which
		appear to be universal and, therefore, can serve to validate models of
		financial markets; and
		\item To explain some market behaviours by studying a decentralised economy as a complex adaptive system, where interactions between heterogeneous individuals may result in emergent
		properties, that may or may not lead to equilibria in the long run.
	\end{itemize}
	
	In using ABM, and thus capturing aspects of complex phenomena through an
	appealing model, we aim at understanding the behaviour of that model and its
	consistency with general phenomena or the statistical properties of
	financial time-series.
	ABMs are often executed as Monte-Carlo simulations and usually
	generate time-series of variables both on the individual and the macro level.
	Since there is a potentially infinite set of possible realisations, to gain an
	understanding of the model's operation and to check its consistency these
	time-series are analysed using econometric methods.
	
	In our model individual financial institutions' choices depend on individual's
	expectations of the future and their attitudes toward risk and losses, however
	it also depends on prices and their volatility which are not an individual
	element but determined through many market interactions. These
	emergent elements can have feedback effects in the agents population, altering individuals' behaviour.
	
	Our ABM is built using the Java Agent Based Modelling (JABM) toolkit
	\citep{Phelps2012}. The JABM toolkit is a framework used to build agent-based
	models employing a discrete-event simulation framework and the entities of the
	simulation model are represented using objects. JABM uses the
	dependency-injection design pattern that can be used to implement highly
	configurable simulation models, with different randomly-drawn values for free
	parameters, which are executable as Monte-Carlo simulations \citep{Phelps2012}.
	For facilitating our experiments in agent-based computational economics we used
	JASA (Java Auction Simulator API) \citep{Phelps2007}, which is a
	high-performance auction simulator built on top of the JABM toolkit. JASA is
	highly extendable and implements variants of an order-driven market, which is a
	market in which buyers and sellers meet via a limit order-book, a place where
	buy and sell orders are matched as they arrive over time, subject to some
	priority rules \citep{Abergel2016}. Some models have introduced the hypothesis that the mechanics of the order-book play
	an important role in explaining some of the stylised facts \citep{Lillo2004}.
	This approach goes back to the work of \citep{Gode1993} and the more recent works
	of \citep{Chiarella2002}, \citep{LEBARON2007}, \citep{LeBaron2008} and \citep{Chiarella2009b}.
	Our ABM extends JASA as necessary to adapt the market structure to our model and
	the implementation of the regulatory framework\footnote{The original
		class diagram of JASA can be found at \url{http://jasa.sourceforge.net/doc/api/}. Archived at:\url{https://web.archive.org/web/20220221170206/https://jasa.sourceforge.io/doc/api/}}.\\

	%\section{Literature Review}
	%\label{sec:Literature Review}
	\subsection{Methodology}
	\label{subsec:Methodology}
	
	%\section{Methodology}
	%\label{sec:Methodology}
	
	Our ultimate goal is the development of an abstract model corresponding to an
	hypothesis that yields valid and meaningful explanations about certain
	phenomena. Empirical evidence is vital in building the abstract model and in testing its validity. Only factual evidence alone can
	show whether this abstraction of reality has a meaningful empirical counterpart,
	whereby the model can be taken to be an adequate representation of the ``real
	world'' and if the model can thus be accepted as valid or rejected. An hypothesis should aim to be a sufficiently good approximation for the purpose
	in hand but not fully descriptive. In ABM, as in other scientific disciplines,
	solutions based on simplified cases have allowed scientific explanation and
	understanding to move forward \citep{{LuxThomas1999}, {SornetteD2003},
		{Ghoulmie2005}, {AlfaranoSimone2005}, {Alfarano2007},
		{Alfi2009part1}, {Alfi2009part2}, {Platt2016}}. For example, in
	our model we only make the necessary assumptions about the agents and their
	behaviour. In other words, the model works if it yields only sufficiently
	accurate explanations and the evidence for such an hypothesis always consists of
	its repeated failure to be contradicted. However, some authors defend the notion that introducing complexity into the
	model may be crucial to replicate most of the stylised facts \citep{Ghoulmie2008}.
	Our model contradicts this conclusion. In our view a more complex model is not
	necessary to reproduce financial stylised facts. Other authors nevertheless agree that
	given the simplicity of their models it would not be easy to reproduce many of
	the stylised facts \citep{Chiarella2006}. The success of our relatively simple model
	demonstrates that simplicity should not be a justification for the failure
	to replicate most of the stylised facts.
	% (\hl{add reference to simplicity
	% covered in chapter 2. - QUESTION: is their model as different from mine? What
	% differences justify the number of SF found? Westerhoff (2004) replicates some of
	% the stylised facts in a multi-asset framework. Often these models contain a
	% large number of parameters and variables that make an analysis of the quality of
	% these models rather difficult. Add rationality readings:
	% Goettler (2005)}).
	
	An important aspect that has contributed to some reluctance in accepting ABM as
	a well-established economic theory is the perceived lack of robustness of
	agent-based modelling, namely in the way empirical validation is conducted
	\citep{{Fagiolo2007}, {windrum2007}}. There are several approaches to
	empirical validation in ABM. One of these approaches is indirect calibration
	which, firstly, allows model generated data output validation through the
	identification and replication of a set of stylised facts and, secondly,
	calibrates the model using parameters that are consistent with output
	validation \citep{{Fagiolo2007}, {windrum2007}}. As it is difficult to
	determine how ABM should be empirically validated, we follow a methodology that
	has been successfully used in the past in many fields of science, including
	economics and agent-based modelling \citep{Pruna2020}. Firstly, we build an abstraction of the
	real world, the model, that generates synthetic data, the model generated data
	output, through simulations. Secondly, as in indirect calibration, we test the
	validity of our model by checking whether the model is an adequate
	representation of the portion of reality we are investigating. The degree of
	approximation to the ``real world'' is evaluated by comparing the simulated data
	to empirical observations of the ``real world''. Contrary to the indirect
	calibration approach we do not attempt to calibrate the model. The validation of the model is independent of any particular ethical position or
	normative judgments. The validation is objective and deals with ``what is'', the
	empirically observed facts, and not with ``what ought to be''. Only the
	empirical evidence reveals whether or not this abstraction of reality, our
	model, has a meaningful empirical counterpart.
	
	\section{The Model}
	\label{sec:The Model}
	
	\subsection{Experimental Design}
	\label{sec:Experimental Design}
	
	In this section we describe the experimental design of the ABM model\footnote{The model code can be found at \url{https://www.comses.net/codebase-release/7c016b59-2506-4750-8745-354ab6cd84a0/}} adapted from existing models in the literature \citep{{Chiarella2002}, {Chiarella2009b}}. This approach consists of modelling financial markets, with and without financial regulation, as a population of agents identified by their decision rules, which can be considered as a mapping from agents' information set to the set of possible actions: buy, sell or hold.
	\label{sec:Experimental Design1}
	
	If financial regulation exists, then agents have to adapt their
	behaviour to a mandatory minimum risk-based capital requirement by applying Value-at-Risk (VaR) or Expected Shortfall (ES) as a market risk metric. We implement a model where comparable treatments share the same initial conditions and free parameters remain constant. This procedure guarantees that the initial conditions are identical, which eliminates the effect of these potential sources of variability.
	
	\subsection{Model Market Structure}
	\label{sec:Model Market Structure}
	
	We use an ABM of a financial market in which heterogeneous agents can invest in
	both risky and risk-free assets \citep{{BROCK1998}, {Chiarella2002},
		{Chiarella2009b}, {Hermsen2010}, {OECD2012}}. If agents only consider their demand for shares in isolation, in a single-asset model, without modelling the agents' wider portfolio optimisation problem and risk management strategy, the model would not be suitable for exploring the implications of Basel III since agents would not balance their capital against risk-weighted assets.
	
	The ABM here presented consists of a population of agents, in our case financial
	institutions, $n_{a}$, trading in an order-driven market with continuous
	clearing, over a period of time corresponding to two years, with no official
	market maker, in which orders are submitted in a double auction and executions
	follow price/time priority.
	
	We restrict our world to one in which financial institutions construct a
	portfolio consisting of two assets: a risky asset, stocks, and a risk-free asset,
	cash. Therefore, we use equity positions as a proxy for market risk factors.
	Financial institutions are considered to be risk sensitive which makes them
	rebalance their portfolio every time they place an order in the market. All
	financial institutions have heterogeneous expectations about the expected
	returns, and transaction costs and taxes are assumed to be zero.
	
	Financial institutions can post two types of orders: buy or sell. Every time a
	financial institution $i$ is chosen to enter the market this financial
	institution $i$ can submit a limit order, that is an order to trade a certain
	quantity of stocks at a given price. These orders are submitted sequentially to an
	electronic trading system, matched and executed automatically. This is known as
	the limit-order book, where the lowest price for which there is an outstanding
	limit sell order, which is called the ask price, matches the highest buy price,
	which is called the bid price. If agents submit an order before their previous
	order gets executed, the latest order works as a cancellation order and
	overrides the previous one.
	
	In our model agents can place orders of size larger than one which allow us to
	explore the implications of regulatory proposals, such as Basel III, for
	portfolio dediversification and market instability.
	
	Each financial institution receives an initial endowment of
	cash, $c^{i}_{0}$, and an initial quantity of stocks, $s^{i}_{0}$. All
	agents know the fundamental price, $p^{f}_{t}$, which follows a geometric
	Brownian motion (GBM), as in \citep{Chiarella2009b}:
	\begin{equation}
		\label{eq:Fundamental Price GBM}
		\frac{\Delta{S}}{S}=\mu\Delta{t}+\sigma\epsilon\sqrt{\Delta{t}}
	\end{equation}
	where $\Delta{S}$ is the change in the stock price S in a small
	time interval $\Delta{t}$ and $\epsilon$ has a standard normal distribution. The
	parameter $\mu$ is the drift and $\sigma$ is the volatility of the fundamental
	price.
	
	\comment{``Even Mandelbrot and
		Hudson [280] acknowledge that the martingale property is usually not contradicted by
		empirical evidence'', in Ehrentreicht 2007}
	
	The price at time $t$, $p_{t}$, is determined by the market and is given by the
	price at which transactions occur. If no transactions occur at a given moment
	in time then the price is determined by the last transaction price. If no bids or
	asks are listed in the book then a proxy of the price is given by the previous
	traded or quoted price. The risk-free rate, $r_{f}$, is assumed to be constant
	over time and the same for all agents.
	
	Despite the fact that we investigate the potential occurrence of defaults, in
	our model there is no actual default, which means that agents stay in the market
	even if they cannot participate due to technical default, i.e.
	when they fail to: 1) fulfil an obligation to repay a loan in case of
	leverage, or 2) buy-back the stock at some point in the future in case of
	short-selling. In a situation of technical default agents stay in the market, even if they cannot temporarily
	participate, as a potential increase in stocks prices can generate positive
	changes in agents' balance sheet and put them actively back into the market.
	This possible scenario shows the importance of oscillations in the balance
	sheet, even in the absence of trading, and the endogenous risk \citep{{shin2010},
		{Beale2011}, {Zhou2013}}.
	In our model there is no lending/borrowing between financial institutions, which
	means that any systemic effect we might see in the model cannot be attributed to
	financial networks or interconnections. Instead, spillover effects operate
	through financial institutions' behaviour and impact on market prices, rather
	than direct exposure between them.
	
	\subsubsection{Financial Institutions' Expectations}
	\label{subsec:Financial Institutions' Expectations}
	
	Economic agents form expectations and act on the basis of predictions generated
	by these expectations \citep{Arthur1997}. Agents' intrinsic strategies are
	partially modelled based on their expectations of future prices and consist of
	three components: fundamentalist, chartist and noise-induced. Financial
	institution $i$ time horizon, $\tau^{i}$, depends on its components.
	Long term investors typically give more weight to fundamentalist strategies with
	longer time horizons, whilst day traders give more weight to chartist rules.
	Hence, the time horizon is a function of the probability of each agent entering
	the market, $\lambda^{i}$, and determines the interval ($t+\tau(\lambda^{i})$)
	while the agent's expectation about the return will prevail.
	
	Every time an agent $i$ is chosen to enter the market, this
	financial institution $i$ forms an expectation in time $t$ about the return in
	time $t+\tau(\lambda^{i})$, $\widehat{r}^{i}_{t,t+\tau(\lambda^{i})}$. Financial
	institutions make their expectations about returns based on the following equation:
	\begin{equation}
		\label{eq:Return Expectations}
		%\widehat{r}^{i}_{t,t+\tau^{i}}=g^{i}_{1}\ln{\frac{p^{f}_{t}}{p_{t}}}+g^{i}_{2}\overline{r}_{L_{i}}+n_{i}\epsilon_{t}
		\widehat{r}^{i}_{t,t+\tau(\lambda^{i})}=g^{i}_{1}\log({\frac{p^{f}_{t}}{p_{t}}})+g^{i}_{2}\overline{r}_{t,L^{i}}+n^{i}\epsilon^{i}_{t}
	\end{equation}
	where $g^{i}_{1}$, $g^{i}_{2}$ and $n^{i}$ represent the weights given to
	fundamentalist, chartist and noise-induced components, respectively. The sign of
	$g^{i}_{2}$ indicates a trend chasing strategy if $g^{i}_{2}>0$ and a contrarian
	if $g^{i}_{2}<0$. All financial institutions use a linear combination of these
	components.
	
	The fundamentalist component is assumed to have a stabilising effect on prices,
	whereas the chartist component has the opposite effect and tends to have a
	destabilising effect generating large price jumps and driving asset prices away
	from the intrinsic value of the asset. The average return over the interval used
	by the chartist component is given by
	\begin{equation}
		\label{eq:Chartist rule}
		\overline{r}_{t,L^{i}}=\frac{1}{L^{i}}\sum_{j=1}^{L^{i}}\log\frac{p_{t-j}}{p_{t-j-1}}.
	\end{equation}
	$L^{i}$ is uniformly and independently distributed across financial institutions over the
	interval $(1,L_{max})$. The noise component is randomly assigned
	across financial institutions, $\epsilon^{i}_{t}\sim N(0,1)$. The price expected at ${t+\tau(\lambda^{i})}$ by financial institution $i$ is
	given by
	\begin{equation}
		\label{eq:expectedPrice}
		\widehat{p}^{i}_{t,t+\tau(\lambda^{i})}={p}_{t}e^{\widehat{r}^{i}_{t,t+\tau(\lambda^{i})}}.
	\end{equation}
	
	\subsubsection{Model Constraints}
	\label{subsec:Model Constraints}
	
	Financial institutions' wealth is constituted by cash and stocks and all
	financial institutions are given an initial endowment of cash and stocks. Thus
	the wealth expression for financial institution $i$ at time $t$ is represented
	by:
	\begin{equation}
		\label{Wealth}
		W^{i}_{t}=c^{i}_{t}+s^{i}_{t} \times p_{t}
	\end{equation}
	where $c^{i}_{t}$ represents the amount of cash, $s^{i}_{t}$ the quantity
	of stocks and $p_{t}$ the current price. If equation \ref{Wealth} is negative
	then agent $i$ is in technical default.
	
	Financial institutions' behaviour can be restricted by two types of constraints:
	a budget constraint and/or regulatory constraints, depending on the
	treatment.
	
	What determines the optimal demand for assets in investors' portfolios depends
	on how the maximisation problem is set up, subject to the investor's constraints. The behaviour of economic agents in the face of uncertainty involves balancing
	expected risks against expected rewards. The classical mean-variance (M-V) framework introduced by \citep{Markowitz1952} and \citep{Markowitz1959} is the first proposed model of the reward-risk type and
	popularly referred to as Modern Portfolio Theory (MPT). \citep{Markowitz1952} suggested that the portfolio
	choice is based on two criteria: the expected portfolio return and the variance
	of the portfolio return, the latter used as a proxy for risk. Markowitz's M-V formulation equally penalises overperformance -- positive
	deviation from the mean -- and underperformance -- negative deviation from the
	mean, which may lead to inferior solutions suggested by the models using
	it. Nevertheless, not only does the M-V analysis remain a well adopted tool in
	the industry, as it is intuitive and easy to apply in practice and correctly
	describes investors' choices, or sufficiently well approximated choices, through
	quadratic utility functions \citep{Levy1979}.

	In our model financial institutions maximise the utility function
	\begin{equation}
		\label{utilityFunction}
		U=E(r_{c})-\frac{1}{2}A\sigma^{2}_{c}
	\end{equation}
	Equation \ref{utilityFunction} depends only on the mean and variance of the
	return on that portfolio. 
	%However, using a quadratic utility function poses the
	%problem that marginal utility is negative for levels of wealth above the bliss
	%point. Also, assuming normality for returns, e.g. as in \cite{Chiarella2009b},
	%it may not provide an accurate representation of reality. The use of a normal
	%distribution or a quadratic utility function has been criticised
	%(\cite{Levy1979}). These deficiencies regarding the choice of utility functions
	%are overcome by assuming that financial institutions are only concerned with the
	%M-V maximand and, consequently, the choice of a specific utility function can be
	%avoided (\cite{Cuthbertson2004}).
	Financial institutions portfolio construction is considered to be analogous to standard M-V
	optimisation \citep{{Markowitz1952}, {Levy1979}}, which only involves the first two
	statistical moments and higher moments are not considered.

	% In optimisation theory there are two types of optimisation problems depending on
	% whether the set of feasible solutions is constrained or unconstrained, as we
	% show in the next section.
	
	When leverage is not allowed, which is represented by a
	maximum leverage of 1, all agents' trading is limited by a budget constraint.
	However, when leverage or short-selling are permitted, agents can choose an
	optimal proportion of the risky asset above 1 or below 0, respectively. 
	\label{subsec:Model Constraints2}
	%In the
	%next section we explain how this budget constraint influences agents' trading
	%behaviour.

	\section{Stylised Facts}
	\label{sec:Stylised Facts}
	
	%From the aforementioned ABM literature, 
	Agent-based models allow us to replicate and explain statistically
	regular features of financial time-series. Despite the inherent complexity of
	financial markets, these appear to exhibit stylised facts which make the
	financial markets susceptible of a more rigorous analysis
	\citep{Vijayaraghavan2011}. Since ABM are abstractions from reality, and do not
	try to simulate reality as such, it has been standard practice to measure the
	validity of the ABM by investigating whether or not the model exhibits stylised
	facts. Despite criticism, simple nonlinear ABM have been shown to
	successfully replicate important empirically-observed stylised facts of financial time-series
	data. For example, \citep{Chen2012} and \citep{Pruna2020} identified several statistical properties of
	financial time-series that are replicated through agent-based models, and most
	of ABM's success has been attributed to its ability to correctly reproduce
	stylised facts \citep{Panayi2013}.
	\label{sec:Stylised Facts1}
	
	The reason for investigating these statistical properties is in order to
	ascertain if the model in use is well suited to replicate the stylised facts of real financial
	markets. In other words, if the model can be considered as an adequate and valid
	representation of the ``real world''. 
	%After validating the model, we carry out consistency checks with respect to particular real phenomena that we want to explain. 
	As a result this allows us to confirm whether or not the designed model
	is an appealing one and consistent with what we would have anticipated the model to
	produce. Therefore, in this paper we compare the results obtained in our
	simulation to those statistical properties that appear to be universal with
	respect to different markets, financial instruments and time periods
	\citep{{Ding1993}, {Pagan1996}, {Guillaume1997}, {Cont2001}}.
	
	This empirical validation has acquired the status of a benchmark
	\citep{Coolen2005} and this method of validation is considered a solid starting
	point despite the existent challenges \citep{Platt2016}. Indeed, most of the
	ABM, simple or more complex, are able to replicate at least some of the stylised
	facts. Some authors, e.g. \citep{Lux2000}, \citep{Chen2001}, \citep{Bouchaud2009}, \citep{Platt2016}, conclude that the most common statistical properties of the
	time-series of returns (e.g. heavy tails and volatility clustering) appear as
	emergent phenomena as a consequence of the trading process itself between
	heterogeneous agents, such as the order flow and the response of prices to
	individual orders. 
	
	An extremely rich set of stylised facts is simultaneously replicated for the first time by a single model. In the next sections we match most of the statistical properties of the financial time-series of returns, trading volume, trading duration, transaction size and bid-ask spread, using an ABM.
	
	\subsection{Returns}
	\label{subsection:Returns SF}
	
	We analyse the properties of the distribution of logarithmic asset returns,
	which are defined as:
	\begin{equation}
		\label{eq:returns}
		r_{\Delta{t}}=\log(p_{(t+\Delta{t})})-\log(p_{t})
	\end{equation}
	where $p_{t}$ is the price at time $t$ and $\Delta{t}$ is the sampling time
	interval. In our model we calculate returns as in equation \ref{eq:returns}. Nevertheless,
	and for simplicity, in the following sections we interchangeably
	mention returns and log-returns when referring to returns calculated as described above.
	
	\subsubsection{Moments of the Returns Distribution}
	\label{subsubsec:Moments of the Returns Distribution SF}
	
	The analysis of moments of the returns distribution is used both in theoretical
	and empirical finance. Some agent-based models, e.g.
	\citep{maymin2009}, \citep{Feldman2017}, investigated statistical properties of
	financial time-series which include the first four moments of the returns
	distribution: mean, standard deviation, skewness and kurtosis, and resemble the
	S\&P 500 and the major European indices.
	
	\subsubsection{Aggregational Gaussianity}
	\label{subsubsec:Aggregational Gaussianity SF}
	
	The empirical literature shows that the distribution of returns tends to be
	non-Gaussian, sharp peaked and heavy tailed, and as we move from higher to lower
	frequencies, the degree of leptokurtosis diminishes and the empirical
	distributions of returns tend to approximate a Gaussian distribution
	\citep{{Gopikrishnan1999}, {Cont2001}, {Antypas2013}}. 
	One way of quantifying the deviation from the normal distribution is by using
	the kurtosis of the distribution of log-return, a measure of how outlier-prone
	a distribution is. The kurtosis of the normal distribution is 3. Leptokurtic distributions
	that deviates from the normal distribution have kurtosis greater than 3. The
	kurtosis of a distribution is defined as
	\begin{equation}
		\label{eq:Aggregational Gaussianity}
		\text{kurtosis}=\frac{E(x-\mu)^{4}}{\sigma^{4}}
	\end{equation}
	where $\mu$ is the mean of $x$, $\sigma$ is the standard deviation of $x$, and
	$E(t)$ represents the expected value of the quantity $t$. Kurtosis computes a
	sample version of this population value.
	
	\subsubsection{Bubbles and Crashes}
	%\label{subsubsec:Bubbles and Crashes SF}
	
	An asset market (negative) bubble is a period during which agents are willing to
	pay (less) more for an asset than the asset's fundamental value due to the
	abnormally important influence of future asset price expectations on the
	valuation of assets and, thus, leads to deviations of prices from their
	fundamentals. Historical accounts suggest that an asset price crash becomes more
	likely as the relationship between asset prices and their fundamental value
	grows more extreme, usually upward \citep{{Rosser1997}, {vanNorden1999}}.
	Some authors \citep{{vanNorden1999}, {Phillips2015}}, show the existence
	of a correlation between stock returns and deviation from the fundamental
	price, a long-run general equilibrium price.
	\label{subsubsec:Bubbles and Crashes SF}
	
	The actual price of the asset, $p_t$, may deviate from the fundamental price,
	$p_{t}^{f}$, according to the following relationship \citep{{Rosser1997},
		{vanNorden1999}, {Phillips2015}}:
	\begin{equation}
		\label{eq:Price with bubbles}
		p_{t}=p_{t}^{f}+b_{t}+\epsilon_{t}
	\end{equation}
	where $b_{t}$ is the bubble component at period $t$, and $\epsilon_{t}$ is a zero
	mean, constant variance error term that contains the unexpected innovation of
	both the bubble term and of the fundamental component. Setting the error term
	equal to its expected value of zero, the bubble component, $b_{t}$, is simply
	the difference between the actual price and the fundamental price
	\citep{{vanNorden1999}, {Anderson2011}, {Phillips2015}}. From equation
	\ref{eq:Price with bubbles}, the relative bubble size is given by:
	\begin{equation}
		\label{eq:Relative bubble size}
		B_{t}=\frac{b_{t}}{p_{t}}=\frac{p_{t}-p_{t}^{f}}{p_{t}}
	\end{equation}
	Equation \ref{eq:Relative bubble size} indicates that the market price, $p_{t}$,
	deviates from its fundamental value, $p_{t}^{f}$, by $b_{t}$, the value
	corresponding to the rational bubble.
	
	Contrary to what other studies suggest, e.g. \citep{Flood1990}, the existence of
	bubbles in our model cannot be justified by the misspecification of
	fundamentals, since the fundamental price is public and known to all agents.
	Hence, the possible deviation from the fundamental price is due to model
	microstructure, e.g. the speculative behaviour originating from chartist and
	noise trading components, or the impact of regulatory shocks on market price.
	\citep{Anderson2011} assume the bubble component to have an evolutionary process
	that causes the systematic divergence of actual prices from their fundamental
	values. According to these authors, the correlation between the relative size of
	the bubble and the asset returns in the next period is positive.
	\citep{vanNorden1999} show that deviations from the fundamental price have
	significant predictive power for the distribution of stock returns, exhibiting a
	highly significant but nonlinear relationship between the bubbles and returns.
	Hence, the size of the bubble reflects behaviour of the asset returns.
	\citep{vanNorden1999} conclude that the degree of apparent overvaluation
	influences expected returns but at a much smaller magnitude in the simulations
	than in the actual data.
	
	An estimate of the cross-correlation is calculated as follows \citep{Box2015}:
	\begin{equation}
		\label{eq:cross-correlation}
		r_{xy}(k)=\frac{c_{xy}(k)}{s_{x}s_{y}}; k=0, \pm{1}, \pm{2},...
	\end{equation}
	where the sample cross-covariance function is an estimate of the
	covariance between the time-series of the relative size of the bubbles,
	$x$, and log-return, $y$, at lags k = 0, $\pm{1}$, $\pm{2}$,...
	
	For data pairs ($x_{1}$,$y_{1}$),
	($x_{2}$,$y_{2}$),\ldots,($x_{n}$,$y_{n}$), an estimate $c_{xy}(k)$ of the
	cross-covariance coefficient at lag $k$ is provided by
	\begin{equation}
		c_{xy}(k)=\begin{cases}
			\frac{1}{n}\sum^{n-k}_{t=1}{(x_{t}-\overline{x})(y_{t+k}-\overline{y})}
			\text{, } k=0,1,2,\ldots\\
			\frac{1}{n}\sum^{n+k}_{t=1}{(y_{t}-\overline{y})(x_{t-k}-\overline{x})}
			\text{, } k=0,-1,-2,\ldots\\
		\end{cases}
	\end{equation}
	where $\overline{x}$ and $\overline{y}$ are the sample means of the $x_{t}$
	series and $y_{t}$ series, respectively.
	
	The sample standard deviations of the series are:
	\begin{equation}
		s_{x}=\sqrt{c_{xx}(0)}, \text{ where } c_{xx}(0)=Var(x).
	\end{equation}
	\begin{equation}
		s_{y}=\sqrt{c_{yy}(0)}, \text{ where } c_{yy}(0)=Var(y).
	\end{equation}
	\label{subsubsec:Bubbles and Crashes SF2}

	\subsubsection{Heavy Tails of Return Distribution} 
	\label{subsubsec:Heavy Tails of Return Distribution SF} 
	
	According to the literature, the unconditional empirical distribution of
	log-return is leptokurtic and belongs to the class of so-called heavy-tailed
	distributions, with the tails of the distribution of log-return, $r_{t}$,
	following approximately a power-law, with a tail index which is finite, usually
	higher than two and less than five but, nonetheless, the precise form of the
	tails is difficult to determine \citep{{Jansen1991}, {Cont2001},
		{Lux2002b}, {Cont2007}}:
	\begin{equation}
		F(|r_{t}|>x) \approx cx^{-\alpha}
	\end{equation}
	
	Measuring the tail index of a distribution gives a measure of how heavy the tail
	is \citep{Cont2001}. The tail index $\alpha$ of a distribution may be defined
	as the order of the highest absolute moment which is finite. For a Gaussian or
	exponential tail with $\alpha=+\infty$, all moments are finite, while for a
	power-law distribution with exponent $\alpha$, the tail index is equal to
	$\alpha$. The higher the tail index, the greater the similarity of the tail
	with a Gaussian distribution.
	
	To calculate the left (right) tail index, the log-returns are first arranged in
	ascending (descending) order
	$X_{n}>X_{n-1}>\text{\ldots}>X_{n-k}>\text{\ldots}>X_{1}$, where $k$ denotes the
	number of observations located in the respective tail of the distribution. We
	estimate the left and right tail index using the Hill estimator \citep{hill1975} that has become a standard tool for estimation of the tail index \citep{{Lux2002a},
		{danielsson2016}}:
	\begin{equation}
		\label{eq:Hill estimator}
		\hat\alpha=\Bigg[\frac{1}{k}\sum^{k-1}_{i=0}{log|(X_{n-i,n})|-log|(X_{n-k,n})|}\Bigg]^{-1}.
	\end{equation}
	
	The lower the Hill estimator, the lower the stability of the financial market,
	since more extraordinary events, including losses, occur.
	
	\subsubsection{Conditional Heavy Tails}
	\label{subsubsec:Conditional Heavy Tails SF}
	
	Even after correcting log-returns for volatility clustering via a generalised
	autoregressive conditional heteroscedastic (GARCH) model, the residual
	time-series from an estimated GARCH model will still exhibit a non-Gaussian,
	leptokurtic distribution and heavy tails \citep{{Pagan1996}, {Cont2001}}. 
	%A GARCH model is an extension of the original Engle's ARCH model for variance
	%heteroscedasticity, and 
	If a series exhibits volatility clustering, this
	suggests that past variances might be predictive of the current variance.
	\label{subsubsec:Conditional Heavy Tails SF2}
	We estimate the parameters of a conditional specification, $z_{t}$, which is an
	independent and identically distributed standardised Gaussian process. The
	estimation process infers the innovations from the returns, $\epsilon_t$, and
	gives the corresponding conditional standard deviations, $\sigma_t$:
	\begin{equation}
		\label{eq:conditional heavy tails}
		z_t=\frac{\epsilon_t}{\sigma_t}
	\end{equation}

	\subsubsection{Gain/Loss Asymmetry}
	\label{subsubsec:Gain/Loss Asymmetry SF}
	
	\citep{Cont2001} states that one observes large drawdowns in prices but not
	equally large upward movements. According to \citep{Johansen2000}, these
	drawdowns, defined as the loss from the last maximum within some time horizon
	(local maximum) to the next minimum within some time horizon (local minimum),
	offer a more natural measure of real market risks than the variance, VaR or
	other measures based on fixed time scale distributions of returns.
	
	\citep{Donangelo2006} show that the market as a whole, as monitored by the Dow
	Jones Industrial Average (DJIA), exhibits a fundamental gain-loss asymmetry.
	However, a similar asymmetry is not found for any of the individual stocks that
	compose the DJIA. Other indices, such as S\&P 500 and NASDAQ, also show this
	asymmetry, while, for instance, foreign exchange data do not.
	\citep{Donangelo2006} and \citep{Johansen2006} conclude that an asymmetry between
	gains and losses is not found for individual stocks but only for indices.
	
	Inverse statistics, as introduced in econophysics, determines the distribution
	of waiting times for a given, asset specific, return level. In the context of
	economics, it was recently suggested, partly inspired by earlier work in
	turbulence, that as an alternative the distribution of waiting times needed to
	reach a fixed level of return should be studied. These waiting times were termed
	investment horizons, and the corresponding distributions the investment horizon
	distributions. Furthermore, it was shown that for positive levels of return the
	distributions of investment horizons had a well-defined maximum followed by a
	power-law tail scaling. The maximum of this distribution signifies the optimal
	investment horizon for an investor aiming for a given return
	\citep{Jensen2003}. Therefore, what is the smallest time interval needed for an
	asset to cross a fixed return level, $\rho$?
	
	Given a fixed log-return barrier, $\rho$, of a stock, the corresponding time
	span is estimated for which the log-return of the stock or index for the first
	time reaches the level $\rho$. This can also be called the first passage time
	through the level, or barrier, $\rho$. As the investment date runs through the
	past price history of the stock, the accumulated values of the first
	passage times form the probability distribution function of the investment
	horizons for the smallest time period needed in the past to produce a log-return
	of at least magnitude $\rho$. The maximum of this distribution determines the
	most probable investment horizon which therefore is the optimal investment
	horizon for that given stock.
	
	As the empirical logarithmic stock price process is known not to be Brownian, we
	used a generalised Gamma distribution, as in \citep{Simonsen2002}, of the
	form:
	\begin{equation}
		p_{t}=\frac{\nu}{\Gamma(\frac{\alpha}{\nu})}\frac{\beta^{2\alpha}}{(t-t_{0})^{\alpha+1}}exp\{-(\frac{\beta^{2}}{t+t_{0}})^{\nu}
		\}
	\end{equation}
	The investment horizon, $\tau_{\rho}(t)$, at time $t$, for a return level $\rho$ is defined at
	the smallest time interval, $\Delta{t}$, that satisfies the relation
	$r_{\Delta{t}}\geq\rho$, or in mathematical terms:
	$\tau_{\rho}(t)=inf{\{\Delta{t}>0 \vert r_{\Delta{t}}\geq\rho}\}$.
	
	\subsubsection{Equity Premium Puzzle}
	\label{subsubsec:Equity Premium Puzzle SF}
	
	According to \citep{Mehra1985} and \citep{Kocherlakota1996}, the equity premium
	puzzle consists in a historical (period from 1889 to 1978) average return on
	equity (average real annual yield on the S\&P 500 Index was nearly 7 percent)
	that exceeds the average return on risk-free asset (average yield on short-term debt
	was 0.8 percent). \citep{Mehra2003} shows that stocks and bonds pay off in
	approximately the same states of nature or economic scenarios, and hence they
	should command approximately the same rate of return, or, on average, should
	command, at most, a 1 percentage point return premium over bills.
	However, empirical data shows that the mean premium on stocks over bills is
	considerably and consistently higher. This puzzle underscores the inability of
	standard paradigms of financial economics to explain the magnitude of the risk
	premium.
	
	\subsubsection{Excess Volatility}
	\label{subsubsec:Excess Volatility SF}
	
	\citep{shiller1989} defines excess volatility as the difference between an over
	large variability of price movements given the relatively low variability of
	fundamentals. \citep{Shiller1981} and \citep{LeRoy1981} show that there is
	evidence of excess price volatility, particularly in the stock market. The
	volatility of the news arrival process is quantified by $\sigma(r^{f})$, which
	is the standard deviation of the fundamental log-return, whereas the volatility
	of the market can be measured a posteriori as the standard deviation of
	log-return, $\sigma(r)$. The order of magnitude of the volatility of log-return
	may be quite different from that of the input noise representing news arrivals
	reflected in the fundamental value, expressed by the inequality
	$\sigma(r)>\sigma(r^{f})$ \citep{{Shiller1981}, {LeRoy1981},
		{Westerhoff2003}, {Cont2007}, {Zhu2009}}.
	
	\citep{Cont2007} states that it is difficult to justify the volatility in asset
	log-return by variations in fundamental economic variables. Hence, the
	volatility of the arrival of new information on the market cannot explain
	returns volatility.
	
	\subsubsection{Leverage Effect}
	\label{Leverage Effect-LF SF}
	
	Leverage is defined as the correlation, with time lag $\tau$, between future
	volatility and past return of an asset \citep{{Bouchaud2001}, {Cont2001}}.
	Most measures of volatility of an asset are negatively correlated with the returns of that asset \citep{Ahlgren2007}.
	
	The so-called leverage effect, or volatility asymmetry, shows that the amplitude
	of relative price fluctuations, or volatility, of a stock tends to increase when
	its price drops, reflecting a negative volatility-return correlation. The
	correlation of returns with subsequent squared returns is defined by
	\begin{equation}
		\label{eq:leverage effect}
		L(\tau)=\text{corr}(r(t,\Delta{t}),(|r(t+\tau),\Delta{t})|^2)
	\end{equation}
	and it starts from a negative value and decays to zero,
	suggesting that negative returns lead to a rise in volatility.
	However, this effect is asymmetric $L(\tau)=L(-\tau)$ and in
	general $L(\tau)$ is negligible for $\tau<0$ \citep{Cont2001}.
	
	\subsubsection{Linear Autocorrelation}
	%\label{subsubsec:Linear Autocorrelation SF}
	
	Price movements in liquid markets do not exhibit
	any significant autocorrelation and the autocorrelation function
	of the price changes is given by the following equation:
	\begin{equation}
		\label{eq:autocorrelation}
		C(\tau)=\text{corr}(r(t,\Delta{t}),r(t + \tau,\Delta{t}))
	\end{equation}
	and rapidly decays to zero \citep{Cont2001}.
	\label{subsubsec:Linear Autocorrelation SF}
	
	The autocorrelation function measures the correlation between $z_{t}$ and
	$z_{t+k}$, where $k=0,\ldots,K$ and $z_{t}$ is a stochastic process.
	Using the same approach as \citep{Box2015}, the estimate of the $k^{th}$ lag
	autocorrelation $\rho_{k}$ is
	\begin{equation}
		\label{eq:autocorrelation}
		r_{k}=\widehat{\rho}_{k}=\frac{c_{k}}{c_{0}}
	\end{equation}
	where $c_{0}$ is the sample variance of the time-series and
	\begin{equation}
		c_{k}=\widehat{\gamma}_{k}=\frac{1}{N}\sum^{N-k}_{t=1}(z_{t}-\overline{z})(z_{t+k}-\overline{z})
		\quad k=0,\ldots,K
	\end{equation}
	is the estimate of the autocovariance $\gamma_{k}$, $\overline{z}$ is the
	sample mean of the time-series and the values $r_{k}$ in
	equation \ref{eq:autocorrelation} may be called the \textit{sample}
	autocorrelation function.
	
	According to \citep{Fama1970} and \citep{Fama1991}, there is no evidence of
	substantial linear dependence between lagged price changes or returns, and this
	is often cited as support for the ``efficient market hypothesis''. Empirical
	data show that the absence of autocorrelation does not seem to hold
	systematically when the time scale $\Delta{t}$ is increased: weekly and monthly
	returns do however exhibit some autocorrelation. \citep{Fama1965} findings show
	that the first-order (lag 1) autocorrelations of daily returns are positive for
	twenty-two out of thirty DJIA stocks. \citep{Fisher1966}
	results suggest that the autocorrelations of monthly returns on three indices
	(Combination Price Index (0.19), Standard \& Poor's Composite Index (0.11),  and
	the Dow-Jones Industrial Average (0.09)) exhibit positive first serial
	correlation coefficients. \citep{Lo1988} find significant positive first-order
	autocorrelation for weekly holding-period returns. Monthly holding-period
	returns also exhibit significant positive serial correlation. However, given
	that the sizes of the data sets are inversely proportional to $\Delta{t}$ in
	equation \ref{eq:autocorrelation} the statistical evidence is less conclusive
	and more variable from sample to sample \citep{Cont2001}.
	\label{subsubsec:Linear Autocorrelation SF2}
	
	\subsubsection{Long Memory}
	\label{Long Memory SF}
	
	The estimate of the autocorrelation function follows the methodology described
	in \ref{subsubsec:Linear Autocorrelation SF}. Long memory is defined as the
	autocorrelation function of absolute returns, which decays as a function of the
	time lag:
	\begin{equation}
		\label{eq:long memory}
		C(\tau)=\text{corr}((|r(t+\tau),\Delta{t})|,|r(t,\Delta{t})|)
	\end{equation}
	
	\subsubsection{Power Law Behaviour of Returns}
	%\label{subsubsec:Power Law Behaviour of Returns SF}
	
	Mathematically, a quantity $x$ obeys a power law if it is drawn from a
	probability distribution \citep{Clauset2009}
	\begin{equation}
		\label{eq:power law}
		% P(|r_{t}|>x) \sim x^{-\zeta_{r}}
		p(x) \sim x^{-\zeta_{r}}
	\end{equation}
	where $\zeta_{r}$ is a constant parameter of the distribution known as the
	exponent or scaling parameter. Power-law distributions of returns are continuous
	distributions and let $x$ represent the quantity in whose distribution we are
	interested. The probability that a return has an absolute value larger than $x$
	is found to be a continuous power-law distribution empirically described by a
	probability density $p(x)$ such that \citep{{Gabaix2003}, {Clauset2009}}
	\begin{equation}
		\label{eq:Continuous power-law distribution}
		p(x)dx=Pr(x\leq|r|<x+dx)=Cx^{-\zeta_{r}}dx,
	\end{equation} 
	where $r$ is the observed returns and C is a normalisation constant. In
	practice, few empirical phenomena obey power laws for all values of $x$, and
	this density diverges as $x \rightarrow 0$, so equation \ref{eq:Continuous
		power-law distribution} cannot hold for all $x \geq 0$. In such cases a
	power-law applies only for values greater than some minimum $x_{min}$, a lower
	bound to the power-law behaviour, and the tail of the distribution follows a
	power-law distribution \citep{Clauset2009}.
	\label{subsubsec:Power Law Behaviour of Returns SF}
	
	\citep{Clauset2009} define the basic functional form, $f(x)=x^{-\zeta}$, and the
	appropriate normalisation constant $C$ such that ${\textstyle
		\int^{\infty}_{x_{min}} Cf(x)dx=1}$ for the continuous case. 
	The scaling parameter typically lies in the range $2<\zeta<3$,
	although there are occasional exceptions \citep{Clauset2009}. The probability
	that a return has an absolute value larger than $x$ is found empirically to be
	expressed as in equation \ref{eq:power law}, with $\zeta_{r}\approx3$
	\citep{Gabaix2003}. The inverse cubic law distribution of returns
	represented in equation \ref{eq:power law} is considered universal,
	regardless of stock markets, tick size, sizes of stocks, time periods, and also
	applies to different stock market indices \citep{{Gabaix2003},
		{Vijayaraghavan2011}}. 
	
	We find the correct fitting of the power-law to synthetic distribution of returns by estimating the parameters of a power-law distribution\footnote{The software used can be found at \url{http://www.santafe.edu/~aaronc/powerlaws/}. Archived at:\url{https://web.archive.org/web/20211219074932/https://aaronclauset.github.io/powerlaws/}}. Firstly, we estimate $\zeta_{r}$ which requires a value for the lower bound, $x_{min}$. The estimate $\hat{x}_{min}$ is the value of ${x}_{min}$ that minimises the distance between the CDFs of the data and the fitted model:
	\begin{equation}
		\label{eq:estimation minimum x}
		D=\displaystyle \max_{x \geq x_{min}} |S(x)-P(x)|
	\end{equation}
	where $S(x)$ is the CDF of the data for the observations with value at least
	$x_{min}$, and $P(x)$ is the CDF for the power-law model that best fits the data
	in the region $x \geq x_{min}$.
	
	\citep{Clauset2009} use the method of maximum likelihood for
	fitting parameterised models such as power-law distributions to observed data. 
	Assuming that the synthetic data is drawn from a distribution that follows a
	power-law for $x \geq x_{min}$, the maximum likelihood estimator (MLE) for the
	continuous case is \citep{Clauset2009}
	\begin{equation}
		\hat\zeta=1+n\Bigg[\sum^{n}_{i=1}{ln \frac{x_{i}}{x_{min}}}\Bigg]^{-1}
	\end{equation}
	where $x_{i}$, $i=1,\ldots,n$, are the observed values of $x$ such that $x_{i}
	\geq x_{min}$.
	
	After fitting a power-law distribution to our model generated data and finding
	estimates of the parameters $\zeta$ and $x_{min}$, we should know whether the
	power-law is a plausible fit with the data. The approach used by
	\citep{Clauset2009}, and replicated here, is based on the Kolmogorov-Smirnov
	statistic and is used to sample many synthetic data sets from a true power-law
	distribution, measure how far they fluctuate from the power-law form, and
	compare the results with similar measurements on data from our simulations. If
	the data from our simulations varies greatly from the power-law form
	than the typical synthetic one, then the power-law is not a plausible fit with
	the data.
	
	The goodness-of-fit test generates a p-value that quantifies the plausibility of
	the data is drawn from a power-law distribution. The p-value is defined to be
	the fraction of synthetic distances that are larger than the simulation
	distance. If the p-value is large, then the difference between the empirical
	data and the model can be attributed to statistical fluctuations alone; if it is
	small, the model does not provide a plausible fit with the data.
	
	Finally, to quantify the uncertainty in our estimates for $\zeta$ and $x_{min}$
	we use the method of \citep{Clauset2009} to generate a synthetic data
	set with a similar distribution to the original by drawing a new sequence of points
	$x_{i}$, $i=1,\ldots, n$, uniformly at random, from the original data (with
	replacement). Then $x_{min}$ and $\zeta$ are estimated again. By taking the
	standard deviation of these estimates over a large number of repetitions of this
	process, principled estimates of the uncertainty in the original estimated
	parameters can be derived.
	\label{subsubsec:Power Law Behaviour of Returns SF2}
	
	\subsubsection{Power Law Behaviour of Volatility}
	%\label{subsubsec:Power Law Behaviour of Volatility SF}
	
	The same methodology implemented in \ref{subsubsec:Power Law Behaviour of
		Returns SF} was used to study the power law behaviour of volatility. According
	to \citep{Liu1999}, the cumulative distribution of volatility is consistent with
	the following power-law asymptotic behaviour:
	\begin{equation}
		\label{eq:power law volatility}
		P(v_{t}>x) \sim x^{-\zeta_{\sigma}}
	\end{equation}
	
	The volatility is often estimated by calculating the standard deviation of the
	price changes in an appropriate time window. However, one can also use other
	ways of estimating it. We follow the approach used by \citep{Liu1999} and
	estimate volatility as the local average of absolute price change over a
	suitable time window $T=n\Delta{t}$:
	\begin{equation}
		\label{eq:volatility}
		v_{T}(t)\equiv\frac{1}{n}\sum_{t^{'}=t}^{t+n-1}|G(t^{'})|,
	\end{equation}
	where $n$ is an integer and the price change $G(t)$ is defined as the change in
	the logarithm of the price $Z$:
	\begin{equation}
		\label{eq:price change}
		G(t) \equiv \log{Z(t+\Delta(t))}-\log{Z(t)}
	\end{equation}
	
	There are two parameters in this definition of volatility, $\Delta(t)$ and $n$.
	The parameter $\Delta(t)$ represents the sampling time interval for the data and
	the parameter $n$ the moving average window size. 
	
	\subsubsection{Volatility Clustering}
	\label{subsubsec:Volatility Clustering SF}
	
	The absence of autocorrelations in returns, as previously analysed in \ref{subsubsec:Linear Autocorrelation SF}, gave some empirical support for
	random walk models of prices in which the returns are considered to be independent
	random variables. However, the absence of serial correlation does not imply the
	independence of the increments: independence implies that any nonlinear function
	of returns will also have no autocorrelation \citep{Cont2001}.
	
	Different measures of volatility display a positive autocorrelation over several
	days, which quantifies the fact that high-volatility events tend to cluster in
	time \citep{{Cont2001}, {Russell2010}}. One finds almost no autocorrelation
	for raw returns, but simple nonlinear functions of returns, such as absolute or
	squared returns, exhibit significant and persistence positive autocorrelation --
	periods of quiescence and turbulence tend to cluster together. This
	autocorrelation function remains positive and decays slowly providing
	quantitative evidence of volatility clustering. We use the following
	autocorrelation function of the squared returns, which is classically used to
	measure volatility clustering:
	\begin{equation}
		\label{eq:volatility clustering}
		C_{(\tau)}=\text{corr}((r(t+\tau),\Delta(t))^{2},r(t,\Delta(t))^{2})
	\end{equation}
	
	Volatility clustering is one of the most important stylised facts in financial
	time-series data. Whereas price changes themselves appear to be unpredictable,
	the magnitude of those changes, as measured, for example, by absolute or
	squared returns, appears to be partially predictable in the sense that large
	changes tend to be followed by large changes -- of either sign -- and small changes tend to
	be followed by small changes. 
	
	Asset price fluctuations are thus characterised by episodes of low volatility,
	with small price changes, irregularly interchanged with episodes of high
	volatility, with large price changes. Volatility clustering has been shown to be
	present in a wide variety of financial assets including stocks, market indices,
	exchange rates, and interest rate securities \citep{Gaunersdorfer2007}. These
	authors present the clustered arrival of random ``news'' about economic
	fundamentals as an explanation for the existence of volatility clustering, which
	contradicts \citep{Cont2007} who maintains the difficulty of justifying
	volatility in returns by variations in fundamental economic variables.
	
	\subsubsection{Volatility Volume Correlations}
	\label{subsubsec:Volatility Volume Correlations SF}
	
	According to \citep{Cont2007} trading volume is positively correlated with market
	volatility, and trading volume and volatility show the same type of ``long
	memory'' behaviour \citep{Lobato2000}. \citep{Goodhart1997} observe that when
	more information is revealed then asset prices are more volatile. Greater market
	depth and liquidity is often associated to ``good news'' and more information,
	and, on this basis, one can explain a positive correlation between volume and
	volatility.
	
	A methodology identical to the one implemented in \ref{subsubsec:Bubbles and Crashes SF} was used. \citep{Brock1996} observe that
	the crosscorrelation function of absolute returns is approximately zero with
	past and future volumes but is positive for absolute returns with current
	volumes. \citep{Engle2000} also observes that larger volume predicts rising
	volatility.
	
	\subsubsection{Unit Roots}
	\label{subsubsec:Unit Roots SF}
	
	The unit-root hypothesis of the financial data is another well-established
	stylised fact of financial markets \citep{{deVries1994}, {Alfarano2007},
		{alexander2008}}. Hence, we will investigate if the model generated
	time-series of log-return are stationary.
	We apply two different unit root tests using low and high-frequency returns:
	the augmented Dickey- Fuller (ADF) test and the Phillips-Perron (PP) test. 
	
	The
	PP test differs from the ADF test mainly in how it deals with serial
	correlation and heteroskedasticity in the errors. The PP test allows errors to be dependent
	with heteroscedastic variance and ignores any serial correlation in the test
	regression, while the ADF test uses a parametric autoregression to approximate
	the ARMA structure of the errors in the test regression. Since returns on
	financial assets often have conditional heteroscedasticity, the PP test has
	become popular and is generally favoured in the analysis of financial
	time-series \citep{{zivot2007}, {alexander2008}}.
	The PP test assess the null hypothesis of a unit root in a time-series of
	returns, $r$. The test uses the model:
	\begin{equation}
		r_{t}=c+ \delta t+ar_{t-1}+e(t)
	\end{equation}
	where $c$ is the drift, $\delta$ is the deterministic trend, and $a$ is the
	autoregressive coefficient.
	The null hypothesis restricts $a=1$. The PP unit root test has stationarity in
	the alternative hypothesis. The test uses modified Dickey-Fuller statistics to
	account for serial correlations in the innovations process $e(t)$.
	
	Additionally, we perform a stationarity test: the KPSS (Kwiatkowski, Phillips,
	Schmidt and Shin) test. Unit root tests cannot distinguish highly persistent
	stationary processes from nonstationary processes clearly and the ADF and PP
	tests have very low power against I(0) alternatives that are close to being
	I(1) \citep{zivot2007}.
	The KPSS test assesses the null hypothesis that a univariate time-series is
	stationary against the alternative that it is a nonstationary unit root process.
	The test uses the structural model:
	\begin{equation}
		y_{t}=c_{t}+ \delta t+u_{1t}
	\end{equation}
	and
	\begin{equation}
		c_{t}=c_{t-1}+u_{2t},
	\end{equation}
	where $c$ is the random walk term, $\delta$ is the trend coefficient, $u_{1t}$
	is a stationary process and $u_{2t}$ is an independent and identically distributed process with mean 0
	and variance $\sigma^{2}$.
	The KPSS test of the null hypothesis against the alternative reverses the
	strategy of the unit root tests:
	\begin{equation}
		H_{0}: \sigma^{2}=0 \text{ vs } H_{1}:\sigma^{2}>0
	\end{equation}
	where the null hypothesis implies that $c_{t}$ is constant and acts as the model
	intercept, and $\sigma^{2}>0$ introduces the unit root.
	
	\subsection{Trading Volume}
	\label{subsection:Trading Volume SF}
	
	While the inverse cubic law distribution of price returns, mentioned above in \ref{subsubsec:Power Law Behaviour of Returns SF}, seems to be
	universal, there is less consensus as to the universality of, for example, the
	distributions of trading volume and whether the volume distribution is
	L\'{e}vy-stable \citep{Vijayaraghavan2011}.
	\label{subsection:Trading Volume SF1}
	
	\subsubsection{Power Law Behaviour of Trading Volume}
	\label{Power Law Behaviour of Trading Volume SF}
	
	The universality of the distributions for volatility, trading volume and number
	of trades is of interest because it may help in understanding the statistical
	relationship between returns and market activity. However, the estimation
	of the tail exponent is a delicate matter, and the universality of these
	distributions is not consensual.
	
	According to \citep{Gabaix2003} some empirical studies show that the distribution
	of trading volume, $V_{t}$, obeys a power law:
	\begin{equation}
		\label{eq: power law behaviour of trading volume}
		P(V_{t}>x) \sim x^{-\zeta_{V}}
	\end{equation}
	with $\zeta_{V} \approx 1.5$.
	These authors also tested the universality of equation \ref{eq: power law behaviour of trading volume} by analysing stocks data from the Paris Bourse over
	the period 1994--1999 and data from the US stock market. \citep{Gabaix2003}
	conclude that equation \ref{eq: power law behaviour of trading volume} holds
	for both markets, consistent with the possibility of universality.
	However, some authors rule out the claim of universality and the possibility
	that this distribution could be L\'{e}vy-stable after studying other markets, such
	as the Korean \citep{Hong2009}, the
	Chinese \citep{Qiu2009}, and the Indian \citep{Vijayaraghavan2011}. 
	
	Some authors \citep{{Gopikrishnan2000}, {Plerou2001}} analyse the statistics
	of the number of shares traded in a time interval $\Delta{t}$ and conclude that
	the probability distribution in equation \ref{eq: power law behaviour of
		trading volume} has a tail that decays as a power-law with an exponent within
	the L\'{e}vy-stable domain $0<\zeta_{V}<2$, where $\zeta_{V}$ has the average
	value $\zeta_{V}=1.7\pm0.1$.
	
	Other authors \citep{Eisler2006} hold that the shape of this distribution and
	the explanation of the exponent in terms of the inverse cubic law of stock
	returns are much debatable. These authors found a significantly higher exponent,
	around 2.2, for the same data set (in most cases greater than 2) and concluded
	that the distribution of traded volume in fixed time windows is not
	L\'{e}vy-stable.
	
	\subsubsection{Long Memory of Volume}
	\label{Long Memory of Volume SF}
	
	Long memory is a form of extreme persistence in a time-series. Trading volume
	time-series are highly persistent and exhibit autocorrelations that decay slowly
	as one moves to longer lags \citep{{Bollerslev1999}, {Lobato2000},
		{LeBaron2008}, {Fleming2011}}. \citep{Russell2010} identify long sets
	of positive autocorrelation for the log of volume spanning many transactions.
	\label{subsection:Trading Volume SF2} 
	
	\subsection{Trading Duration}
	\label{subsection:Trading Duration SF}
	
	High-frequency data are irregularly time-spaced and could be statistically
	interpreted as point processes. Duration is commonly defined as the time
	interval between consecutive events, e.g. the time spells between financial
	transactions. The duration between two consecutive transactions in finance is
	important, for it may signal the arrival of new information, and is inversely
	related to trading intensity, which in turn depends on the arrival of new
	information. Trading durations are associated with the behaviour of informed
	traders, since trading intensity reflects the existence of news. Hence, the
	dynamic behaviour of durations thus contains useful information about market
	activities. Long durations are likely to be associated with no news
	and lower volatility, while a cluster of short durations and high trading
	activity are an indication of the existence of new information and are
	associated with large quote revisions and strong autocorrelations of trades
	\citep{{Dufour2000b}, {Tsay2009}}. \citep{Aldridge2012} observes that the
	variation in duration between subsequent trades may also be due to low levels of
	liquidity, trading halts on exchanges or the strategic motivations of traders.
	
	Following \citep{Engle1998} description of a point process, we consider a
	stochastic process that is simply a sequence of times ${t_{0}, t_{1},\ldots, t_{n},\ldots}$
	with $t_{0}<t_{1}<\ldots<t_{n}<\ldots$. Simultaneous trades exist, equivalent to
	zero trade durations, but since the smallest time increment is the tick, orders
	executed within a single tick are aggregated \citep{Engle1998}. Hence, only the
	unique times are considered and consequently all zero durations are removed.
	This is consistent with interpreting a trade as a transfer of ownership from one
	or more sellers to one or more buyers at a point in time \citep{Engle2000}, and
	this procedure uses the microstructure argument that simultaneous observations
	correspond to split-transactions, i.e. large orders broken into smaller orders
	to facilitate faster execution \citep{Pacurar2008}.
	
	Let $N(t)$ represent the number of events that have occurred by time
	$t\in[0,T]$. Then, $t_{N(T)}=T$ is the last observed point of the sequence and
	$0=t_{0} \leq t_{1} \leq \ldots \leq t_{N(T)}=T $ corresponds to the observed
	point process. Let $t_{i}$ be the time at which the $i^{th}$ trade occurs and let
	$x_{i}=t_{i}-t_{i-1}$ denote the duration between trades \citep{{Engle2000},
		{Pacurar2008}}.
	
	\subsubsection{Clustering of Trade Duration}
	\label{subsubsec:Clustering of Trade Duration SF}
	
	Clustering of trade durations can be defined as long (short) durations that tend
	to be followed by long (short) durations. Duration clustering is theoretically
	attributable to the presence of either informed traders or liquidity traders
	\citep{Pacurar2008} and these phenomena may be due to new information arising
	in clusters.
	
	A quantity commonly used to measure the clustering of trade durations is the
	autocorrelation function of the squared transactions duration:
	\begin{equation}
		\label{eq: clustering trade durations}
		C_{(\tau)}=\text{corr}((x(t+\tau),\Delta(t))^{2},x(t,\Delta(t))^{2})
	\end{equation}
	
	\citep{Engle1998} studied the clustering of transactions in IBM transaction data
	and identify large autocorrelations in the time intervals between trades.
	The same authors observe that the clustering of transactions occurs both due to
	the bunching of informed traders and to the clustering of liquidity traders,
	when spreads are small. \citep{Engle1998} calculate the autocorrelations and partial
	autocorrelations in the waiting times between events and conclude that the
	autocorrelations and partial autocorrelations are far from zero and that all
	signs are positive. These authors examined the Ljung-Box statistic and concluded
	that the null hypothesis that the first 15 autocorrelations are 0 can be very
	easily rejected. The highly significant positive autocorrelations generally start at a
	low value and then decay slowly, indicating that persistence is an important
	issue when analysing trade durations \citep{Pacurar2008}. \citep{Engle1998}
	observe that long sets of positive autocorrelations are what one finds for
	autocorrelations of squared returns which show that volatility clustering and
	duration clustering exhibit similarities.
	
	\subsubsection{Long Memory of Trade Duration}
	\label{subsubsec:Long Memory of Trade Duration SF}
	
	Long memory reflects long run dependencies between transaction durations and it
	is a concept related to the clustering of trade duration. A measure of long
	memory of trade duration is the autocorrelation function of the duration of
	transactions defined by
	\begin{equation}
		\label{eq:long memory trade duration}
		C(\tau)=\text{corr}((x(t+\tau),\Delta{t}),x(t,\Delta{t}))
	\end{equation}
	
	\citep{Russell2010} observe that the transaction rates exhibit strong
	temporal dependence and the autocorrelations for the durations between trades
	exhibit long sets of positive autocorrelation spanning many transactions.
	A slowly decaying autocorrelation function may be associated with a long-memory
	process and evidence for long sets of positive autocorrelations for trade
	durations and long memory have been reported in the literature
	\citep{{Engle1998}, {Bauwens2004}}. \citep{Engle1998} identify similarities
	between the autocorrelation function of trade duration and the autocorrelations
	of squared returns.
	
	\subsubsection{Overdispersion}
	\label{subsubsec:Overdispersed SF}
	
	Overdispersion is defined as the ratio of standard deviation to mean
	\citep{Pacurar2008}. Some literature reports that trade durations are
	overdispersed \citep{{Engle1998}, {Dufour2000a}, {Bauwens2004}},
	i.e. the standard deviation is greater than the mean. However, some studies find
	underdispersion for some stocks \citep{{Ghysels1998}, {Bauwens2006}}.
	
	\subsection{Transaction Size}
	\label{subsection:Transaction Size SF}
	
	According to \citep{Dufour2000b}, trades in asymmetric information models convey
	information held by informed traders and observed in trading activity. Hence,
	market changes, namely change in prices, depend on the characteristics of
	trades, including the number of transactions. The importance of studying this power
	law resides in the fact that, at the aggregate daily level, the number of trades
	is the component of aggregate volume that best explains daily price volatility
	\citep{Dufour2000b}.
	
	\subsubsection{Power Law Behaviour of Trades}
	\label{subsection:Power Law Behaviour of Trades SF}
	
	There has been a long-running debate about whether the distributions of trading
	volume (v. equation \ref{eq: power law behaviour of trading volume}) and number
	of trades, occurring in a given time interval $\Delta{t}$, are universal
	\citep{Vijayaraghavan2011}. According to \citep{Gabaix2003}, the distribution of
	the number of trades, $N_{t}$, obeys a power law:
	\begin{equation}
		\label{eq: power law behaviour of number of trades}
		P(N_{t}>x) \sim x^{-\zeta_{N}}
	\end{equation}
	and is as universal as equations \ref{eq:power law} and \ref{eq: power law behaviour of trading volume}, with $\zeta_{N} \approx 3.4$. However,
	\citep{Vijayaraghavan2011} observe that the evidence for the invariance of
	this distribution seems less unequivocal.
	
	\citep{Plerou2001} show that $N_{\Delta t}$ contrasts with a
	Gaussian time-series and is inconsistent with Gaussian statistics, and
	displays an asymptotic power-law decay, with a mean value
	$\zeta_{N}=3.40\pm0.05$. This value of $\zeta>2$ is outside the L\'{e}vy-stable and
	is inconsistent with a stable distribution for $N_{t}$.
	
	\subsection{Bid-ask spread}
	\label{subsubsection:Bid-ask spread SF}
	
	The spread is defined as \citep{{Hasbrouck1991}, {tsay2002}}
	\begin{equation}
		\label{eq:bid-ask spread}
		s_{t}=p_{t}^{a}-p_{t}^{b}
	\end{equation}
	where $p_{t}^{a}$ is the ask price, $p_{t}^{b}$ is the bid price and the
	difference $p_{t}^{a}-p_{t}^{b}$ is call the bid-ask spread. Typically, the
	bid-ask spread is small in magnitude in relation to the stock price
	\citep{tsay2002}.
	
	\subsubsection{Spread Correlated with Price Change}
	\label{subsubsec:Spread Correlated with Price Change SF}
	
	\citep{Hasbrouck1991} observes that price impact is associated with wide spreads
	and \citep{Engle2000} concludes that higher bid-ask spreads predict rising
	volatility. \citep{Amihud1987} also observe that the greater the spread, the
	greater the close-to-close return variance. \citep{Amihud1986} suggest
	that higher-spread assets yield higher expected returns and that the
	market-observed average returns are an increasing function of the spread.
	Equally, \citep{Goodhart1997} observe that a positive association between
	volatility and the spread would normally be expected, with positive correlation
	between volatility and spread and the main direction of causality running from
	volatility to spread. Since greater volatility is associated with the revelation
	of more information and incorporates the `bounce' between bid and ask prices,
	then a higher spread will feedback into greater volatility. Also
	\citep{Bollerslev1994} observe that spreads rise as volatility
	increases, showing a strong positive relationship between volatility and
	spreads.
	
	\subsubsection{Thinness and Large Spread}
	\label{subsubsec:Thinness and Large Spread SF}
	
	\citep{McInish1992} observe that there is an inverse relationship between
	spreads and trading activity.
	\citep{Bessembinder1994} suggests that increased expected volume is likely to be
	associated with decreased spreads. \citep{Muranaga1996} also conclude that
	infrequently traded stocks are characterized by large bid-ask spreads.
	\citep{Goodhart1997} observe that a negative association between market depth
	and the spread would normally be expected. For example, \citep{Easley1996}
	observe that on the London Stock Exchange, spreads for the most active ``alpha''
	stocks average 1 percent, while the spreads for the least active ``delta''
	stocks average 11 percent. In an extreme case, in the absence of trading during
	an interval, potentially perceived by traders as ``bad news'', the spreads might be expected
	to subsequently worsen.
	\label{sec:Stylised Facts2}
	
	\section{Results and Model Validation}
	\label{sec:Results and Model Validation}
	
	In this section we use econometric properties to analyse the data generated from
	a relatively simple ABM (v. \ref{sec:Experimental Design1}--\ref{subsec:Model Constraints2}), and see whether they
	are able to display a number of previously described empirical features
	frequently observed in real financial time-series and identified in the
	literature (v. \ref{sec:Stylised Facts1}--\ref{sec:Stylised Facts2}).
	
	In the following subsections we investigate if our ABM is able to replicate for the first time a
	comprehensive list of stylised facts regarding returns, trading volume, trading
	duration, transaction size and bid-ask spread. We also investigate how low- or
	high-frequency data influences the distributional properties of the time-series
	and the replication of their statistical properties.
	
	The stylised facts that analyse seasonalities or depend on time (e.g. calendar
	effects, periodic effects, bursts, U shape, turn-of-the-year decline) cannot be
	investigated as ``real time'' seasonalities and intraday variations (e.g.
	opening day, closing day, lunch time, market behaviour in different time zones)
	and are not captured by our model. The analyses presented in subsequent sections
	of this chapter use data from unregulated simulations only, unless otherwise
	stated. When only one unregulated experimental treatment is presented it refers
	to that with initial conditions of the ES treatment.
	
	Finally, \citep{Cont2001} observes that the interpretation of autocorrelation
	functions for heavy-tailed time-series can be problematic and might not
	adequately depict the dependence structure of the nonlinear, non-Gaussian
	time-series due to the unreliability of the estimators of the autocorrelation function and the large confidence intervals associated with them. Hence, any
	conclusions regarding the autocorrelation function should be carefully drawn.
	All autocorrelation and cross-correlation over 100 simulations were computed by
	generating 100 independent realisations of our model, computing either the
	autocorrelation or cross-correlation function for each realisation, and then
	taking the average and/or the median of the autocorrelation or cross-correlation
	and respective upper and lower confidence bounds.
	
	On each of the boxplots in this chapter, the central red mark indicates the
	median, the blue dot indicates the mean, and the bottom and top edges of the box
	indicate the $25^{th}$ and $75^{th}$ percentiles, respectively. The whiskers
	extend to the most extreme data points not considered as outliers, and the
	outliers are plotted individually using the red `+' symbol.
	
	\subsection{Returns}
	\label{subsection:Returns}
	
	In the subsequent analysis of different stylised facts, $\Delta{t}$ in equation
	\ref{eq:returns} represents one tick. When analysing low-frequency data each
	tick represents one day and for high-frequency each tick represents one
	transaction event.
	
	\subsubsection{Moments of the Returns Distribution}
	\label{subsubsec:Moments of the Returns Distribution}
	
	The statistical properties under investigation include the first four moments of
	the returns distribution, in case the following moments exist: mean, standard
	deviation, skewness and kurtosis. If these moments of the returns distributions of the
	simulated data broadly match those found in the real data, it shows that the
	theoretical model is consistent with these stylised facts
	\citep{Cuthbertson2004}.
	
	Table \ref{table:Moments of the Returns Distribution} shows the first four
	moments of the daily log-return of the S\&P 500 computed from adjusted closing
	prices for both dividends and splits, compared with the closing prices of all
	investigated treatments: unregulated, VaR and ES.
	
	%\begin{table}[!t]
	%	\centering
	%	\begin{tabular}{????}
	%	\toprule
	% 	% first line
	%	\midrule
	%	% tale body	
	%	\bottomrule			
	%	\end{tabular}
	%	\caption{}
	%	\label{tab:????}	
	%\end{table}
	
	\begin{table}[!hbpt]
		\centering
		\begin{threeparttable}
			\caption{Moments of log-return distribution}
			\label{table:Moments of the Returns Distribution}
			\begin{tabular}{c c c c c c c}
				\toprule
				{} & {Mean} & {Standard} & {Minimum} & {Maximum} & {Skewness} & {Kurtosis}  \\
				{} & {} & {Deviation} & {} & {} & {} & {}\\
				\midrule
				Unregulated & $\convert{0.00002797794371587781}$ & 0.020 &
				-0.005 & 0.005 & -0.168  & 13.156  \\
				VaR & $\convert{0.0008248343224457544}$ & 0.024 &  -0.007 & 0.007 &
				-0.212 & 10.505 \\
				ES & $\convert{0.0005572359257501435}$  & 0.022 & -0.006 & 0.006 &
				0.005 & 9.980
				\\
				S\&P 500 & 0.067 & 0.166 & -0.229 & 0.110 & -1.008 & 28.924 \\
				\bottomrule
			\end{tabular}
			\begin{tablenotes}
				\scriptsize
				\item Note: The data describe the moments of the average daily log-return
				distribution over 100 simulations for each of the treatments (unregulated,
				VaR and ES) versus the S\&P 500 returns from 1967 to 2016. All prices are
				closing day prices and S\&P 500 prices are adjusted closing day prices for both dividends and splits. Mean is the annualised average log-return.
				Standard deviation is the annualised standard deviation log-return. S\&P data
				were downloaded from yahoo!finance website. All four moments are our own
				calculations.
			\end{tablenotes}
		\end{threeparttable}
	\end{table}
	
	%\begin{table}[!hbpt] \centering
	%	\begin{threeparttable}
	%		\caption{Moments of log-return distribution}
	%		\label{table:Moments of the Returns Distribution}
	%		\footnotesize
	%		\begin{tabular}{c c c c c c c} \hline\hline
	%			& Mean & St. Deviation & Minimum & Maximum & Skewness & Kurtosis  
	%			\\\hline
	%			Unregulated & $\convert{0.00002797794371587781}$ & 0.020 &
	%			-0.005 & 0.005 & -0.168  & 13.156  \\
	%			VaR & $\convert{0.0008248343224457544}$ & 0.024 &  -0.007 & 0.007 &
	%			-0.212 & 10.505 \\
	%			ES & $\convert{0.0005572359257501435}$  & 0.022 & -0.006 & 0.006 &
	%			0.005 & 9.980
	%			\\
	%			S\&P 500 & 0.067 & 0.166 & -0.229 & 0.110 & -1.008 & 28.924 \\\hline\hline 
	%		\end{tabular}
	%		\begin{tablenotes}
	%			\scriptsize
	%			\item Note: The data describe the moments of the average daily log-return
	%			distribution over 100 simulations for each of the treatments (unregulated,
	%			VaR and ES) versus the S\&P 500 returns from 1967 to 2016. All prices are
	%			closing day prices and S\&P 500 prices are adjusted closing day prices for both dividends and splits. Mean is the annualised average log-return.
	%			Standard deviation is the annualised standard deviation log-return. S\&P data
	%			were downloaded from yahoo!finance website. All four moments are our own
	%			calculations.
	%		\end{tablenotes}
	%	\end{threeparttable}
	%\end{table}
	
	The results obtained broadly resemble those obtained in other agent-based models \citep{{maymin2009}, {Feldman2017}}, and the S\&P 500, except in their
	magnitude. The standard deviation is much greater than the mean and the mean is
	positive and very small. The smaller magnitude of all the moments of the
	log-returns may be explained by the fact that the baseline treatments without
	leverage or short-selling are relatively stable. As in the S\&P 500, skewness is
	negative, except for the ES treatment (0.005), and kurtosis is greater than 3, which reflects a distribution more
	outlier-prone than the normal distribution.
	
	We can conclude then that our model is broadly consistent with the moments of
	the return distribution. However, other experimental treatments with leverage or short-selling that introduce more
	volatility into the model might present moments with greater magnitude than the
	one observed in the table.
	
	\subsubsection{Aggregational Gaussianity}
	\label{subsubsec:Aggregational Gaussianity}
	
\begin{figure}[!hbpt]
		\centering
		\begin{subfigure}[h]{0.4\textwidth}
			\includegraphics[width=\textwidth]{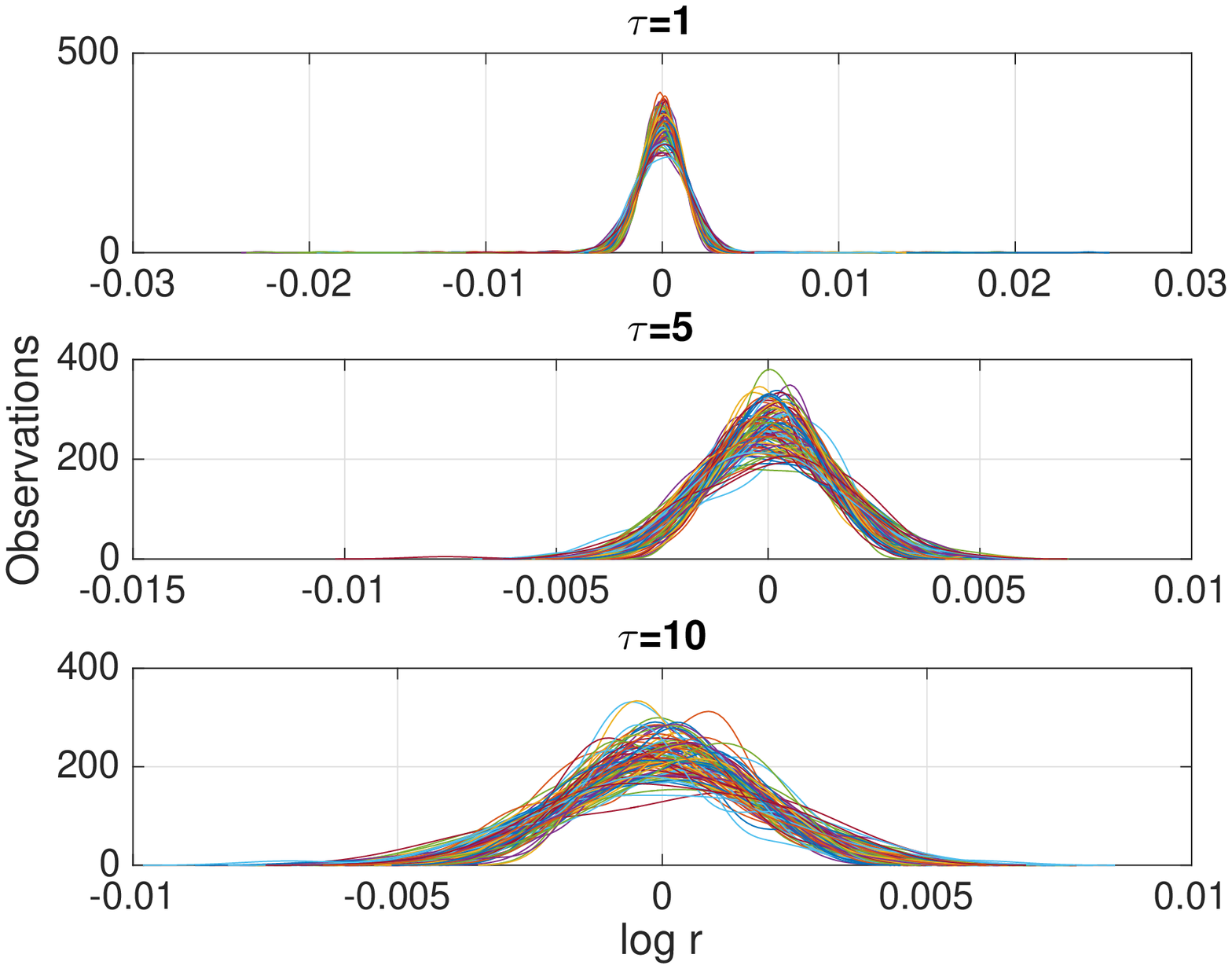}
			\footnotesize
			\caption{Low-frequency}
			\label{fig:Kernel estimator of the density of returns LF}
		\end{subfigure}
		\hspace{2cm}
		\begin{subfigure}[h]{0.4\textwidth}
			\includegraphics[width=\textwidth]{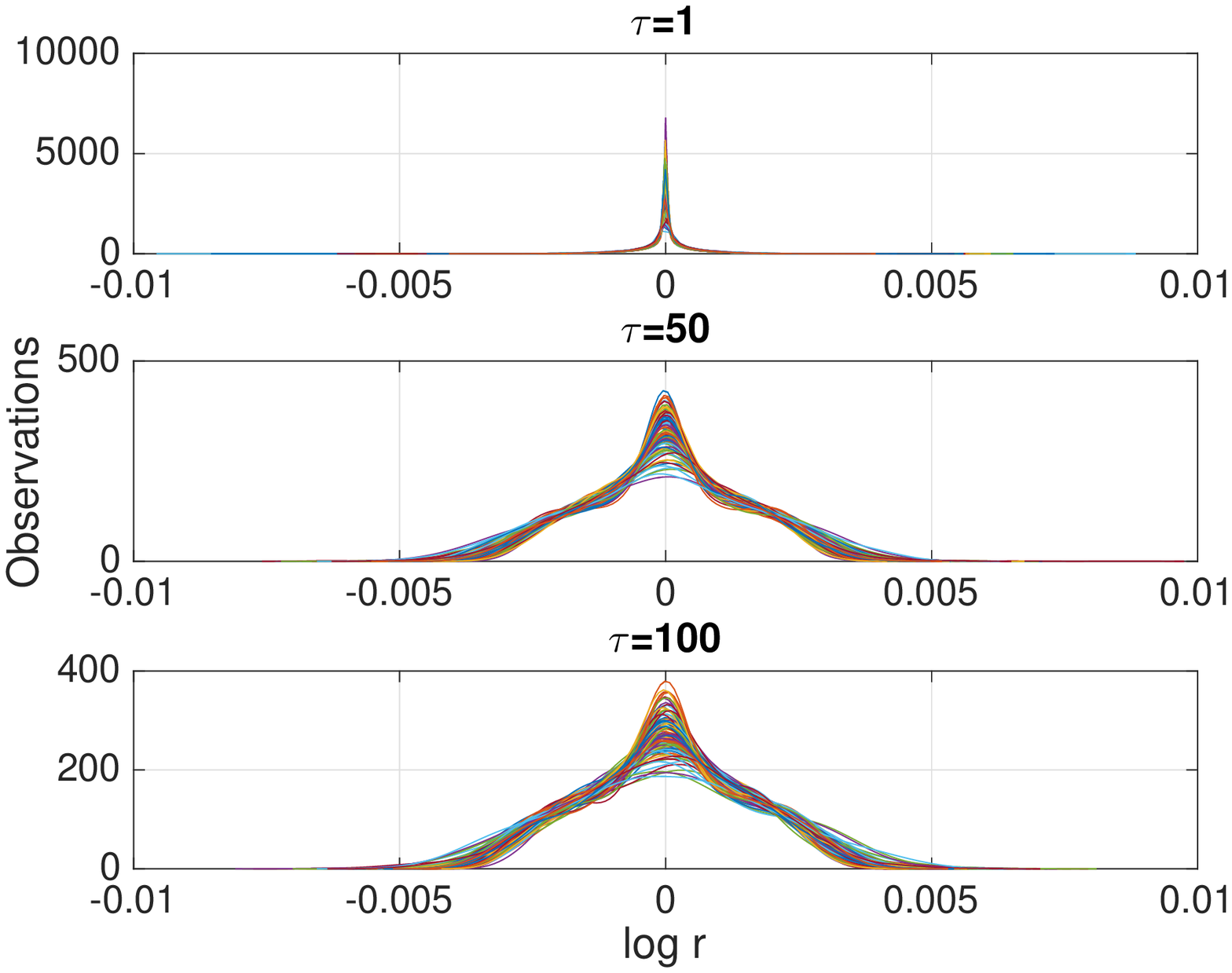}
			\footnotesize
			\caption{High-frequency}
			\label{fig:Kernel estimator of the density of returns HF}
		\end{subfigure}
		\floatfoot{\scriptsize{Note: $\tau$ represents the time scale, days for
				low-frequency data and transaction events for high-frequency data.}}
		\caption{Kernel estimator of the density of log-return for 100 simulations}
		\label{fig:Kernel estimator of the density of returns over 100 simulations.}
	\end{figure}

	Figures \ref{fig:Kernel estimator of the density of returns LF} and
	\ref{fig:Kernel estimator of the density of returns HF} show that the results
	from our model regarding the shape of the distribution are not the same at
	different time scales, which is consistent with the conclusions from the
	literature regarding empirical data.
	
	%\begin{figure}[!t]
	%\centering
	%\includegraphics[width=0.8/textwidth]{????}
	%\caption{}
	%\label{fig:????}
	%\end{figure}
	
\begin{figure}[!hbpt] \centering
		\begin{subfigure}[h]{0.4\textwidth}
			\includegraphics[width=\textwidth]{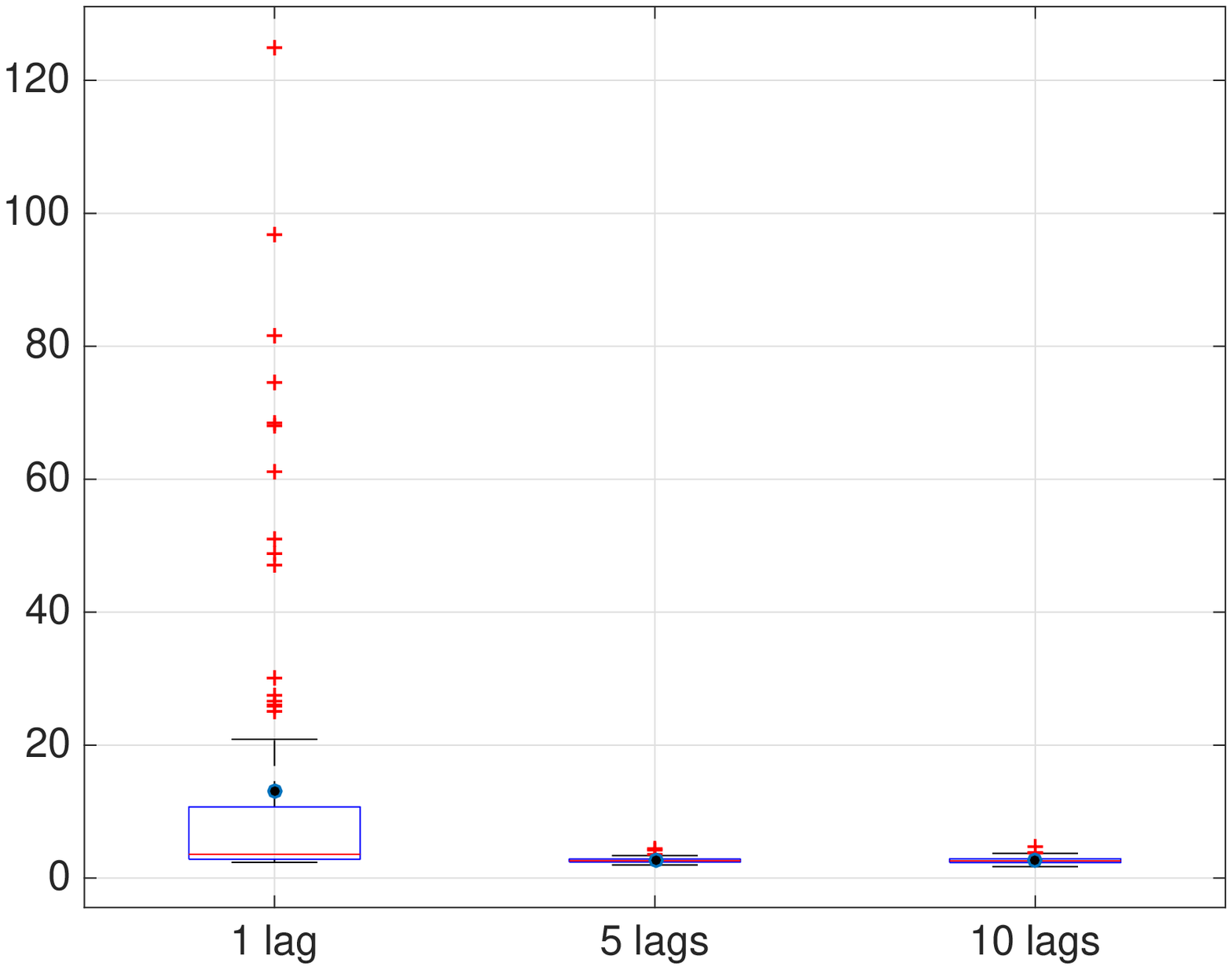}
			\footnotesize
			\caption{Low-frequency}
			\label{fig:Kurtosis of Returns LF}
		\end{subfigure}
		\hspace{2cm}
		\begin{subfigure}[h]{0.4\textwidth}
			\includegraphics[width=\textwidth]{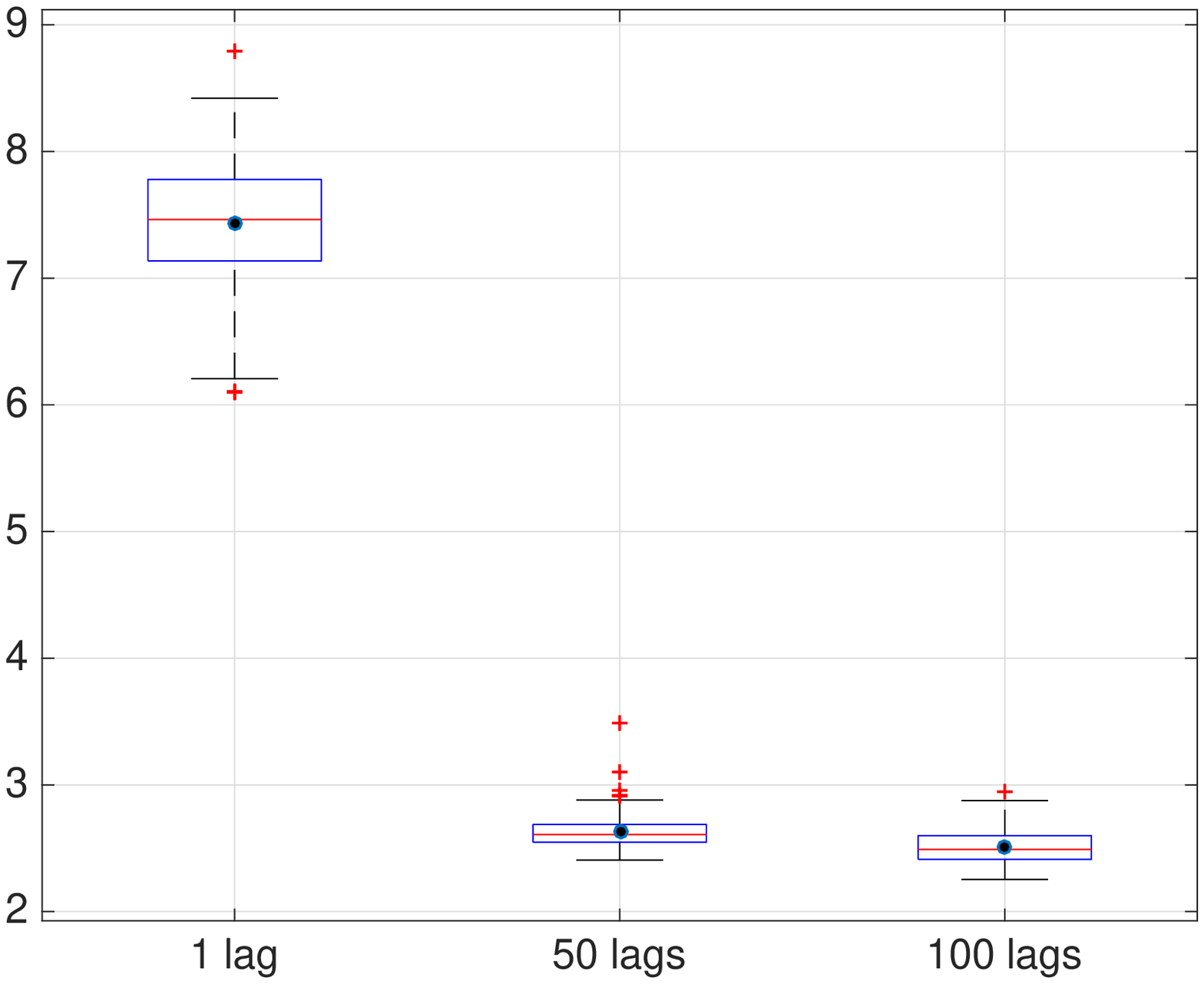}
			\footnotesize
			\caption{High-frequency}
			\label{fig:Kurtosis of Returns HF}
		\end{subfigure}
		\caption{Kurtosis of log-return over 100 simulations}
		\label{fig:Kurtosis of Returns over 100 simulations}
	\end{figure}

	Figures \ref{fig:Kurtosis of Returns LF} and \ref{fig:Kurtosis of Returns HF}
	show how kurtosis decreases to values below 3 as the time scale increases, a
	sign of aggregational gaussianity.
	
	\subsubsection{Bubbles and Crashes}
	\label{subsubsec:Bubbles and Crashes}
	
	We use low-frequency data to investigate the occurence of bubbles and crashes as
	the fundamental price is a concept applied to low-frequency only.
	Figure \ref{fig:Cross-correlation between bubbles and log-return over 100
		simulations.} confirms the existence of significant positive correlation in lag
	0 between asset returns and the relative size of the bubble, as suggested by
	the empirical evidence (v. \ref{subsubsec:Bubbles and Crashes SF}).
	
	In cross-correlation confidence bounds are calculated as
	$\frac{[-numSTD;numSTD]}{\sqrt{N}}$, where $numSTD$ is the number of standard
	deviations for the sample cross-correlation estimation error assuming variables
	are uncorrelated, and $N$ is the length of the time-series. We use $numSTD=2$
	which corresponds to approximately 95 percent confidence bounds and plots
	estimation error bounds 2 standard deviations away from 0. We use these
	calculations in all cross-correlations published.
	
	\begin{figure}[!hbpt] \centering
		\begin{subfigure}[h]{0.4\textwidth}
			\includegraphics[width=\textwidth]{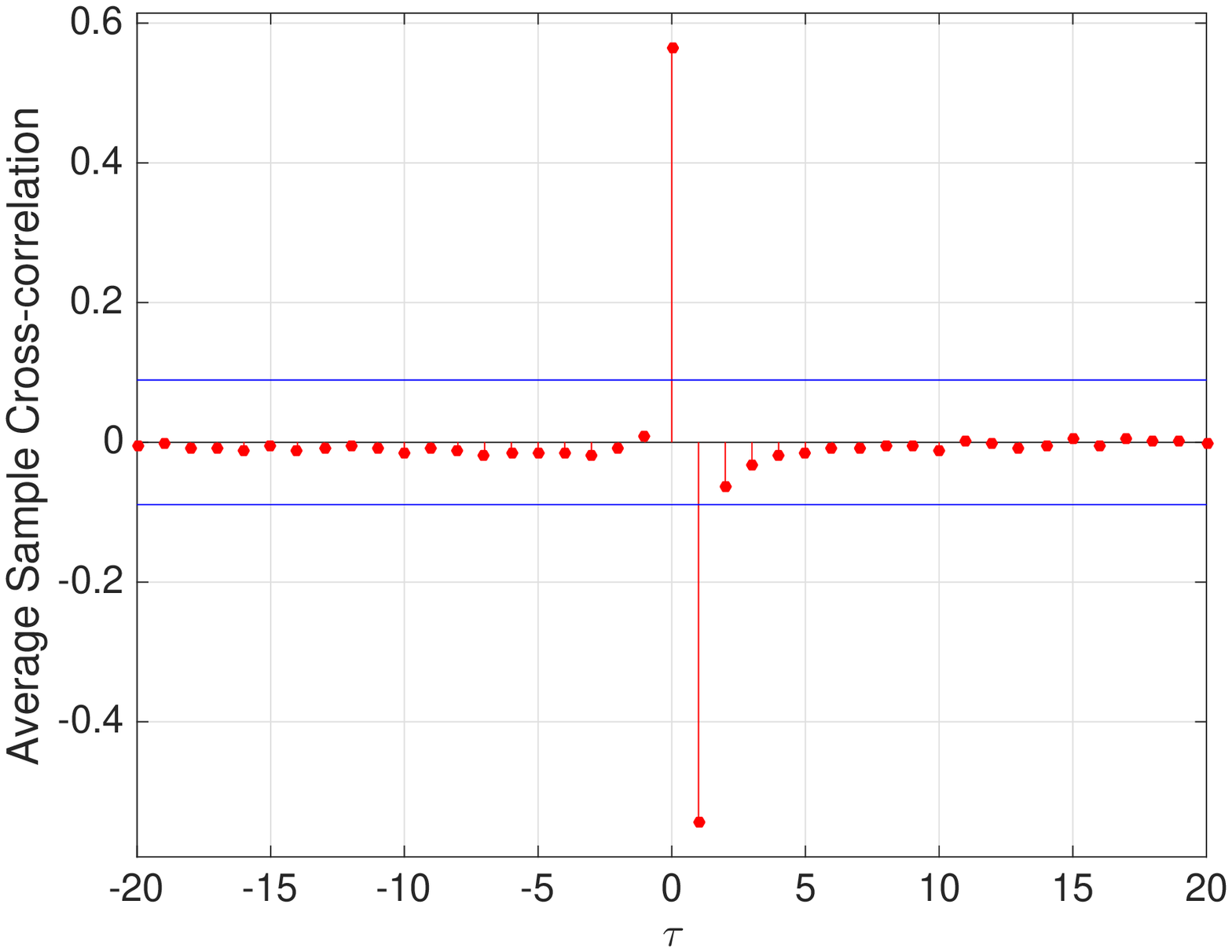}
			\footnotesize
			\caption{Average cross-correlation}
			\label{fig:Average Crosscorrelation}
		\end{subfigure}
		\hspace{2cm}
		\begin{subfigure}[h]{0.4\textwidth}
			\includegraphics[width=\textwidth]{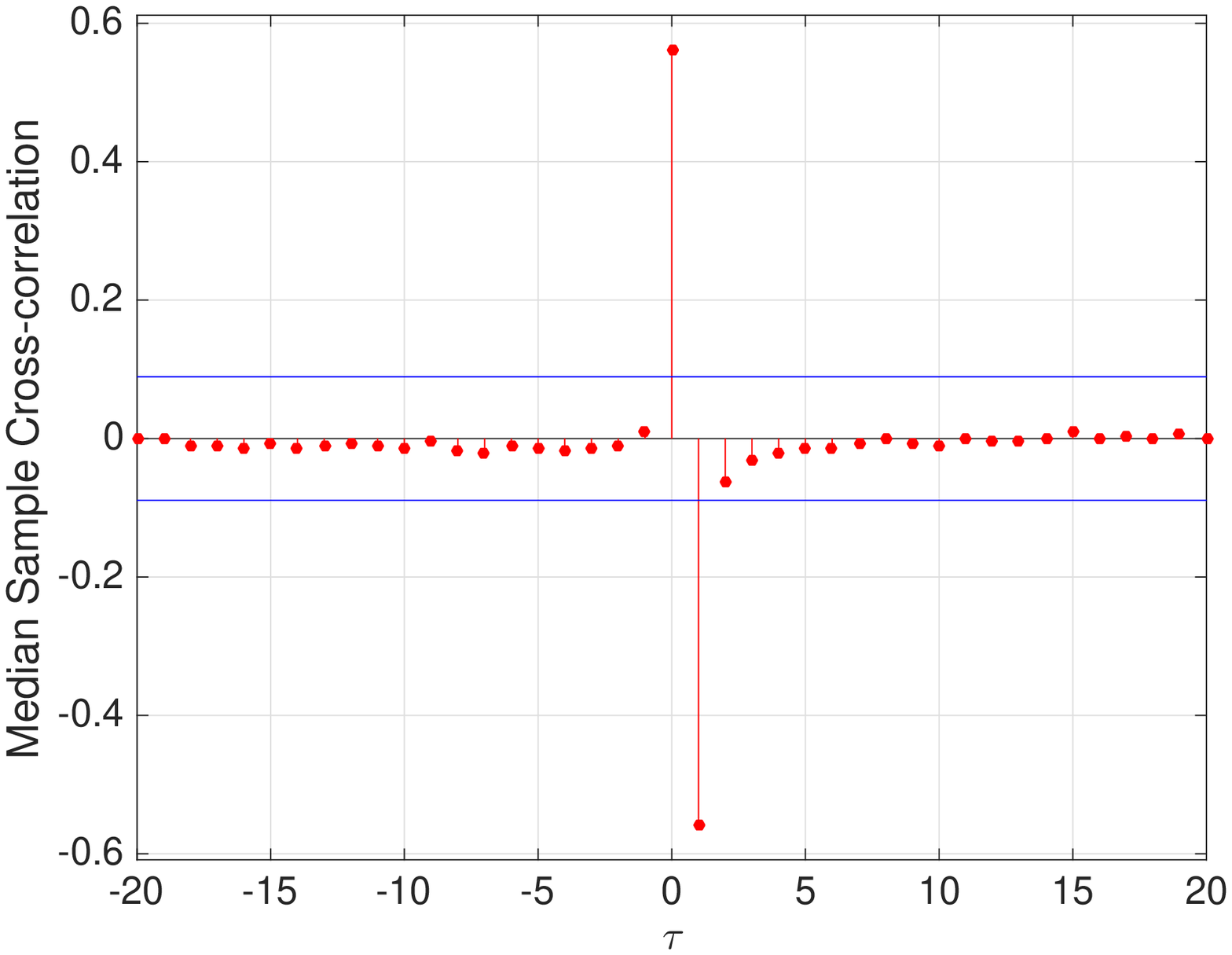}
			\footnotesize
			\caption{Median cross-correlation}
			\label{fig:Median Crosscorrelation}
		\end{subfigure}
		\caption{Cross-correlation between bubbles and log-return over 100
			simulations} 
		\floatfoot{\scriptsize{Note: The blue horizontal lines represent the approximate
				upper and lower confidence bounds $[-0.0892;0.0892]$, assuming bubbles and log-return are
				uncorrelated.}}
		\label{fig:Cross-correlation between bubbles and log-return over 100 simulations.}
	\end{figure}
	
	However the correlation is negative and significant for $\tau=1$, which may
	suggest the correction of the bubble towards the fundamental price immediately
	after the positive bubble in $\tau=0$. This may indicate that when the relative
	size of a bubble and log-return are positively correlated due to a transaction
	price greater than the fundamental price, as in $\tau=0$, there is a market
	correction in the next period. This correction could be explained by the
	fundamentalist component of the agents, which brings the price back to the
	fundamental price after a positive or negative bubble. Also, the non-existence
	of either leverage or short-selling generates more stable time-series, closer
	to the fundamental price, and consequently these corrections are more likely.
	
	The fundamental price is known by all agents who use it to form expectations
	about next period returns, as previously shown in equation \ref{eq:Return
		Expectations}. Financial institutions not only use the information on past
	prices but also actual information on the fundamental price, which is not
	reflected in past prices and helps agents forecast future returns.
	Therefore, as the positive (negative) bubble size grows, the fundamentalist
	component in equation \ref{eq:Return Expectations} integrates this information
	into returns expectations and subsequently the financial institutions adjust the
	direction of the orders down (up) towards the fundamental price. Hence, the
	fundamentalist component of the returns expectation formation works as a
	mean-reverting and stabilising process.
	
\begin{figure}[!hbpt] \centering
		\begin{subfigure}[h]{0.4\textwidth}
			\includegraphics[width=\textwidth]{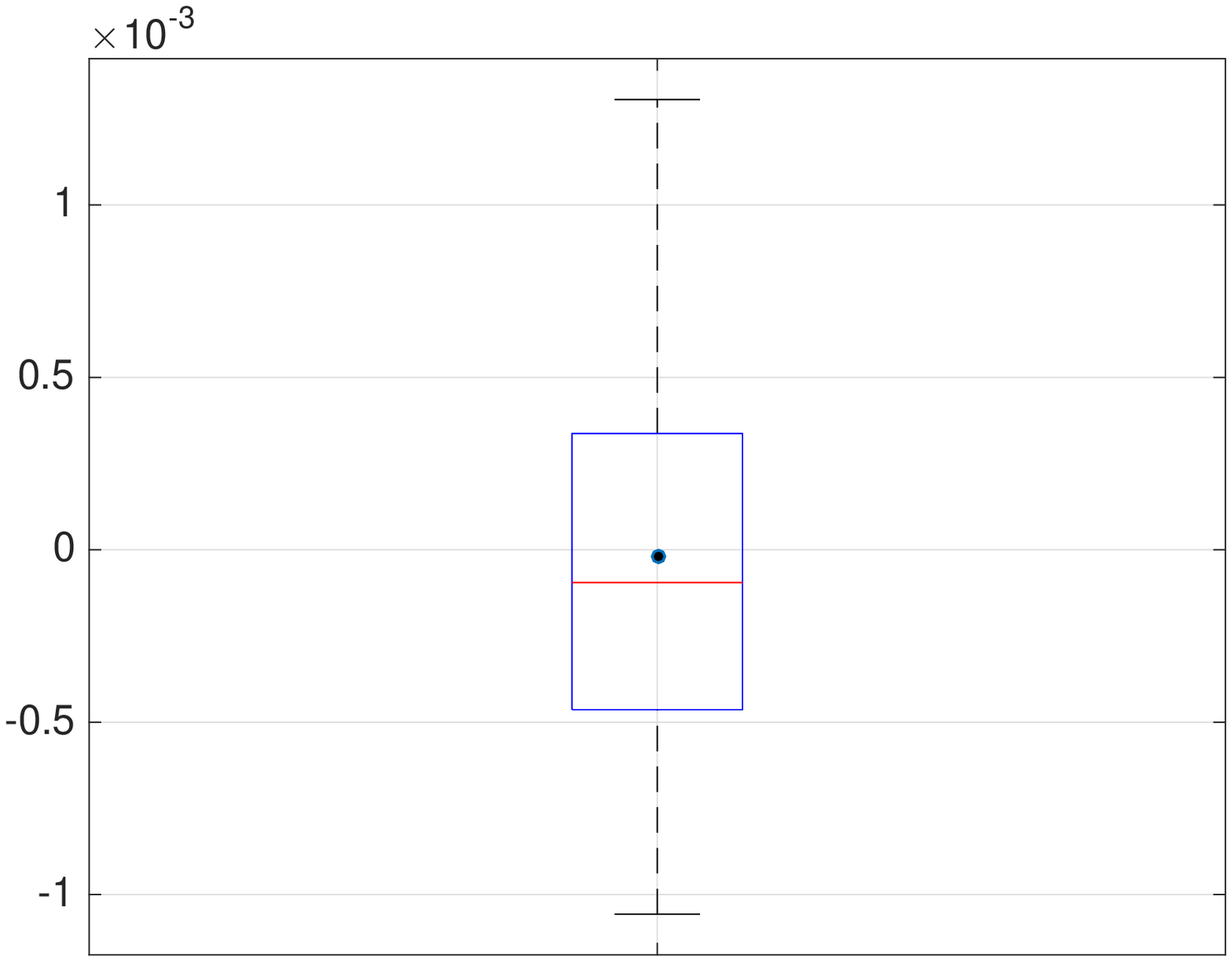}
			\footnotesize
			\caption{Boxplot}
			\label{fig:Boxplot mean size of the bubble}
		\end{subfigure}
		\hspace{2cm}
		\begin{subfigure}[h]{0.4\textwidth}
			\includegraphics[width=\textwidth]{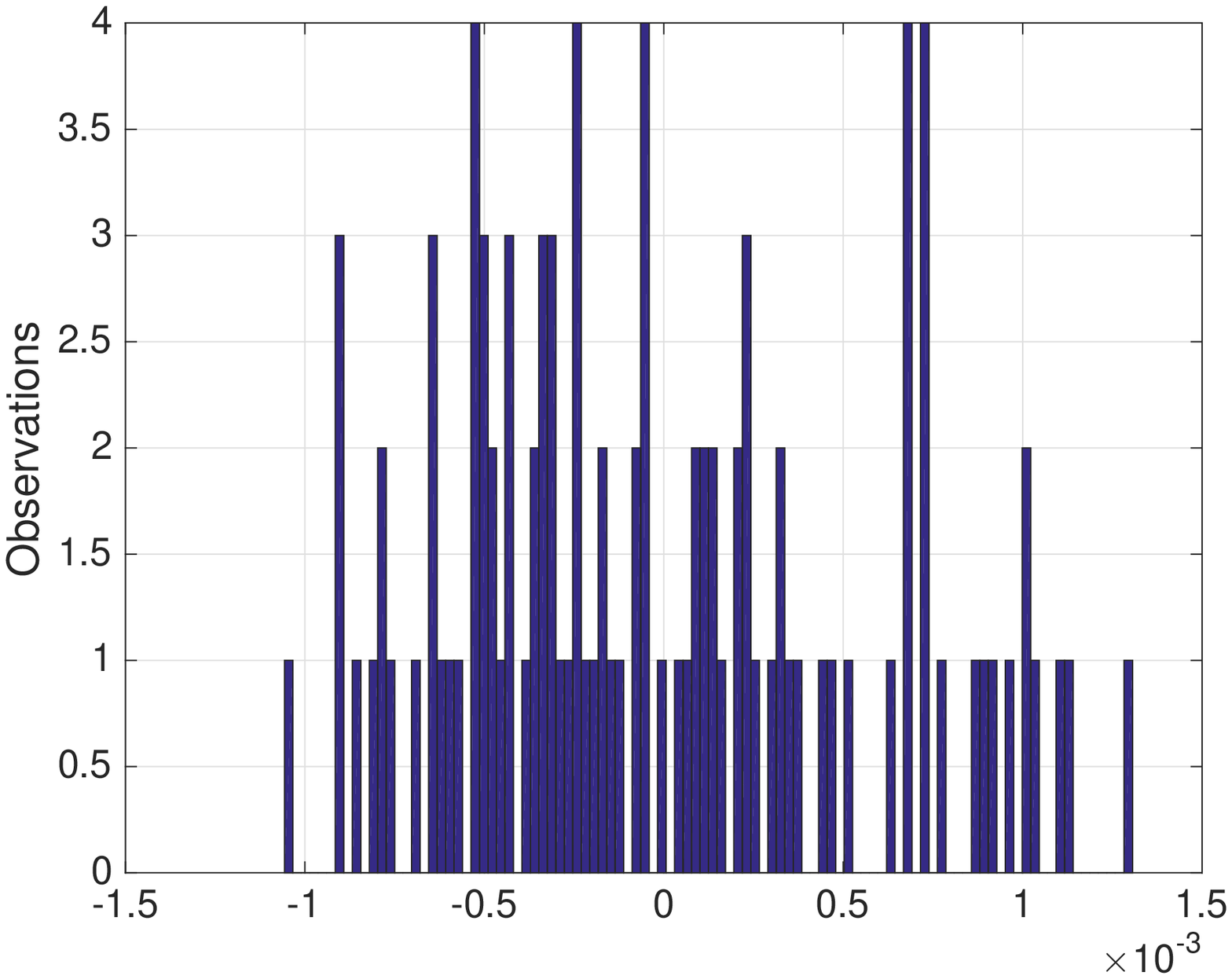}
			\footnotesize
			\caption{Histogram}
			\label{fig:Histogram mean size of the bubble}
		\end{subfigure}
		\caption{Mean size of bubbles over 100 simulations}
		\label{fig:Size of the bubble}
	\end{figure}

	Figure \ref{fig:Size of the bubble} shows that the mean bubble size in the
	unregulated treatment is close to 0, either positive or negative. This reflects
	the stability of the baseline treatment without capital requirements, leverage or short-selling.
	However, the experimental treatment with leverage exhibits large and positive
	bubbles. This suggests that these results could be very different if another
	treatment was used to study the size of bubbles. 
	
	Contrary to what other studies suggest (e.g. \citep{Flood1990}), the existence of
	bubbles in our model cannot be explained by the misspecification of
	fundamentals, since the fundamental price is public and known to all agents.
	Hence, in this case the possible deviation from the fundamental price is due to
	speculative behaviour originating from chartist and noise trading components.
	
	\subsubsection{Heavy Tails of Return Distribution} 
	\label{subsubsec:Heavy Tails of Return Distribution} 
	
\begin{figure}[!hbpt]
		\centering
		\begin{subfigure}[h]{0.4\textwidth}
			\includegraphics[width=\textwidth]{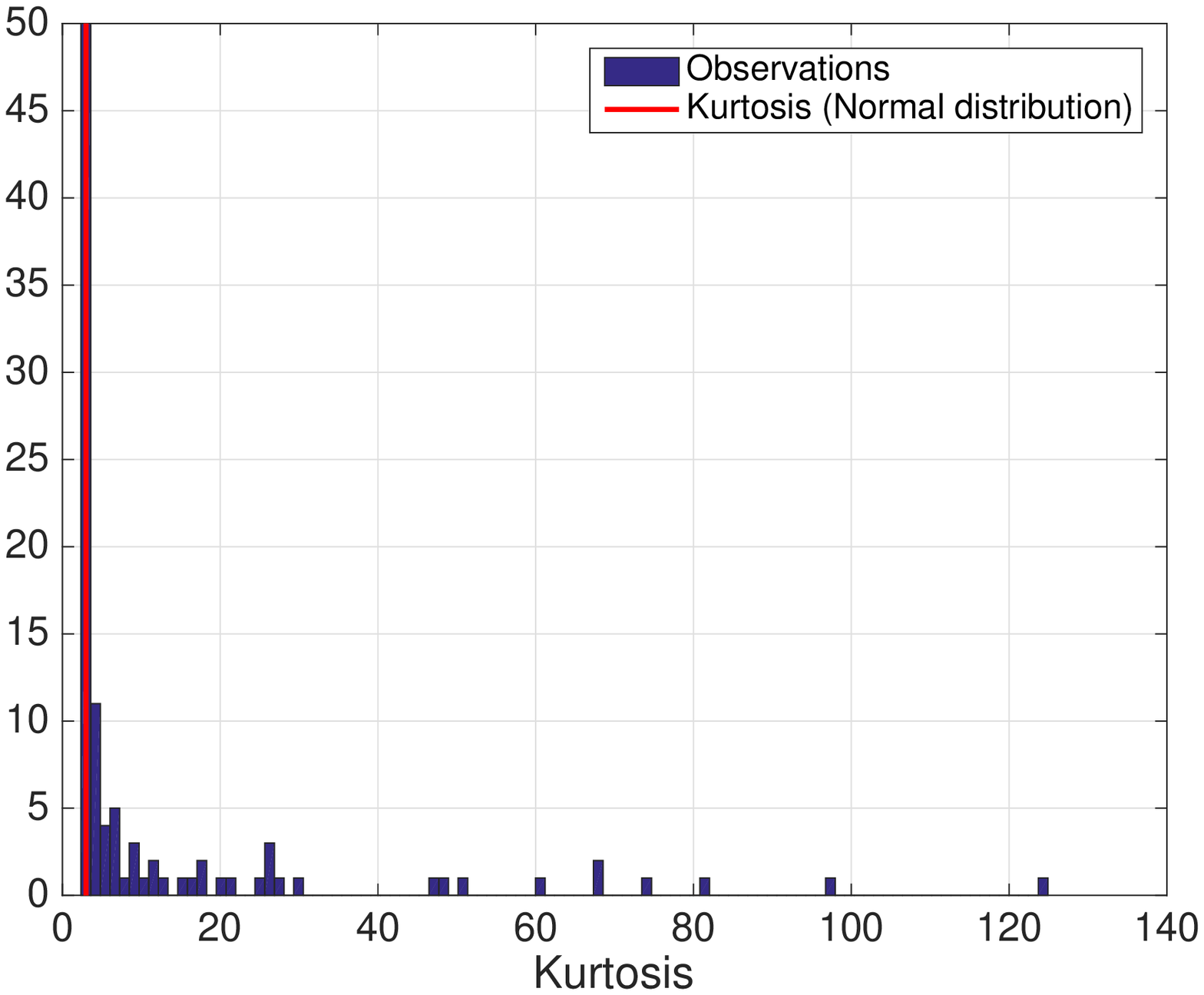}
			\caption{\footnotesize{Low-frequency}}
			\label{fig:Low-frequency Kurtosis over 100 simulation}
		\end{subfigure}
		\hspace{2cm}
		\begin{subfigure}[h]{0.4\textwidth}
			\includegraphics[width=\textwidth]{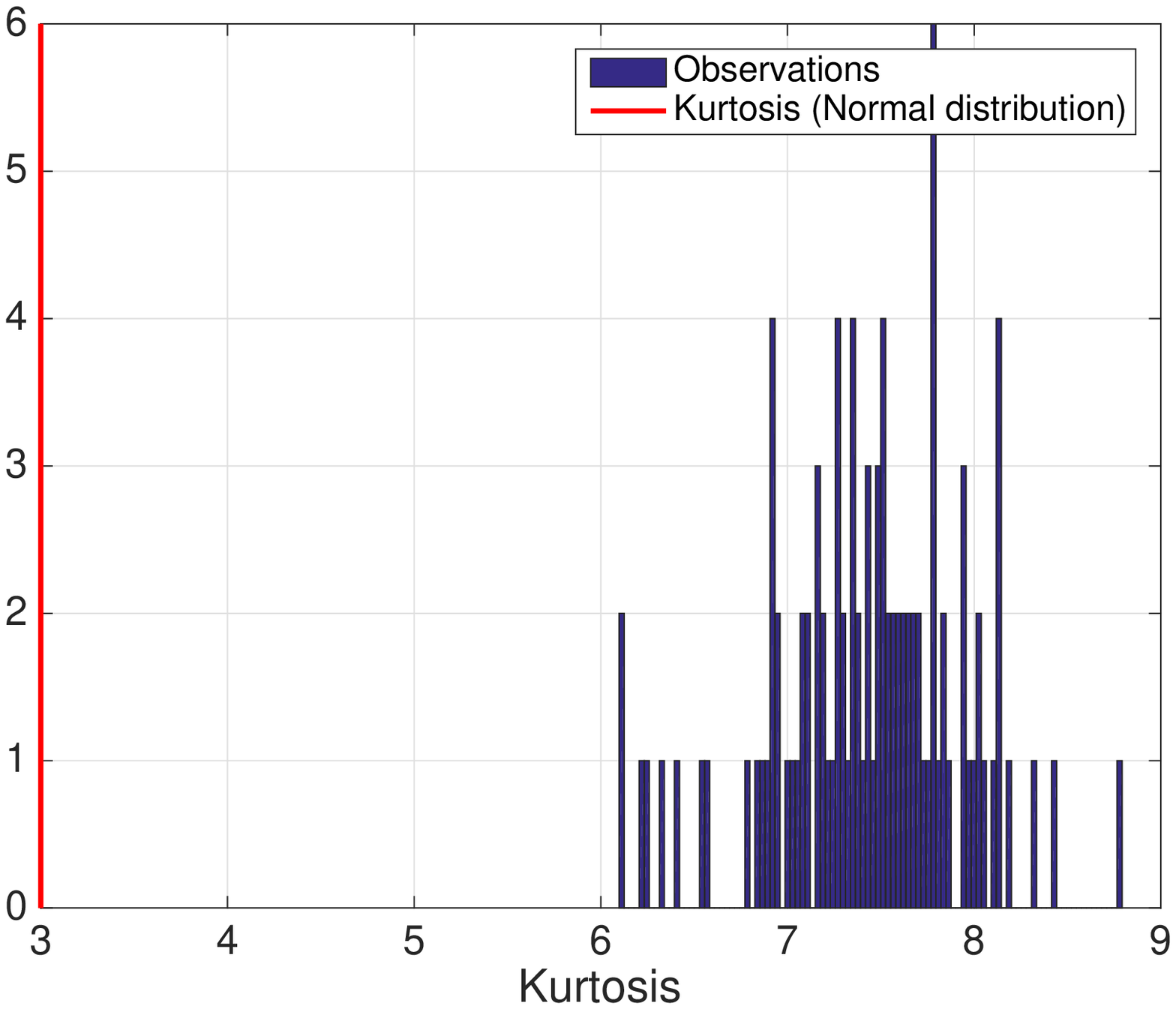}
			\footnotesize
			\caption{High-frequency}
			\label{fig:High-frequency Kurtosis for 100 simulation}
		\end{subfigure}
		\caption{Kurtosis for 100 simulations} 
		\floatfoot{Note: The kurtosis is 3 for the standard Normal distribution.}
		\label{fig:Kurtosis over 100 simulation}
	\end{figure}

	Figure \ref{fig:Kurtosis over 100 simulation} shows that log-return distribution
	has a heavy-tailed distribution in high-frequency data. In low-frequency data,
	half of the observations exhibit a kurtosis of 3, which is the kurtosis for
	normal distributions. Therefore, the higher the frequency of price observations
	the greater the kurtosis.
	
	Despite the fact that half of the daily observations exhibit a kurtosis of 3,
	table \ref{table:Kurtosis-Heavy Tails of Return Distribution} shows that the
	analysis of the mean and median kurtosis for the unregulated treatment indicates
	a kurtosis greater than 3 \citep{Taylor2005}. The shape of the distribution is
	more leptokurtic for mean and median kurtosis in high-frequency data, except for
	the mean of the low-frequency unregulated treatment. The analysis of kurtosis in
	treatments with capital requirements reveals the distributions of returns to
	be more leptokurtic, with particularly high values for high-frequency data.
	
	\begin{table}[!hbpt]
		\centering
		\caption{Kurtosis over 100 simulations}
		\label{table:Kurtosis-Heavy Tails of Return Distribution}
		\footnotesize
		\begin{tabular}{c c c c c}
			%& \multicolumn{4}{c}{Kurtosis}\\\cline{2-5}
			\toprule
			{} & \multicolumn{2}{c}{Low-frequency} &
			\multicolumn{2}{c}{High-frequency}\\
			%		\cline{2-3}\cline{4-5}
			{} & {Mean} & {Median} & {Mean} & {Median}\\
			\midrule
			Unregulated & 13.156 & 3.578 & 7.426 & 7.463 \\
			VaR & 10.505 & 4.288 & 255.824 & 242.133\\
			ES & 9.981 & 4.431 & 259.851 & 213.375\\
			\bottomrule
		\end{tabular}
	\end{table}
	
	Another indicator used to determine the existence of heavy tails is the Hill
	estimator and the analysis of the tail index. The tail size $k$ was not
	specified by \citep{hill1975}, hence we exhibit results for a bandwidth of tail
	sizes extending from 1 percent to a maximum of 10 percent of the size of
	the underlying time-series.
	
	Tables \ref{table:Hill Estimator - Heavy Tails of Return
		Distribution LF} and \ref{table:Hill Estimator - Heavy Tails of Return
		Distribution HF} exhibit the differences between low-frequency and
	high-frequency data.
	Table \ref{table:Hill Estimator - Heavy Tails of Return Distribution LF} shows
	high values of the tail index for low-frequency data. The fourth moment --
	kurtosis -- of the distribution does not exist in low-frequency data but only
	for a tail sample of 10 percent. All other moments exist and are finite. The tail
	index is smaller for treatments with capital requirements than for the unregulated
	treatment. These results show greater instability in the VaR and ES treatments
	relative to the unregulated treatment. The Hill estimator is known to be
	asymptotically normal and consistent \citep{{Hall1982}, {Mason1982},
		{Clauset2009}}. Low-frequency data sample size is $n=503$, which is not
	a large sample. As the Hill estimator tends to overestimate the tail
	exponent of the stable distribution if the sample size is not very large, we
	analyse the true tail behaviour using larger data sets of high-frequency.
	
	%\begin{table}[!hbpt]
	%	\centering
	%	\caption{Kurtosis over 100 simulations}
	%	\label{table:Kurtosis-Heavy Tails of Return Distribution}
	%	\footnotesize
	%	\begin{tabular}{c c c c c}
	%		%& \multicolumn{4}{c}{Kurtosis}\\\cline{2-5}
	%		\toprule
	%		{} & \multicolumn{2}{c}{Low-frequency} &
	%		\multicolumn{2}{c}{High-frequency}\\
	%		%		\cline{2-3}\cline{4-5}
	%		{} & {Mean} & {Median} & {Mean} & {Median}\\
	%		\midrule
	%		Unregulated & 13.156 & 3.578 & 7.426 & 7.463 \\
	%		VaR & 10.505 & 4.288 & 255.824 & 242.133\\
	%		ES & 9.981 & 4.431 & 259.851 & 213.375\\
	%		\bottomrule
	%	\end{tabular}
	%\end{table}
	
	\begin{table}[!hbpt]
		\centering
		\caption{Hill estimator for low-frequency returns over 100 simulations}
		\label{table:Hill Estimator - Heavy Tails of Return Distribution LF}
		\footnotesize
		\begin{tabular}{c c c c c c}
			\toprule
			{} & {} & \multicolumn{4}{c}{Tail index}\\
			{} & {} & \multicolumn{2}{c}{Mean} & \multicolumn{2}{c}{Median}\\
			{} & \raisebox{1.5ex}{Tail sample} & {Left} & {Right} & {Left} & {Right}\\
			\midrule
			& $1\%$ & 8.71 & 8.84 & 7.46 & 7.91    \\
			& $2.5\%$ & 6.28 & 6.25 & 6.00 & 6.37  \\
			Unregulated & $5\%$ & 4.82 & 4.90 & 4.83 & 4.83  \\
			& $10\%$ & 3.52 & 3.71 & 3.52 & 3.64 \\
			\addlinespace
			& $1\%$ & 6.38 & 7.48 & 5.13 & 6.84   \\
			& $2.5\%$ & 5.40 & 5.73 & 4.74 & 5.41  \\
			VaR & $5\%$ & 4.20 & 4.45 & 4.00 & 4.37  \\
			& $10\%$ & 3.16 & 3.37 & 3.14 & 3.43 \\
			\addlinespace
			& $1\%$ & 7.44 & 7.19 & 6.61 & 5.95   \\
			& $2.5\%$ & 5.74 & 5.79 & 5.08 & 5.09  \\
			ES & $5\%$ & 4.43 & 4.60 & 4.34 & 4.55  \\
			& $10\%$ & 3.27 & 3.43 & 3.26 & 3.45 \\
			\bottomrule
		\end{tabular}
	\end{table}
	
	Table \ref{table:Hill Estimator - Heavy Tails of Return Distribution HF}
	for high-frequency data shows that the fourth moment only exists in the
	unregulated treatment with a tail size of 1 percent. As in low-frequency data
	the treatments with capital requirements exhibit smaller tail indices relative to
	the unregulated treatment, reflecting higher instability. However, the tail
	index for high-frequency returns is considerable smaller than for low-frequency,
	which may demonstrate the higher market instability at lower tick
	sizes or the overestimation of the Hill estimator using low-frequency data.
	
	\begin{table}[!hbpt]
		\centering
		\caption{Hill estimator for high-frequency returns over 100 simulations}
		\label{table:Hill Estimator - Heavy Tails of Return Distribution HF}
		\footnotesize
		\begin{tabular}{c c c c c c}
			\toprule
			{} & { }& \multicolumn{4}{c}{Tail index}\\
			{}& {} & \multicolumn{2}{c}{Mean} & \multicolumn{2}{c}{Median}\\
			{}& \raisebox{1.5ex}{Tail sample} & {Left} & {Right} & {Left} & {Right}\\
			\midrule
			& $1\%$ & 5.36 & 5.40 & 5.39 & 5.49    \\
			& $2.5\%$ & 3.85 & 3.90 & 3.85 & 3.89  \\
			Unregulated & $5\%$ & 2.78 & 2.84 & 2.77 & 2.82  \\
			& $10\%$ & 1.82 & 1.89 & 1.80 & 1.88 \\
			\addlinespace
			& $1\%$ & 2.85 & 2.93 & 2.77 & 2.89   \\
			& $2.5\%$ & 2.81 & 2.87 & 2.87 & 2.93  \\
			VaR & $5\%$ & 2.39 & 2.40 & 2.44 & 2.41  \\
			& $10\%$ & 1.79 & 1.76 & 1.77 & 1.74 \\
			\addlinespace
			& $1\%$ & 3.36 & 3.42 & 3.37 & 3.43   \\
			& $2.5\%$ & 3.10 & 3.17 & 3.16 & 3.20  \\
			ES & $5\%$ & 2.53 & 2.56 & 2.55 & 2.55  \\
			& $10\%$ & 1.82 & 1.82 & 1.80 & 1.80 \\
			\bottomrule
		\end{tabular}
	\end{table}
	
	\subsubsection{Conditional Heavy Tails}
	\label{subsubsec:Conditional Heavy Tails}
	
	Tables \ref{table:Hill Estimator Average} (mean) and \ref{table:Hill Estimator
		Median} (median) exhibit results similar to the unconditional heavy tails and
	confirm the differences between treatments with conditional heavy tails (t or
	Gaussian). The implementation of financial regulation reduces the tail index
	and, consequently, increases the occurrence of extreme events.
	
	\begin{table}[!hbpt]
		\centering
		\caption{Conditional Hill Estimator over 100 simulations (mean)}
		\label{table:Hill Estimator Average}
		\footnotesize
		\begin{tabular}{c c c c c c c c}
			\toprule
			{} & {} & \multicolumn{6}{c}{Tail index}\\
			{} & {} & \multicolumn{2}{c}{Unconditional} & 
			\multicolumn{2}{c}{Conditional t} & \multicolumn{2}{c}{Conditional
				Gaussian}\\
			{} & \raisebox{1.5ex}{Tail sample} & {Left} & {Right} & {Left} & {Right} & {Left} & {Right}
			\\
			\midrule
			& $1\%$ & 5.36 & 5.40 & 4.22 & 4.22 & 4.68 & 4.73   \\
			& $2.5\%$ & 3.85 & 3.90 & 3.06 & 3.06 & 3.47 & 3.50  \\
			Unregulated & $5\%$ & 2.78 & 2.84 & 2.30 & 2.33 & 2.59 & 2.67  \\
			& $10\%$ & 1.82 & 1.89 & 1.67 & 1.65 & 1.83 & 1.94 \\
			\addlinespace
			& $1\%$ & 2.85 & 2.93 & 3.86 & 2.83 & 4.13 & 2.87   \\
			& $2.5\%$ & 2.81 & 2.87 & 2.98 & 2.58 & 3.25 & 2.73  \\
			VaR & $5\%$ & 2.39 & 2.40 & 2.33 & 2.15 & 2.58 & 2.31  \\
			& $10\%$ & 1.79 & 1.76 & 1.76 & 1.66 & 1.96 & 1.80 \\
			\addlinespace
			& $1\%$ & 3.36 & 3.42 & 4.04 & 3.21 & 4.26 & 3.25   \\
			& $2.5\%$ & 3.10 & 3.17 & 3.08 & 2.81 & 3.28 & 2.92  \\
			ES & $5\%$ & 2.53 & 2.56 & 2.38 & 2.27 & 2.55 & 2.39  \\
			& $10\%$ & 1.82 & 1.82 & 1.76 & 1.72 & 1.90 & 1.82 \\
			\bottomrule
		\end{tabular}
	\end{table}
	
	\begin{table}[!hbpt]
		\centering
		\caption{Conditional Hill Estimator over 100 simulations (median)}
		\label{table:Hill Estimator Median}
		\footnotesize
		\begin{tabular}{c c c c c c c c} 
			\toprule
			{} & {} & \multicolumn{6}{c}{Tail index}\\
			{} & {} & \multicolumn{2}{c}{Unconditional} & 
			\multicolumn{2}{c}{Conditional t} & \multicolumn{2}{c}{Conditional
				Gaussian}\\
			{} & {} \raisebox{1.5ex}{Tail sample} & {Left} & {Right} & {Left} & {Right} & {Left} & {Right}\\
			\midrule
			& $1\%$ & 5.39 & 5.49 & 4.22 & 4.21 & 4.62 & 4.67  \\
			& $2.5\%$ & 3.85 & 3.89 & 3.05 & 3.07 & 3.41 & 3.46 \\
			Unregulated & $5\%$ & 2.77 & 2.82 & 2.28 & 2.33 & 2.56 & 2.64 \\
			& $10\%$ & 1.80 & 1.88 & 1.49 & 1.74 & 1.81 & 1.93  \\
			\addlinespace
			& $1\%$ & 2.77 & 2.89 & 3.95 & 2.84 & 4.18 & 2.85  \\
			& $2.5\%$ & 2.87 & 2.93 & 2.98 & 2.58 & 3.23 & 2.70 \\
			VaR & $5\%$ & 2.44 & 2.41 & 2.30 & 2.11 & 2.56 & 2.25 \\
			& $10\%$ & 1.77 & 1.74 & 1.74 & 1.64 & 1.97 & 1.78  \\
			\addlinespace
			& $1\%$ & 3.37 & 3.43 & 4.07 & 3.21 & 4.30 & 3.23  \\
			& $2.5\%$ & 3.16 & 3.20 & 3.01 & 2.74 & 3.28 & 2.88 \\
			ES & $5\%$ & 2.55 & 2.55 & 2.32 & 2.18 & 2.55 & 2.35 \\
			& $10\%$ & 1.80 & 1.80 & 1.74 & 1.67 & 1.93 & 1.81  \\
			\bottomrule
		\end{tabular}
	\end{table}
	
	Table \ref{table:Kurtosis} shows that the conditional residual time-series
	generated from our model still exhibit a leptokurtic distribution either with t
	or Gaussian white noise.
	
	\begin{table}[!hbpt]
		\centering
		\caption{Conditional kurtosis over 100 simulations (high-frequency data)}
		\label{table:Kurtosis}
		\footnotesize
		\begin{tabular}{c c c c c c c}
			\toprule
			% & \multicolumn{3}{c}{Kurtosis}\\\hline
			{} & \multicolumn{2}{c}{Unconditional} &  \multicolumn{2}{c}{Conditional t} &
			\multicolumn{2}{c}{Conditional Gaussian}\\
			{} & {Mean} &  {Median} & {Mean} &  {Median} & {Mean} &  {Median} \\
			\midrule
			Unregulated & 7.43 & 7.46 & 10.19 & 10.25 & 8.29 & 8.47 \\
			VaR & 255.82 & 242.13 & 332.28 & 248.66 & 266.38 & 232.49 \\
			ES & 259.85 & 213.38 & 268.29 & 218.30 & 259.85 & 206.65 \\
			\bottomrule
		\end{tabular}
	\end{table}
	
	The returns are heavy-tailed even when applying a model that compensates for the
	time varying volatility. We conclude that our model replicates the stylised fact
	of conditional heavy tales found in empirical data (v. \ref{subsubsec:Conditional Heavy Tails SF2}).
	
	\subsubsection{Equity Premium Puzzle}
	\label{subsubsec:Equity Premium Puzzle}
	
	The equity premium documented in \citep{Mehra1985} is for very long investment
	horizons and it has varied considerably and counter-cyclically over time. Table
	\ref{table:Equity Premium Puzzle} shows the equity premium in our baseline
	treatment. A possible explanation for the negative premium observed in our model
	is the stability of the baseline treatment and mean realised returns of
	approximately zero.
	
	\begin{table}[!hbpt] \centering
		\begin{threeparttable}
			\caption{Equity Premium Puzzle over 100 simulations (unregulated)}
			\label{table:Equity Premium Puzzle}
			\footnotesize 
			\begin{tabular}{c c c}
				\toprule
				\multicolumn{2}{c}{Annual Mean
					Return} & \multicolumn{1}{c}{Equity Premium} \\
				Realised & Risk-free Asset & Realised \\
				\midrule
				% \\\hline $175.60\%$ & $-0.44\%$ & $2.16\%$ & $-2.60$ pps   \\\hline
				$\convert{0.00002797794371587781}$ & 0.0457 & -0.0457 pps \\
				\bottomrule
			\end{tabular}
			\scriptsize
			\begin{tablenotes}
				\item Note: The annual mean return is calculated using the mean
				of the realised returns over 100 simulations. The returns of the
				risk-free asset are fixed.
			\end{tablenotes}
		\end{threeparttable}
	\end{table}
	
	A possible explanation for the small realised returns and, consequently, the
	negative equity premium observed, is the mean-variance portfolio optimisation
	used by the financial institutions. Financial institutions are risk-averse
	however loss aversion, usually identified as a possible explanation for this
	stylised fact \citep{Benartzi1995}, is not considered in this particular experiment.
	\citep{Cuthbertson2004} argued that the equity premium puzzle is one of the
	stylised facts about stock returns that is difficult to explain in conventional
	models. \citep{Benartzi1995} show that loss aversion might explain the equity
	premium puzzle \footnote{Our model generates similar annual returns in the
		baseline treatment (without capital requirements, leverage or short-selling) with Cumulative Prospect Theory-agents. Other Cumulative Prospect Theory (CPT) treatments may exhibit different results.}.
	
	\subsubsection{Excess Volatility}
	\label{subsubsec:Excess Volatility}
	
	Table \ref{table:Excess Volatility of log-return} confirms that annual
	volatility of the market log-return exceeds the annual volatility of the
	fundamental log-return in all treatments. 
	
	In the absence of financial regulations volatility enters the model either
	through the fundamental price or through traders' behaviour. As the annual mean
	volatility of the fundamental price is the same for all treatments and the
	agents' initial conditions are the same in all treatments, the only source of
	volatility that differs between treatments is the implementation of regulation.
	
	\begin{table}[!hbpt] \centering
		\begin{threeparttable}
			\caption{Excess volatility of log-return over 100 simulations}
			\label{table:Excess Volatility of log-return}
			\footnotesize
			\begin{tabular}{c c c}
				\toprule
				{} & \multicolumn{2}{c}{Annual Mean Volatility}\\
				{} & {Market} & {Fundamental} \\
				\midrule
				Unregulated & 0.0197 & 0.006 \\
				VaR & 0.0236 & 0.006 \\
				ES & 0.0216 & 0.006 \\
				\bottomrule
			\end{tabular}
			\scriptsize
			\begin{tablenotes}
				\item Note: The volatility is calculated using the standard
				deviation of the log-return for each of the treatments: Unregulated, VaR and
				ES.
			\end{tablenotes}
		\end{threeparttable}
	\end{table}
	
	\subsubsection{Gain/Loss Asymmetry}
	\label{subsubsec:Gain/Loss Asymmetry}
	
	Table \ref{table:Gain/Loss Asymmetry} shows our analysis of the investment
	horizon distribution $p(\tau_{\rho})$ for a return level of $\rho=0.25$ percent.
	The most likely horizon, which we call the optimal investment horizon, is greater
	for gains than for losses, and accords with the empirical evidence. These
	waiting times are longer in the unregulated treatment as it shows less volatility. The
	more volatile VaR and ES treatments exhibit a smaller time span needed to
	generate a fluctuation or a movement in the price of size $\rho=0.25$ percent.
	
	\begin{table}[!hbpt] \centering
		\begin{threeparttable}
			\caption{Optimal Investment Horizon over 100 simulations}
			\label{table:Gain/Loss Asymmetry}
			\footnotesize 
			\begin{tabular}{c c c c c c c} 
				\toprule
				{} & \multicolumn{3}{c}{Gains}
				&
				\multicolumn{3}{c}{Losses}\\
				{}& {Mean} & {Median} & {Std} & {Mean} & {Median} & {Std} \\
				\midrule
				Unregulated & 8.28 & 6 &
				(6.78) & 6.5 & 5.5 & (5.31) \\
				VaR & 4.22 & 2 & (5.02) & 2.98 & 2 & (2.43) \\
				ES & 5.02 & 2 & (6.97)  & 3.73 & 3 & (2.97) \\
				\bottomrule
			\end{tabular}
			\scriptsize
			\begin{tablenotes}
				\item Note: We analyse the daily closure for all the 100 simulations for each of
				the treatments -- unregulated, VaR and ES. The optimal investment horizons over
				100 simulations were computed by generating 100 independent realisations of our
				model across 504 days, computing the daily return for each realisation, and then calculating the necessary time horizon to
				reach a return level of $\rho=0.25$ percent. The values in parenthesis are
				standard deviations.
			\end{tablenotes}
		\end{threeparttable}
	\end{table}
	
	\subsubsection{Leverage Effect}
	\label{Leverage Effect-LF}
	
\begin{figure}[!hbpt]
		\centering
		\begin{subfigure}[H]{0.4\textwidth}
			\includegraphics[width=\textwidth]{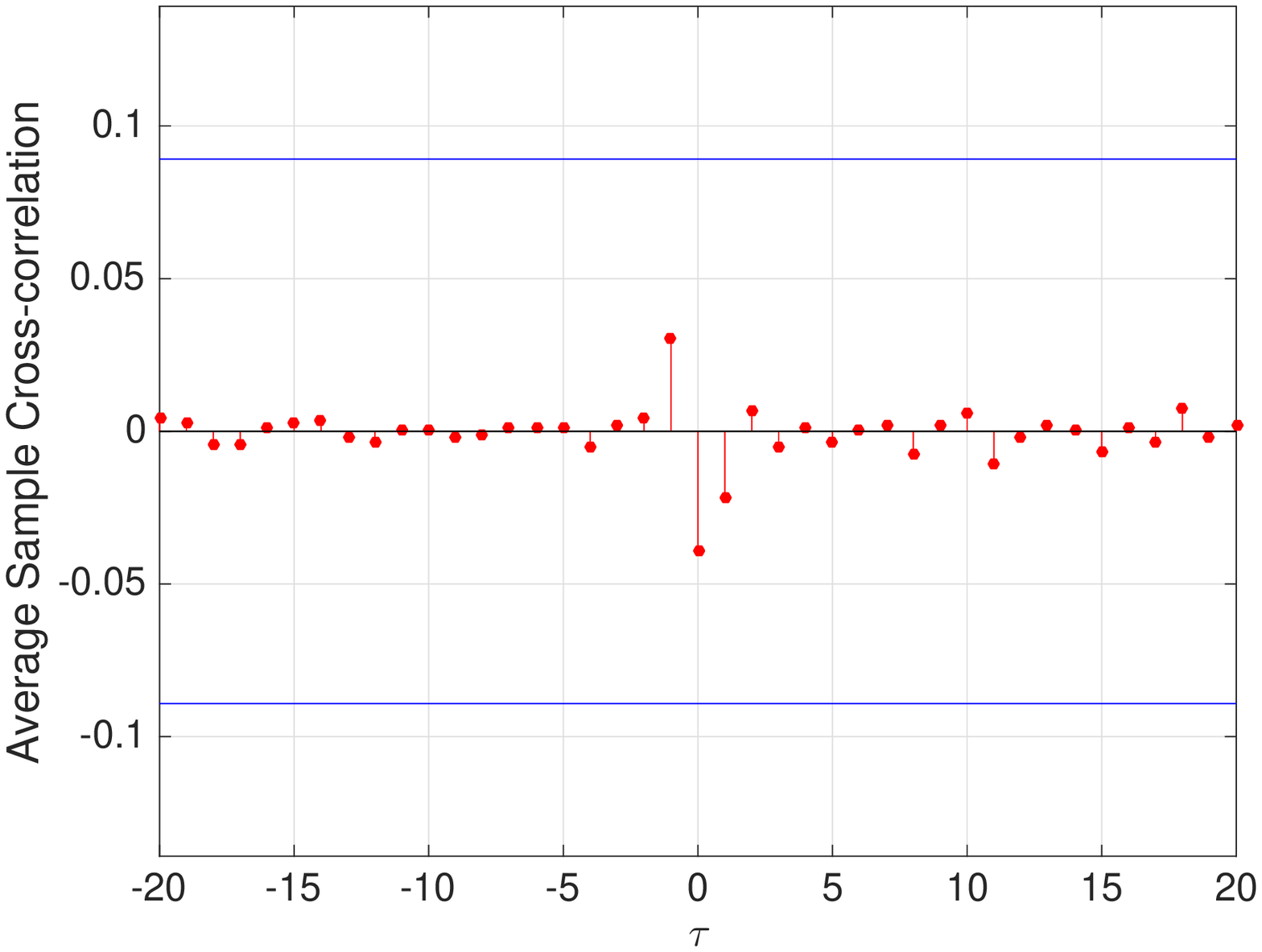}
			\footnotesize
			\caption{Average Returns}
			\label{fig:Average Returns Leverage Effects}
		\end{subfigure}
		\hspace{2cm}
		\begin{subfigure}[H]{0.4\textwidth}
			\includegraphics[width=\textwidth]{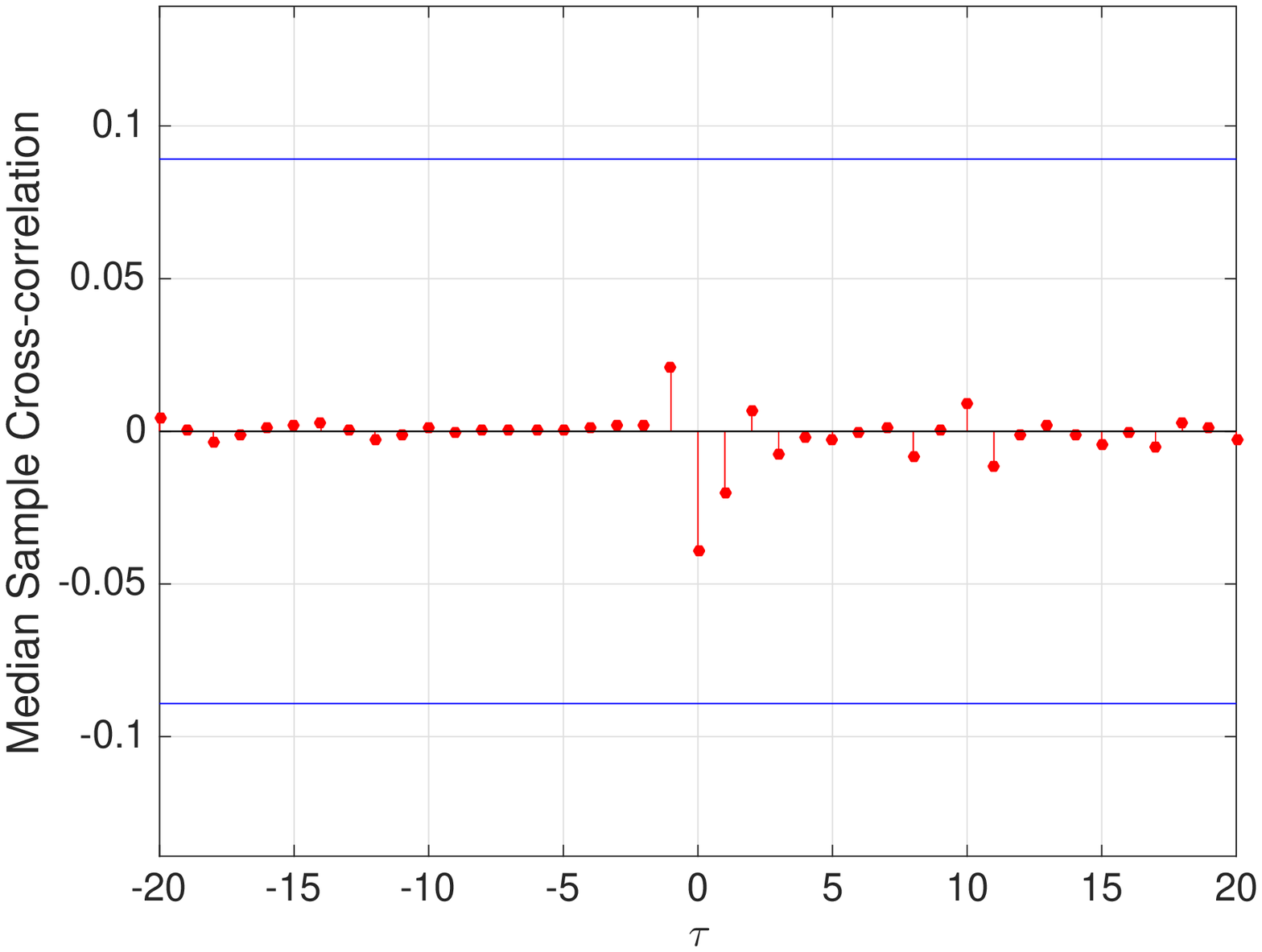}
			\footnotesize
			\caption{Median Returns}
			\label{fig:Median Returns Leverage Effects}
		\end{subfigure}
		\caption{Leverage effects of low-frequency log-return over 100 simulations}
		\label{fig:Leverage Effect LF}
		\floatfoot{\scriptsize{Note: The blue horizontal lines represent the approximate
				upper and lower confidence bounds $[-0.0892;0.0892]$, assuming log-return and volatility
				are uncorrelated.}}
	\end{figure}

	Figure \ref{fig:Leverage Effect LF} shows that the unregulated treatment does
	not generate leverage effects in low-frequency data.
	
\begin{figure}[!hbpt]
		\centering
		\begin{subfigure}[h]{0.4\textwidth}
			\includegraphics[width=\textwidth]{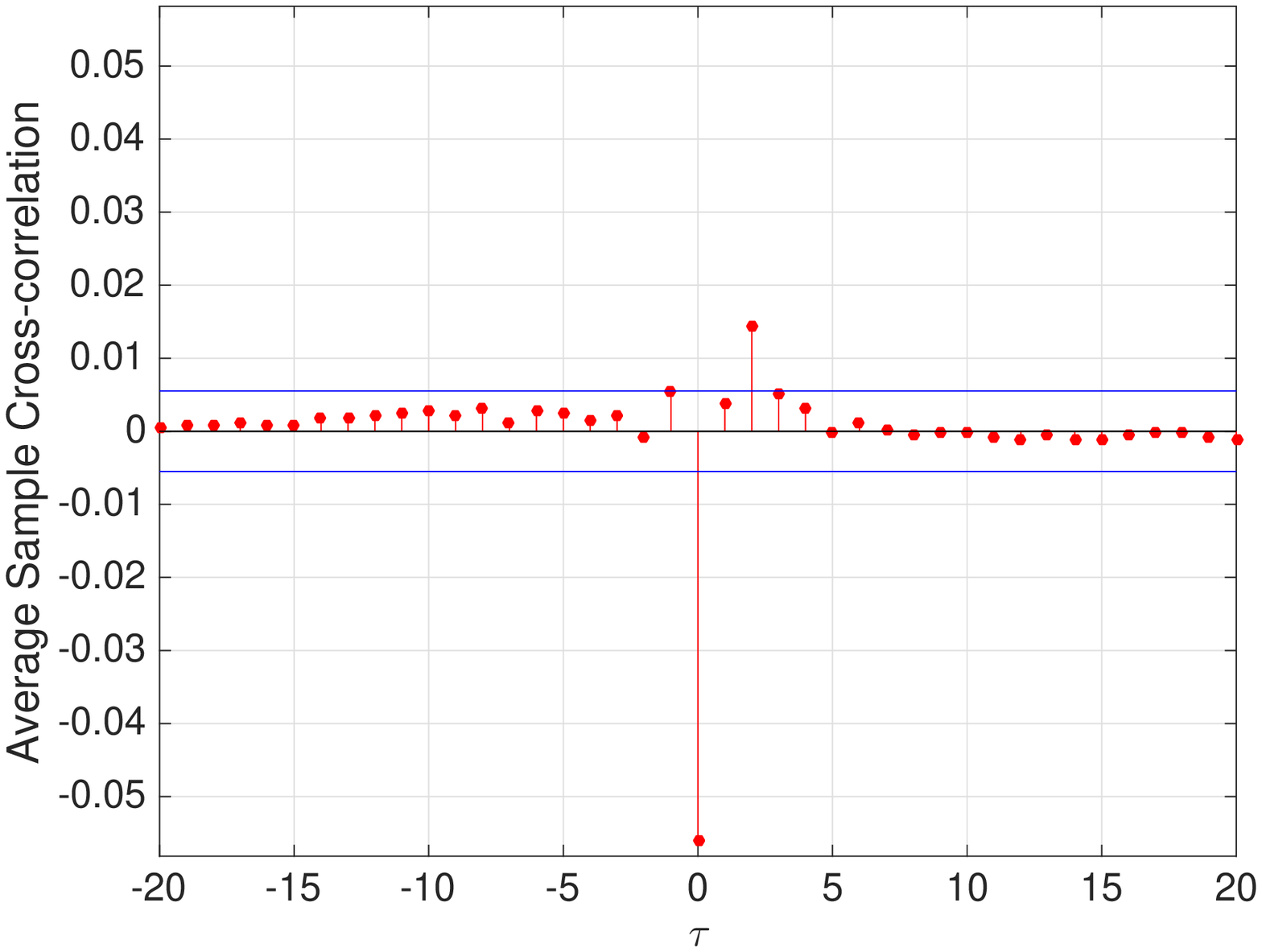}
			\footnotesize
			\caption{Average Returns}
			\label{fig:Average Returns Leverage Effects}
		\end{subfigure}
		\hspace{2cm}
		\begin{subfigure}[h]{0.4\textwidth}
			\includegraphics[width=\textwidth]{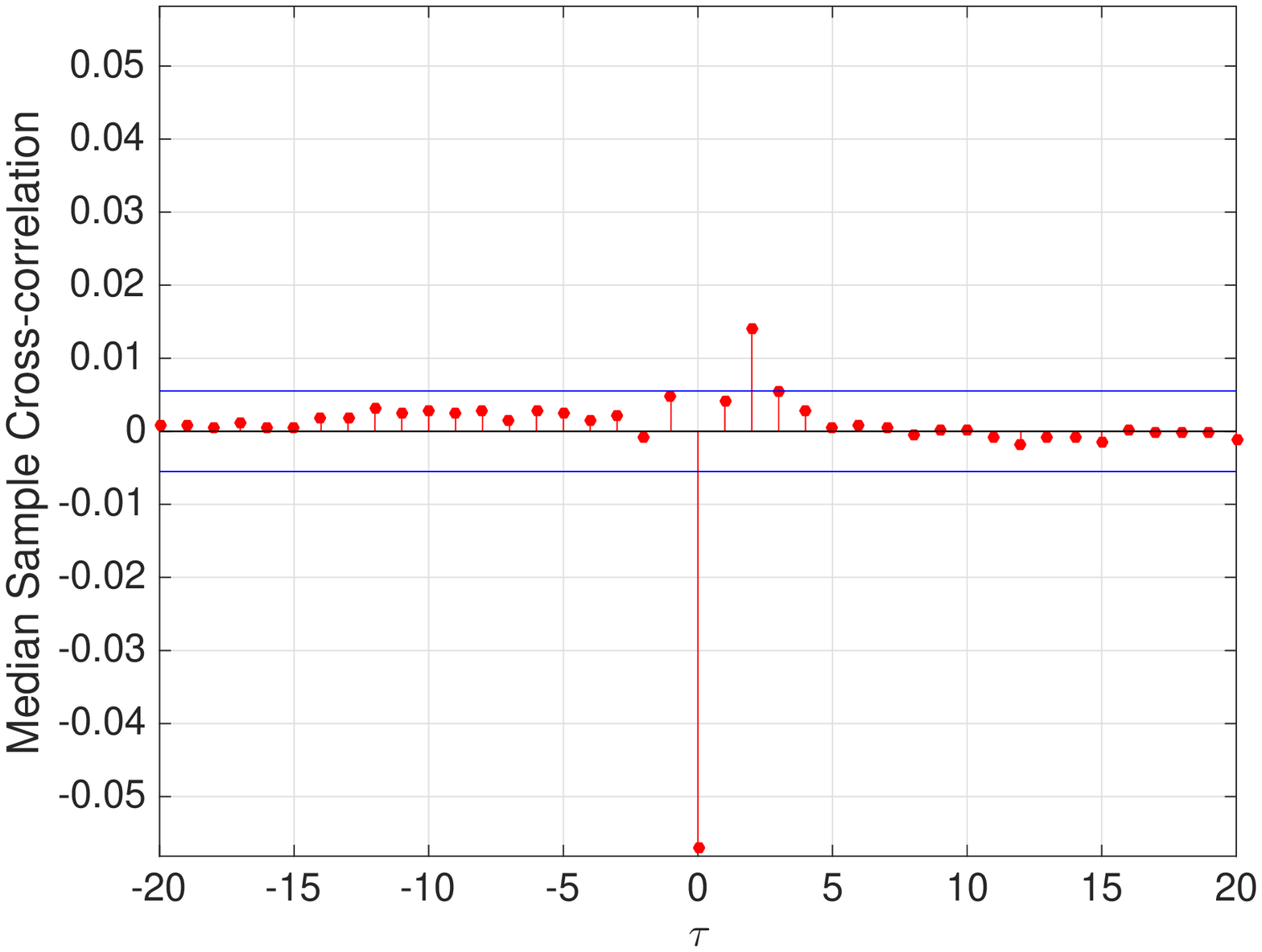}
			\footnotesize
			\caption{Median Returns}
			\label{fig:Median Returns Leverage Effects}
		\end{subfigure}
		\caption{Leverage effects of high-frequency over 100 simulations}
		\label{fig:Leverage Effect HF}
		\floatfoot{\scriptsize{Note: The blue horizontal lines represent the mean of
				approximate upper and lower confidence bounds over 100 simulations $[-0.0063;0.0063]$,
				assuming log-return and volatility are uncorrelated.}}
	\end{figure}

	Figure \ref{fig:Leverage Effect HF} shows that in high-frequency data the
	leverage effect is negative and significant for $\tau=0$. The observed negative
	leverage implies that volatility and returns are negatively correlated: price
	drops increase volatility of an asset, this is the so-called leverage effect.
	
	According to \citep{Bouchaud2001} the leverage effect is much more pronounced for
	indices than single stocks. For both stocks and stock indices, the
	volatility-return correlation is short ranged, with, however, a shorter decay
	time for stock indices than for individual stocks, and the amplitude of the
	correlation is much stronger for indices than for individual stocks.
	
	\subsubsection{Linear Autocorrelation}
	\label{subsubsec:Linear Autocorrelation}

\begin{figure}[!hbpt] \centering
		\begin{subfigure}[h]{0.4\textwidth}
			\includegraphics[width=\textwidth]{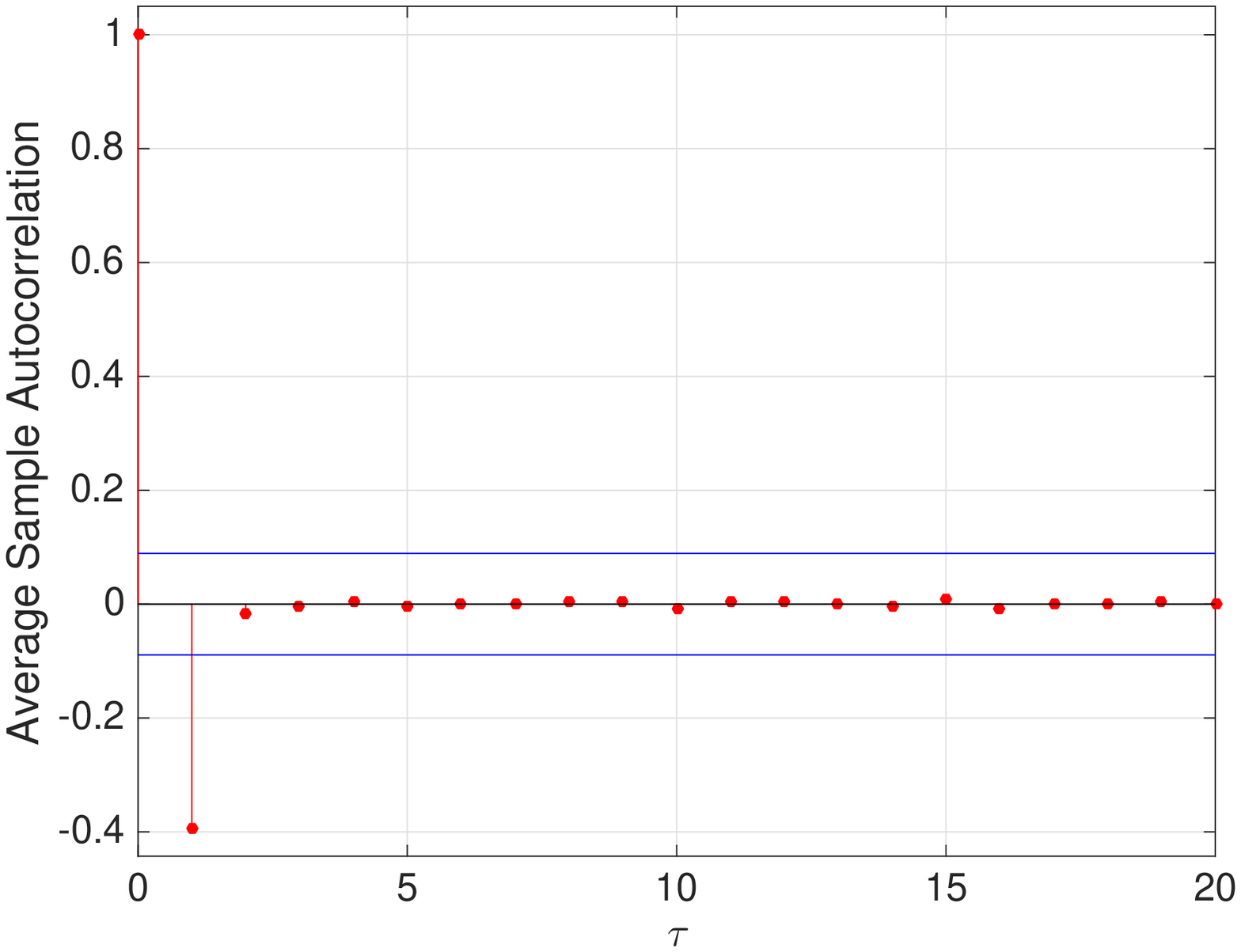}
			\footnotesize
			\caption{Average}
			\label{fig:Absence of Average Low-frequency Autocorrelation}
		\end{subfigure}
		\hspace{2cm}
		\begin{subfigure}[h]{0.4\textwidth}
			\includegraphics[width=\textwidth]{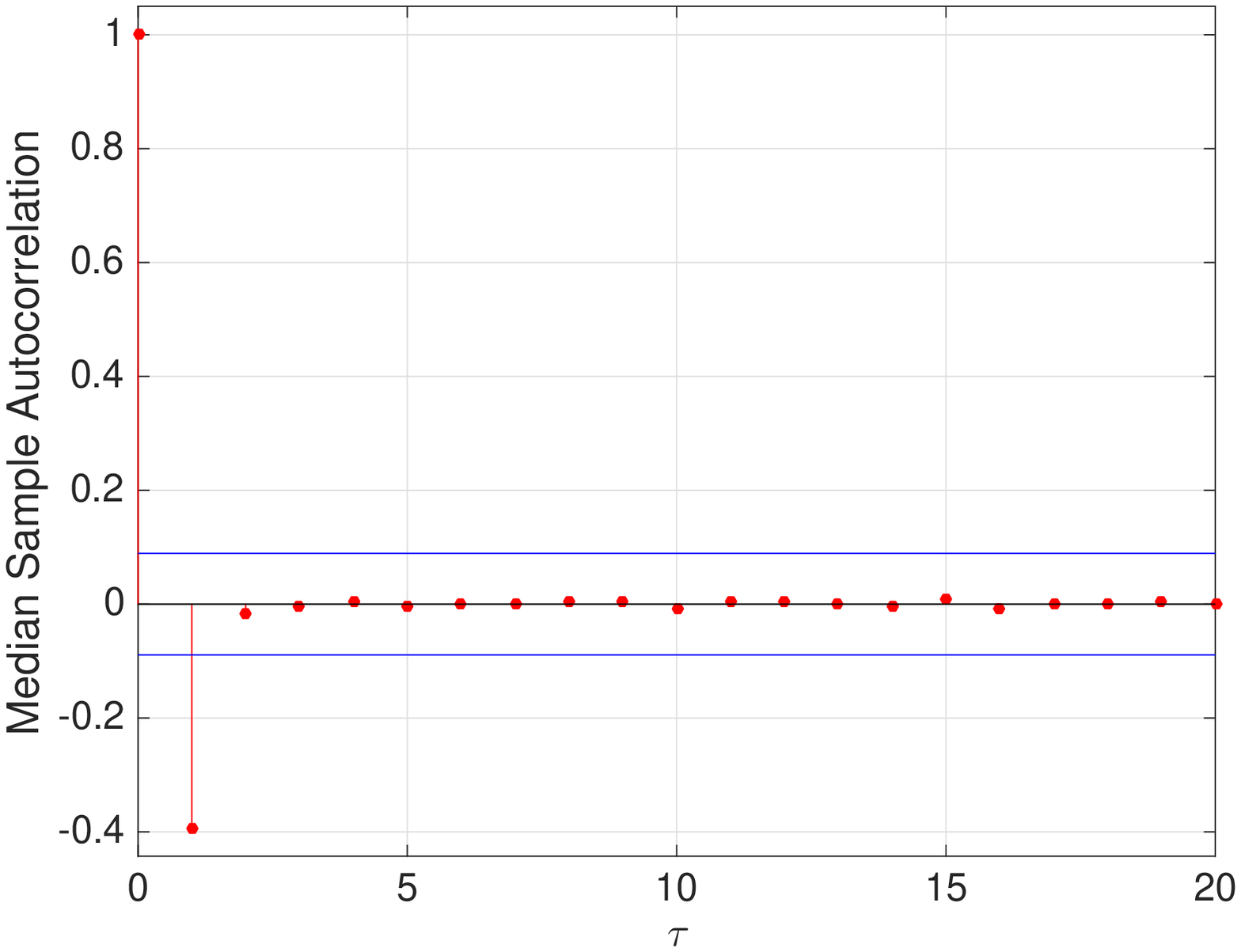}
			\footnotesize
			\caption{Median}
			\label{figure: Absence of Median Low-frequency Autocorrelation}
		\end{subfigure}
		\caption{Autocorrelation in low-frequency log-return over 100 simulations}
		\floatfoot{\scriptsize{Note: The blue horizontal lines represent the approximate
				confidence bounds $[-0.0892;0.0892]$ of the autocorrelation function assuming the time
				serie is a moving average process.}}
		\label{fig:Absence of LF Autocorrelation}
	\end{figure}

	As mentioned in section \ref{subsubsec:Linear Autocorrelation SF2},
	\citep{Fama1965} demonstrates that the first-order autocorrelations of daily
	returns are positive for twenty-two out of thirty stocks of the DJIA, which means that this
	dependence is negative for eight stocks. Figure \ref{fig:Absence of LF Autocorrelation} indicates that daily returns in the
	unregulated treatment exhibit first-order negative autocorrelation.
	
	The estimated standard error for the autocorrelation at lag $k>q$ is
	\begin{equation*}
		SE(r_{k})=\sqrt{\frac{1}{T}(1+2\sum^{q}_{j=1}{r_{j}^{2}})}.
	\end{equation*}
	Confidence bounds are calculated as 
	\begin{equation*}
		SE(r_{k})\times[-numSTD;numSTD]
	\end{equation*} 
	where $numSTD$ is the number of standard deviations for the sample
	autocorrelation function estimation error assuming the theoretical autocorrelation function is 0 beyond
	lag 0. Since we assume that the moving average order that specifies the number
	of lags beyond which the theoretical autocorrelation function is effectively 0
	equals 0, confidence bounds can be expressed as $\frac{[-numSTD;numSTD]}{\sqrt{T}}$. $T$ is the length of the time-series. We
	use $numSTD=2$ which corresponds to approximately 95 percent confidence bounds.
	These calculations are used in all autocorrelations published.
	
	\citep{Fama1965} further demonstrates that the preponderance of positive or
	negative signs in the coefficients for the daily data are partly determined by factors
	peculiar to that asset or industry. However, this author concludes that the
	actual direction of the ``dependence'' varies from study to study. 
	We believe that the lag 1 negative autocorrelation observed in our experimental
	treatment might be caused by the dominance of agents' fundamentalist component that brings the
	price back to the fundamentalist price.

\begin{figure}[!hbpt] \centering
		\begin{subfigure}[h]{0.4\textwidth}
			\includegraphics[width=\textwidth]{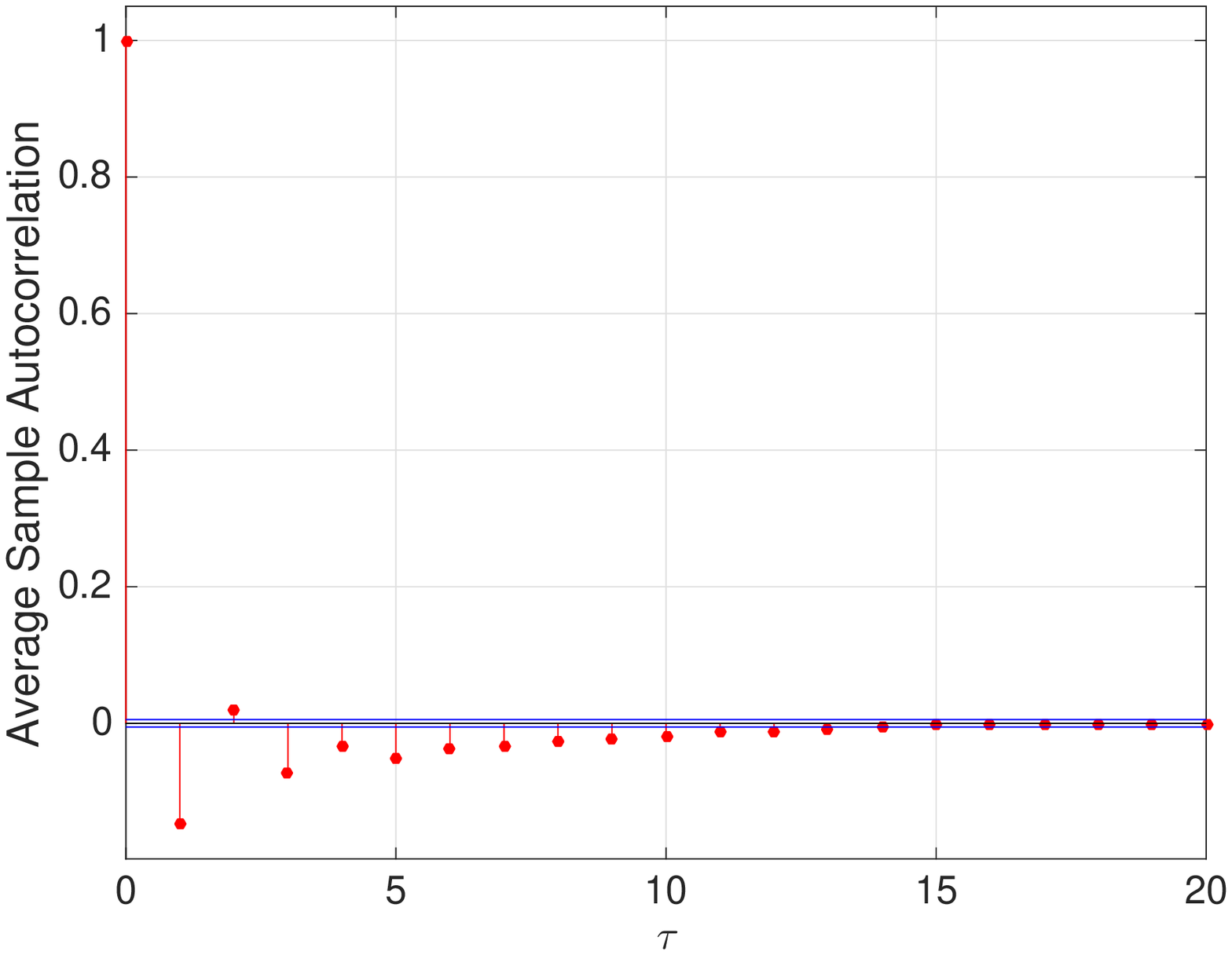}
			\caption{Average Autocorrelation}
			\label{fig:Absence of Average High-frequency Autocorrelation}
		\end{subfigure}
		\hspace{2cm}
		\begin{subfigure}[h]{0.4\textwidth}
			\includegraphics[width=\textwidth]{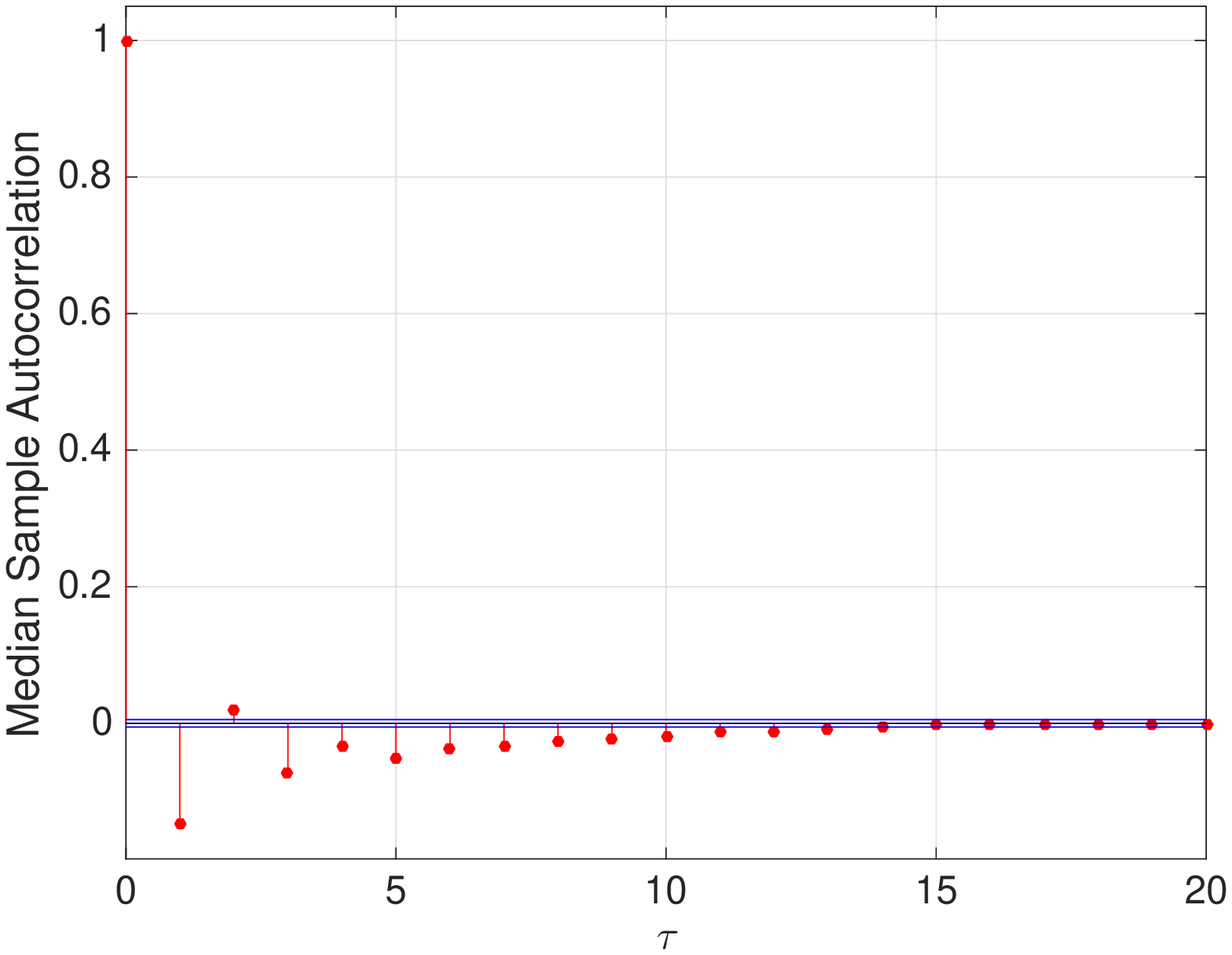}
			\caption{Median Autocorrelation}
			\label{figure: Absence of Median High-frequency Autocorrelation}
		\end{subfigure}
		\caption{Absence of autocorrelation in high-frequency log-return over 100
			simulations}
		\label{fig:Absence of HF Autocorrelation}
		\floatfoot{\scriptsize{Note: The blue horizontal lines represent the approximate
				confidence bounds $[-0.0063;0.0063]$ of the autocorrelation function assuming the time
				serie is a moving average process.}}
	\end{figure}

	Figure \ref{fig:Absence of HF Autocorrelation} shows that intraday returns from
	traded assets are almost uncorrelated, with any important dependence usually
	restricted to a negative correlation between consecutive returns in very small
	intraday time scales \citep{{Cont2001}, {Abergel2016}}. This first-order
	negative autocorrelation is traditionally attributed to microstructure effects,
	as the bid-ask bounce, due to the fact that there is often a spread between the
	price paid by buyer and seller initiated trades and the transaction prices may
	take place either close to the ask or closer to the bid price, which tend to
	bounce between these two limits \citep{Russell2010}.
	
	\subsubsection{Long Memory}
	\label{Long Memory}
	
\begin{figure}[!hbpt] \centering
		\begin{subfigure}[H]{0.4\textwidth}
			\includegraphics[width=\textwidth]{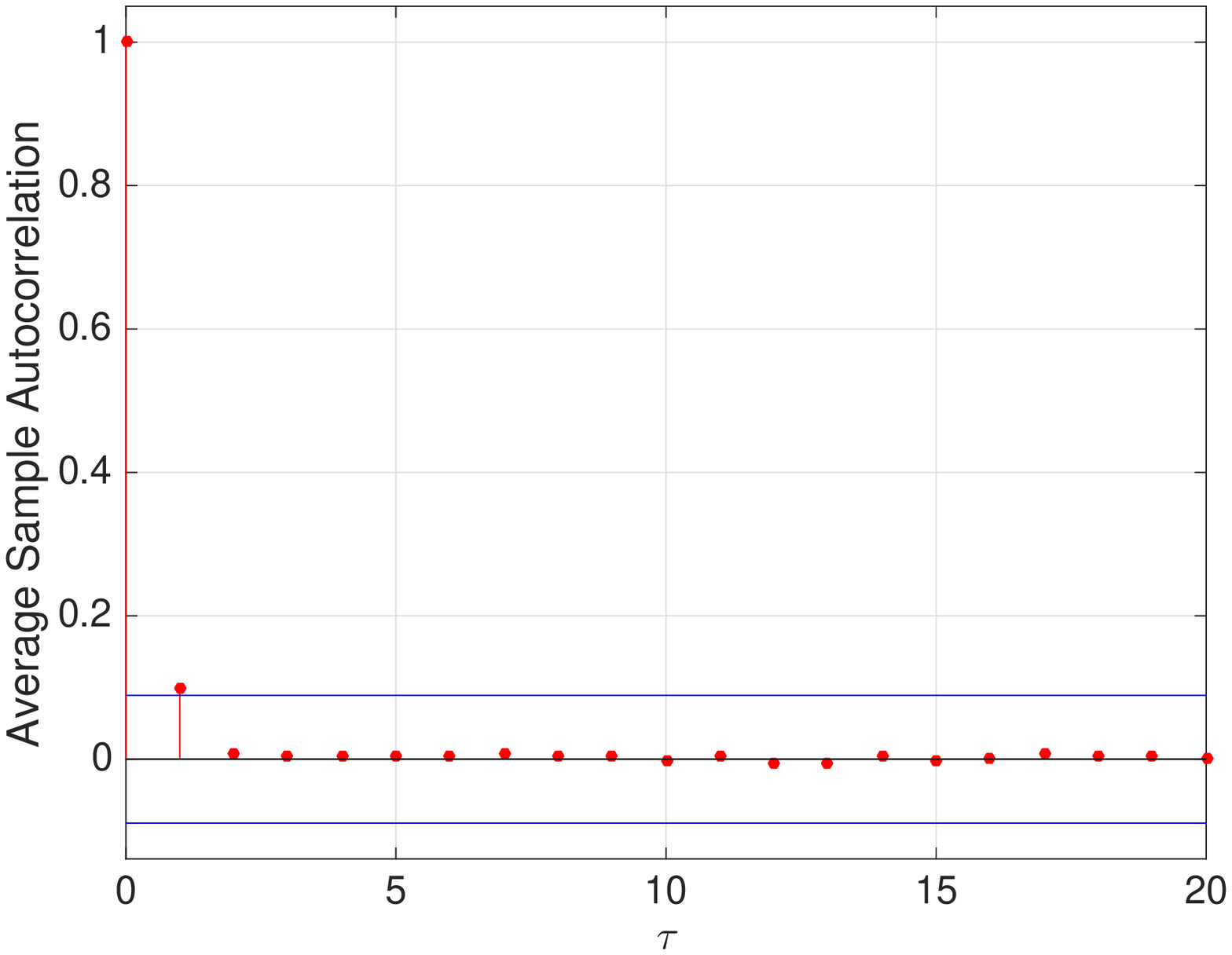}
			\caption{Average}
			\label{fig:Slow decay of autocorrelation in absolute
				returns LF - Average Returns}
		\end{subfigure}
		\hspace{2cm}
		\begin{subfigure}[H]{0.4\textwidth}
			\includegraphics[width=\textwidth]{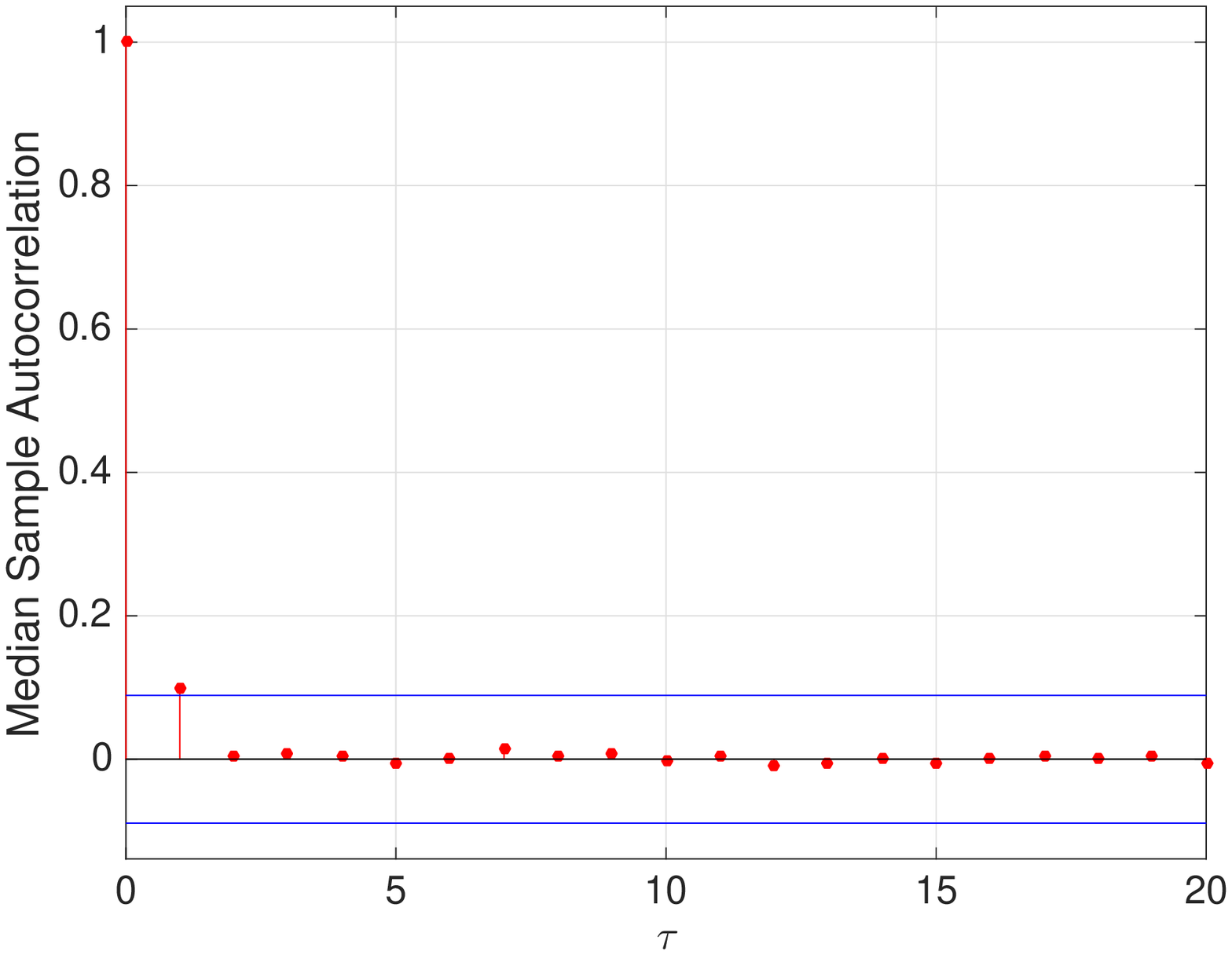}
			\caption{Median}
			\label{fig:Slow decay of autocorrelation in absolute
				returns LF - Median Returns}
		\end{subfigure}
		\caption{Long memory in low-frequency log-return over 100 simulations} 
		\floatfoot{\scriptsize{Note: The blue horizontal lines represent the
				approximate confidence bounds $[-0.0892;0.0892]$ of the autocorrelation function
				assuming the time-series is a moving average process.}}
		\label{fig:Long Memory low frequency}
	\end{figure}

	Figure \ref{fig:Long Memory low frequency} shows that in low-frequency
	time-series there is a short term positive dependence among absolute log-return \citep{Cont2001}, marginally significant only in the first period.
	
\begin{figure}[!hbpt] \centering
		\begin{subfigure}[H]{0.4\textwidth}
			\includegraphics[width=\textwidth]{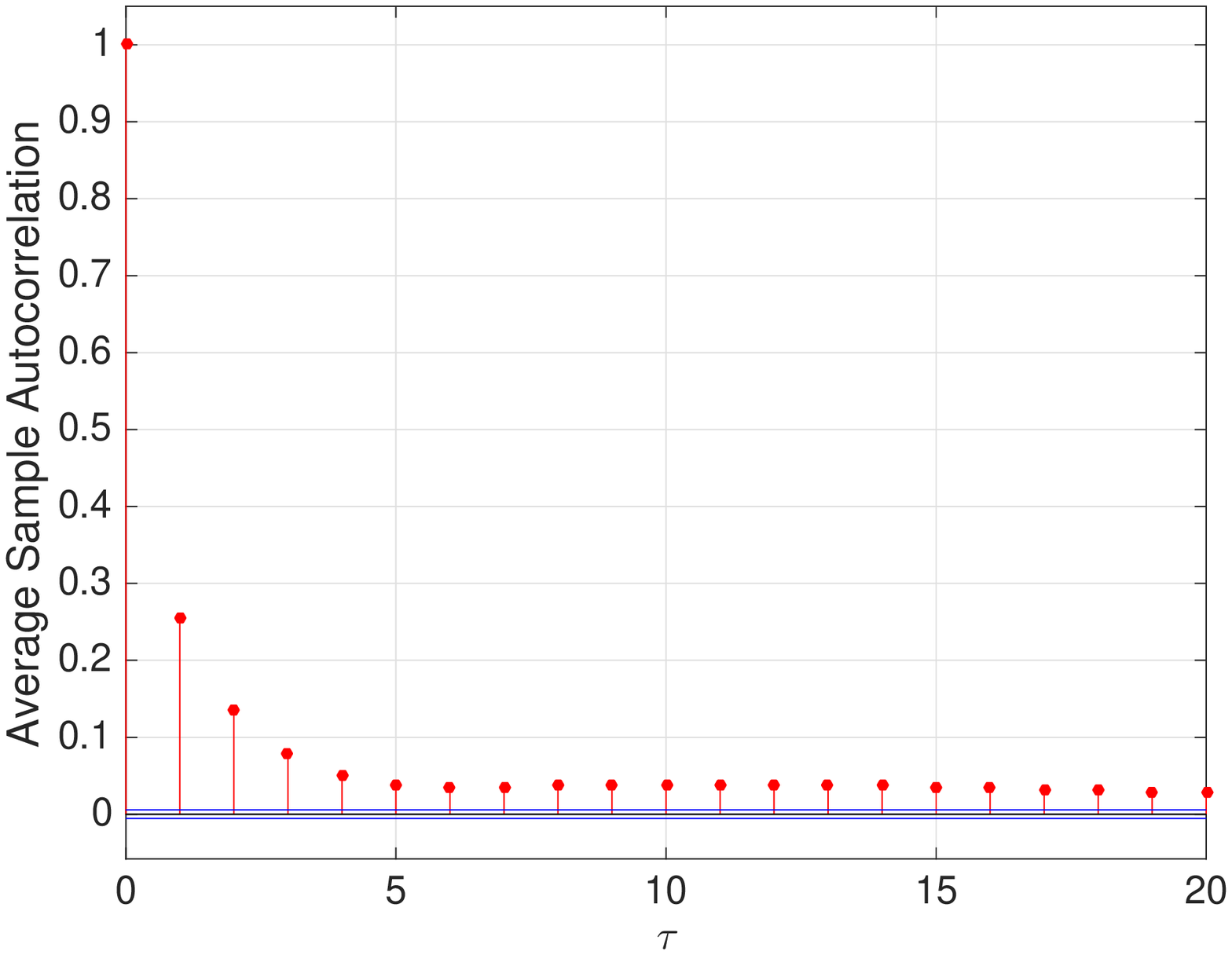}
			\caption{Average}
			\label{fig:Slow decay of autocorrelation in absolute
				returns HF - Average Returns}
		\end{subfigure}
		\hspace{2cm}
		\begin{subfigure}[H]{0.4\textwidth}
			\includegraphics[width=\textwidth]{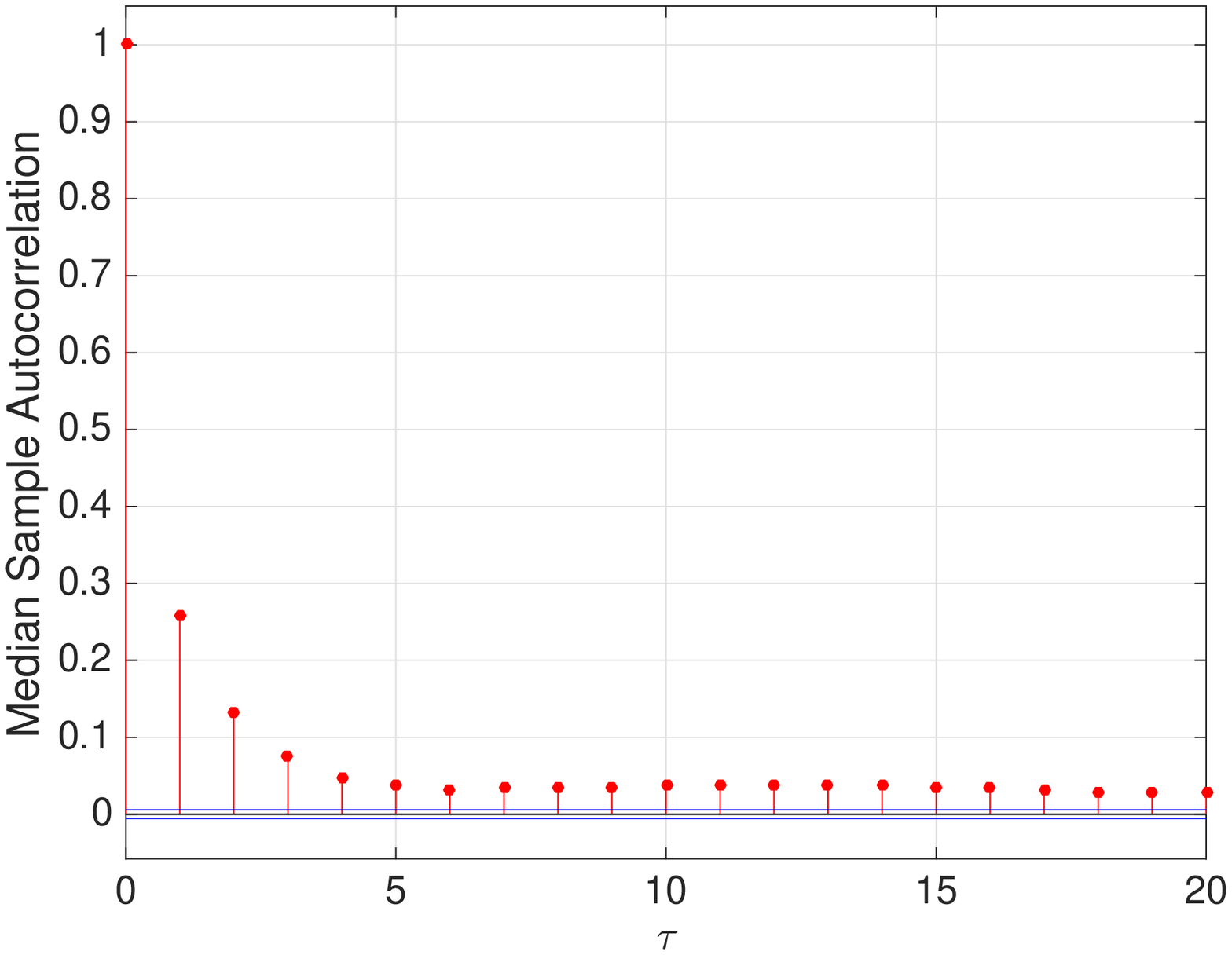}
			\caption{Median}
			\label{fig:Slow decay of autocorrelation in absolute
				returns HF - Median Returns}
		\end{subfigure}
		\caption{Long Memory in high-frequency log-return over 100 simulations} 
		\floatfoot{\scriptsize{Note: The blue horizontal lines represent the
				approximate confidence bounds $[-0.0063;0.0063]$ of the autocorrelation function
				assuming the time-series is a moving average process.}}
		\label{fig:Long Memory high frequency}
	\end{figure}

	However, figure \ref{fig:Long Memory high frequency} shows that in
	high-frequency time-series the autocorrelation function of absolute log-return
	decays slowly and this is sometimes interpreted as a sign of long-range
	dependence.
	
	\citep{Plerou2001} observe that long-range volatility correlations
	arise from trading activity, which we cover in sections
	\ref{subsection:Transaction Size1}--\ref{subsection:Transaction Size2}.
	
	\subsubsection{Power Law Behaviour of Returns}
	\label{subsubsec:Power Law Behaviour of Returns}
	
	Table \ref{table:Power law of returns over 100 simulations} shows the results of
	all the steps described in sections \ref{subsubsec:Power Law Behaviour of Returns
		SF}--\ref{subsubsec:Power Law Behaviour of Returns SF2}. If the calculated p-value is smaller than 0.05 then we rule out the
	power-law hypothesis. Conversely, the hypothesis is a plausible one for the data
	in question. Normally low values of $p$ are considered to be good, since they
	indicate that the null hypothesis is unlikely to be correct. In
	\citep{Clauset2009}, by contrast, the p-value is used as a measure of the
	hypothesis we are trying to verify, and hence high values do not reject the
	power-law hypothesis. We use this analysis when investigating the existence of
	power laws distributions.
	
	\begin{table}[!hbpt] \centering
		\begin{threeparttable}
			\caption{Power law of returns over 100 simulations}
			\label{table:Power law of returns over 100 simulations}
			\footnotesize
			\begin{tabular}{c c c c}
				\toprule
				{} & {Mean} & {Median} & {Standard Deviation}\\
				\midrule
				$\hat{\zeta_{r}}$ & 6.37 & 5.75 & (2.42)\\
				$\hat{x}_{min}$ & 0.0019 & 0.0018 & ($\convert{3.950009987592877e-04}$)\\
				p-value & 0.32 & 0.18 & (0.32)\\
				%D statistic & 0.0646 & 0.0666 & & & \\
				Uncertainty $\hat{\zeta_{r}}$  & 1.91 & 1.58 & (1.18)\\
				Uncertainty $\hat{x}_{min}$  &
				$\convert{0.0003111432989709963}$ &
				$\convert{0.0002912975961896769}$ &
				$\convert{0.00009271194915748616}$\\
				\bottomrule
			\end{tabular}
			\scriptsize
			\begin{tablenotes} 
				\item Note: In the unregulated treatment with
				low-frequency data, 76 out of 100 simulations have a p-value greater than 0.05,
				and therefore consistent with the hypothesis that $x$ is drawn from a
				distribution of the form of equation \ref{eq:power law}.
			\end{tablenotes}
		\end{threeparttable}
	\end{table}
	
	Contrary to what is found in employing the Hill estimator, the applicability of
	the power law estimators used by \citep{Clauset2009} to smaller data sets is not
	considered to be a problem. \citep{Clauset2009} suggest that a sample of $n=50$
	is a reasonable rule of thumb for extracting reliable parameter estimates. Our low-frequency
	data sets are $n=503$, and therefore can be considered as being sufficiently
	large to estimate reliable scaling parameters.
	
	We conclude that 76 out of 100 returns distributions follow a power-law
	distribution and that the scaling parameter of the power-law behaviour of
	returns is greater than the typical inverse cubic, as is usual on longer time
	scales \citep{Ehrentreich2008}. Hence, our experimental treatment confirms the
	scaling behaviour for the majority of simulations, despite the return
	distribution scale lying outside the stable L\'{e}vy-regime.
	\citep{Ehrentreich2008} states that a more Gaussian exponent is
	common in longer time scales, and \citep{VoitJohannes2005} refers the case of
	the DAX as another example of a power law behaviour outside the L\'{e}vy-stable
	range.
	
	\citep{Plerou2000} and \citep{Plerou2001} conclude that the number of trades, as
	covered in sections \ref{subsection:Transaction Size1}--\ref{subsection:Transaction Size2}, cannot alone explain
	the value $\zeta_{r} \approx 3$ and suggest that the pronounced tails of the
	distribution of returns are possibly due to local variance.
	
	\subsubsection{Power Law Behaviour of Volatility}
	\label{subsubsec:Power Law Behaviour of Volatility}
	
	Table \ref{table:Power law of volatility over 100 simulations} shows the
	regression fits of low-frequency volatility for an averaging window T=5 days
	with $\Delta{t}=1$ day. According to \citep{Liu1999}, the larger the choice of
	time interval T, the more accurate is the volatility estimation, but a large
	value for T also implies poor resolution in time. Hence we opted for a time
	interval of 5.
	
	\begin{table}[!hbpt] \centering
		\begin{threeparttable}
			\caption{Power law of volatility over 100 simulations}
			\label{table:Power law of volatility over 100 simulations}
			\footnotesize
			\begin{tabular}{c c c c}
				\toprule
				{} & {Mean} & {Median} & {Standard Deviation}\\
				\midrule
				$\hat{\zeta_{\sigma}}$ & 6.96 & 5.93 & (4.06)\\
				$\hat{x}_{min}$ & 0.0011 & 0.0011 & ($\convert{0.0002141126415718931}$)\\
				p-value & 0.52 & 0.54 & (0.32)\\
				% D statistic & 0.0646 & 0.0666 & & & \\
				Uncertainty $\hat{\zeta_{\sigma}}$  & 2.60 & 1.91 & (1.90)\\
				Uncertainty $\hat{x}_{min}$  & $\convert{0.0001632030535827198}$ &
				$\convert{0.0001597390610122359}$ &
				($\convert{0.00005070285446253053}$)\\
				\bottomrule
			\end{tabular}
			\scriptsize
			\begin{tablenotes}
				\item Note: In the low-frequency unregulated treatment, 92 out of 100
				simulations have a p-value greater than 0.05, and are therefore consistent with
				the hypothesis that $x$ is drawn from a distribution of the form of equation
				\ref{eq:power law volatility}.
			\end{tablenotes}
		\end{threeparttable}
	\end{table}
	
	From the estimates we find that the cumulative distribution of low-frequency
	volatility is consistent with a power law asymptotic behaviour (as in equation
	\ref{eq:power law volatility}) in 92 percent of simulations. The
	results reveal that the estimates of $\zeta_{\sigma}$ for T=5 days with
	$\Delta t=1$ day lie outside the stable L\'{e}vy range of $0<\zeta_{\sigma}<2$.
	
	\subsubsection{Volatility Clustering}
	\label{subsubsec:Volatility Clustering}
	
\begin{figure}[!hbpt] \centering
		\begin{subfigure}[h]{0.4\textwidth}
			\includegraphics[width=\textwidth]{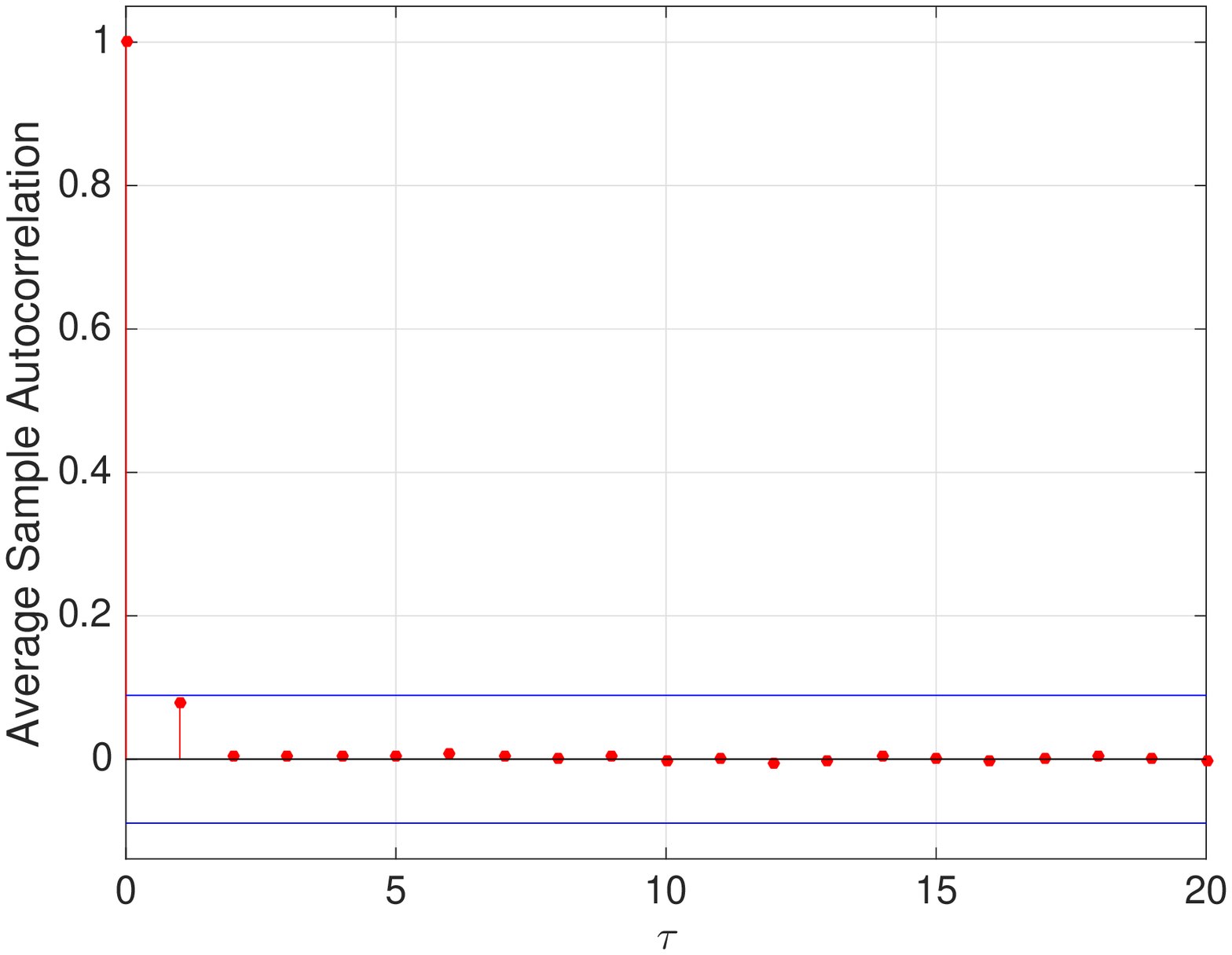}
			\caption{Average Autocorrelation}
			\label{fig:Volatility Clustering - Average Returns}
		\end{subfigure}
		\hspace{2cm}
		\begin{subfigure}[h]{0.4\textwidth}
			\includegraphics[width=\textwidth]{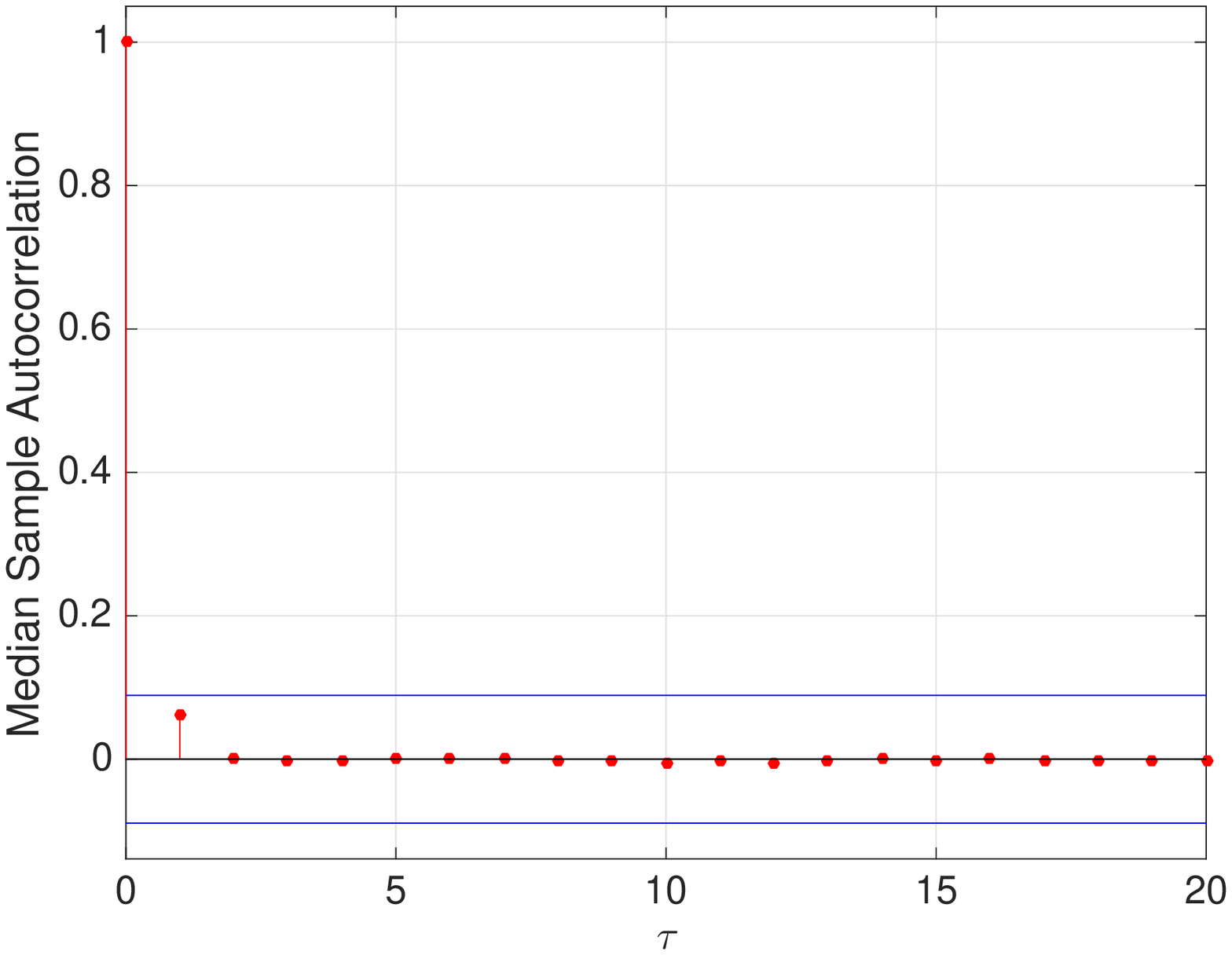}
			\caption{Median Autocorrelation}
			\label{fig:Volatility Clustering - Median Returns}
		\end{subfigure}
		\caption{Volatility clustering in daily squared returns over 100
			simulations}
		\label{fig:Volatility Clustering - low-frequency}
		\floatfoot{\scriptsize{Note: The blue horizontal lines represent the approximate
				confidence bounds $[-0.0892;0.0892]$ of the autocorrelation function assuming the time
				serie is a moving average process.}}
	\end{figure}

	Figure \ref{fig:Volatility Clustering - low-frequency} shows that the magnitude
	of this effect is close to zero in low-frequency returns, except for the
	first-order correlation, which is not significant across all simulations.
	
\begin{figure}[!hbpt] \centering
		\begin{subfigure}[H]{0.4\textwidth}
			\includegraphics[width=\textwidth]{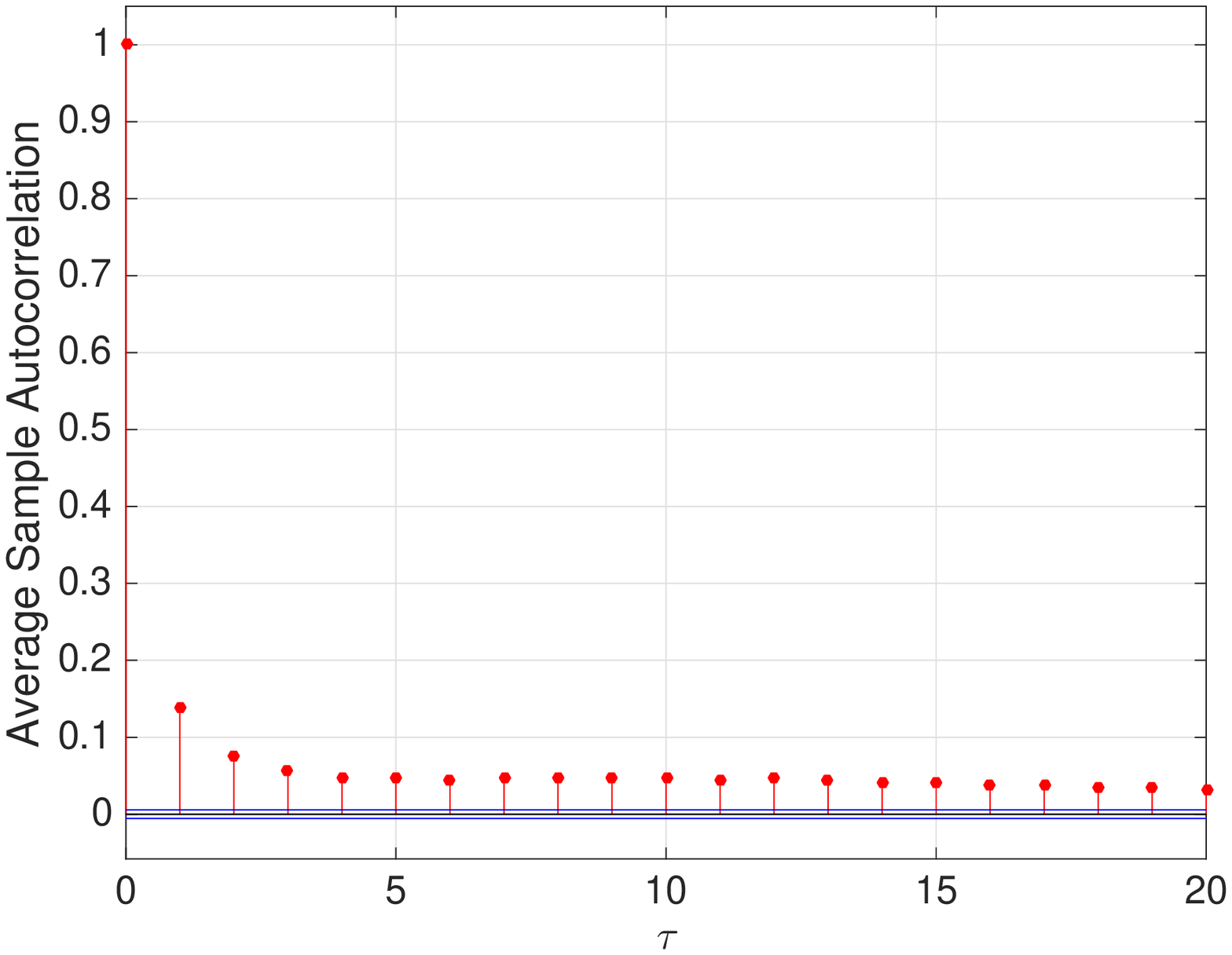}
			\caption{Average Autocorrelation}
			\label{fig:Volatility Clustering - Average Volatility}
		\end{subfigure}
		\hspace{2cm}
		\begin{subfigure}[H]{0.4\textwidth}
			\includegraphics[width=\textwidth]{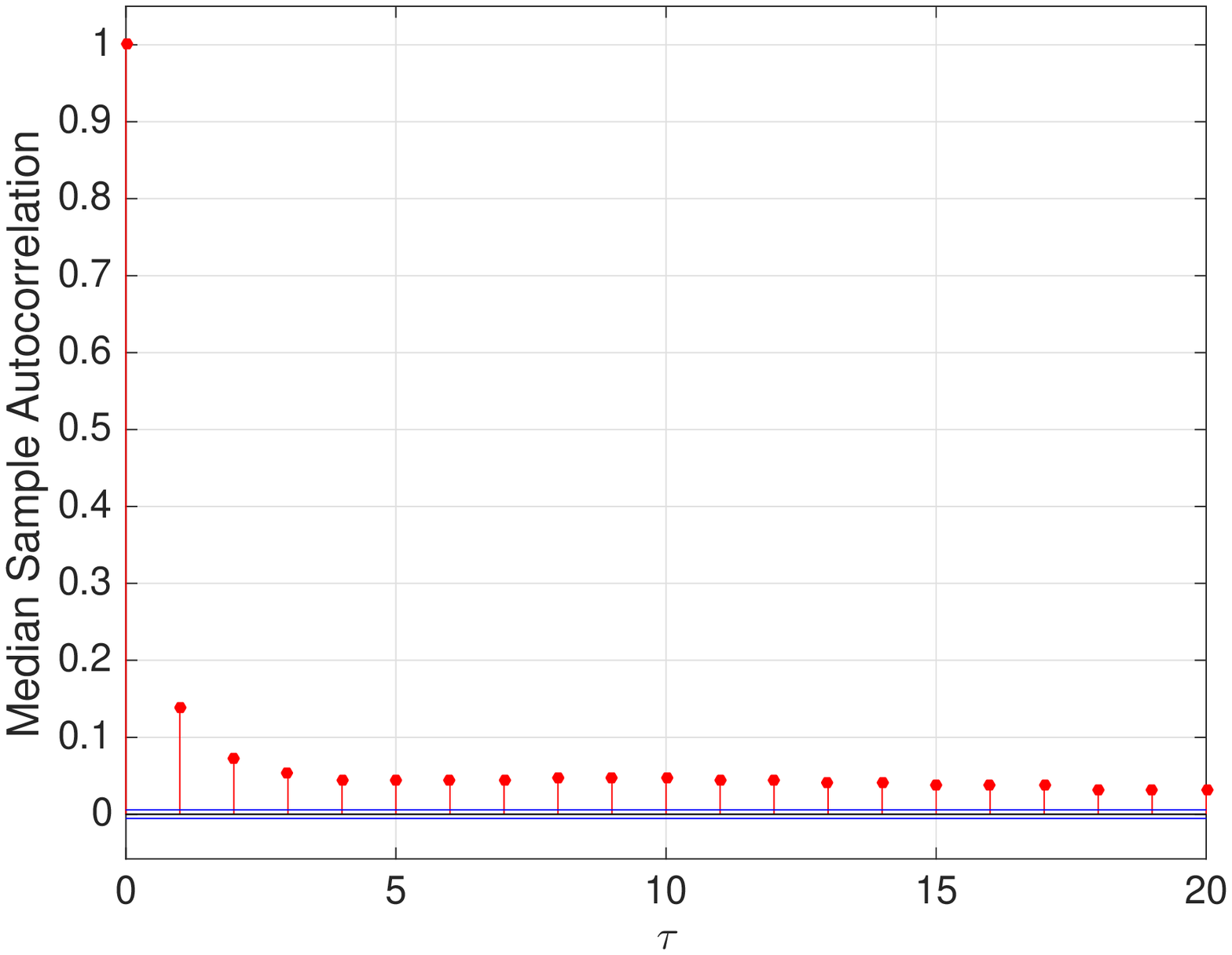}
			\caption{Median Autocorrelation}
			\label{fig:Volatility Clustering - Median Volatility}
		\end{subfigure}
		\caption{Volatility clustering in high-frequency squared returns over 100
			simulations}
		\label{fig:Volatility Clustering over 100 simulations HF}
		\floatfoot{\scriptsize{Note: The blue horizontal lines represent the approximate
				confidence bounds $[-0.0063;0.0063]$ of the autocorrelation function assuming the time
				serie is a moving average process.}}
	\end{figure}

	Figure \ref{fig:Volatility Clustering over 100 simulations HF} shows serial
	correlation effects in high-frequency returns, prolonged and higher than in
	low-frequency returns. The autocorrelation function of volatility decreases
	slowly to zero and is statistically significant for long time periods. The
	positive results obtained for the autocorrelation function of the squared
	returns and their slow decay are verified in empirical financial data and are
	often mentioned as a ``quantitative manifestation" of volatility clustering
	\citep{Cont2007}.
	
	\subsubsection{Volatility Volume Correlations}
	\label{subsubsec:Volatility Volume Correlations}
	
	Figure \ref{fig:Cross-correlation between volatility and trading volume over 100
		simulations LF} exhibits cross-correlation between volatility and volume for
	low-frequency data that is not statistically significant. \label{subsubsec:Volatility Volume Correlations1}

	\begin{figure}[!hbpt] \centering
		\begin{subfigure}[H]{0.4\textwidth}
			\includegraphics[width=\textwidth]{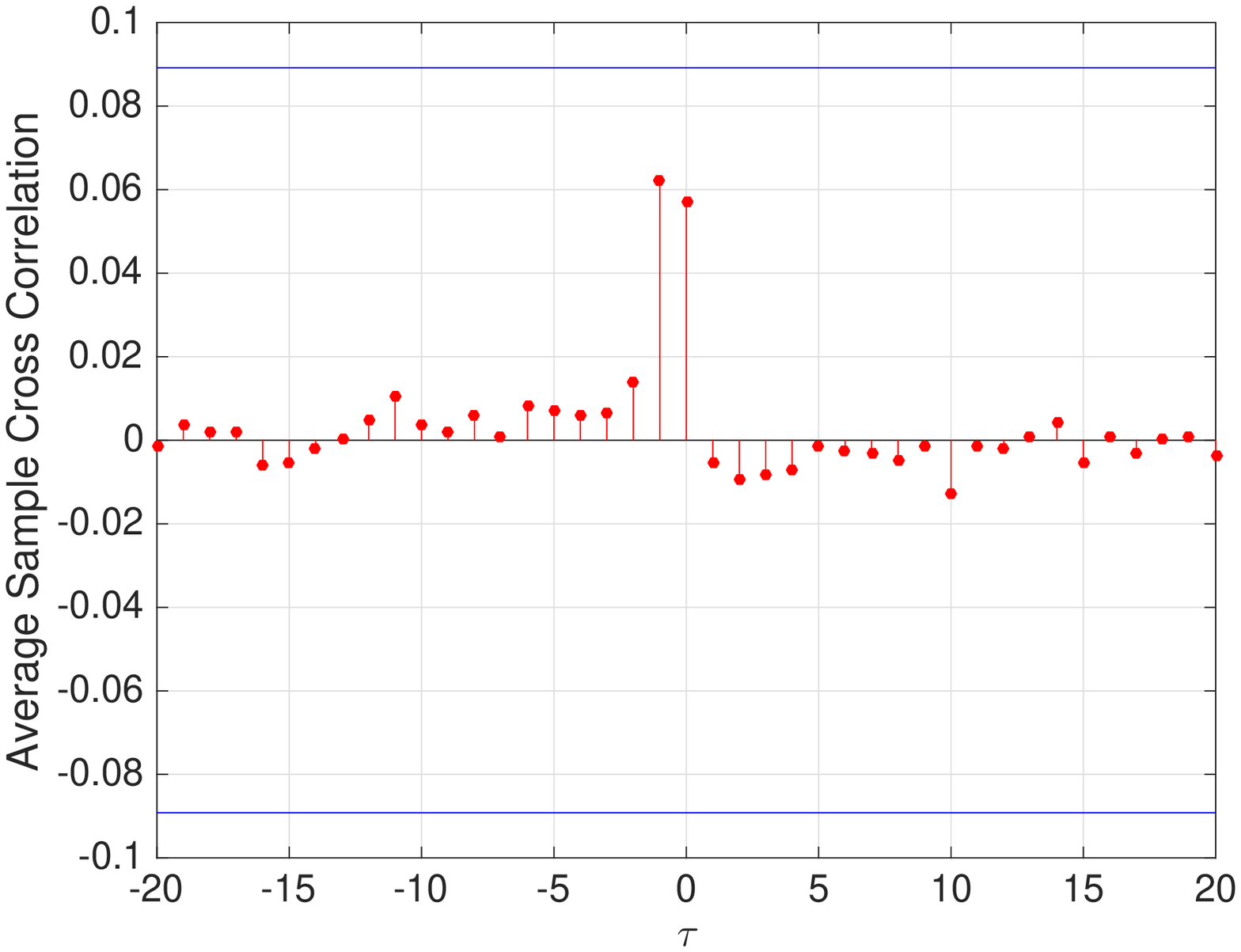}
			\caption{Average cross-correlation}
			\label{fig:Size of transactions - Average LF}
		\end{subfigure}
		\hspace{2cm}
		\begin{subfigure}[H]{0.4\textwidth}
			\includegraphics[width=\textwidth]{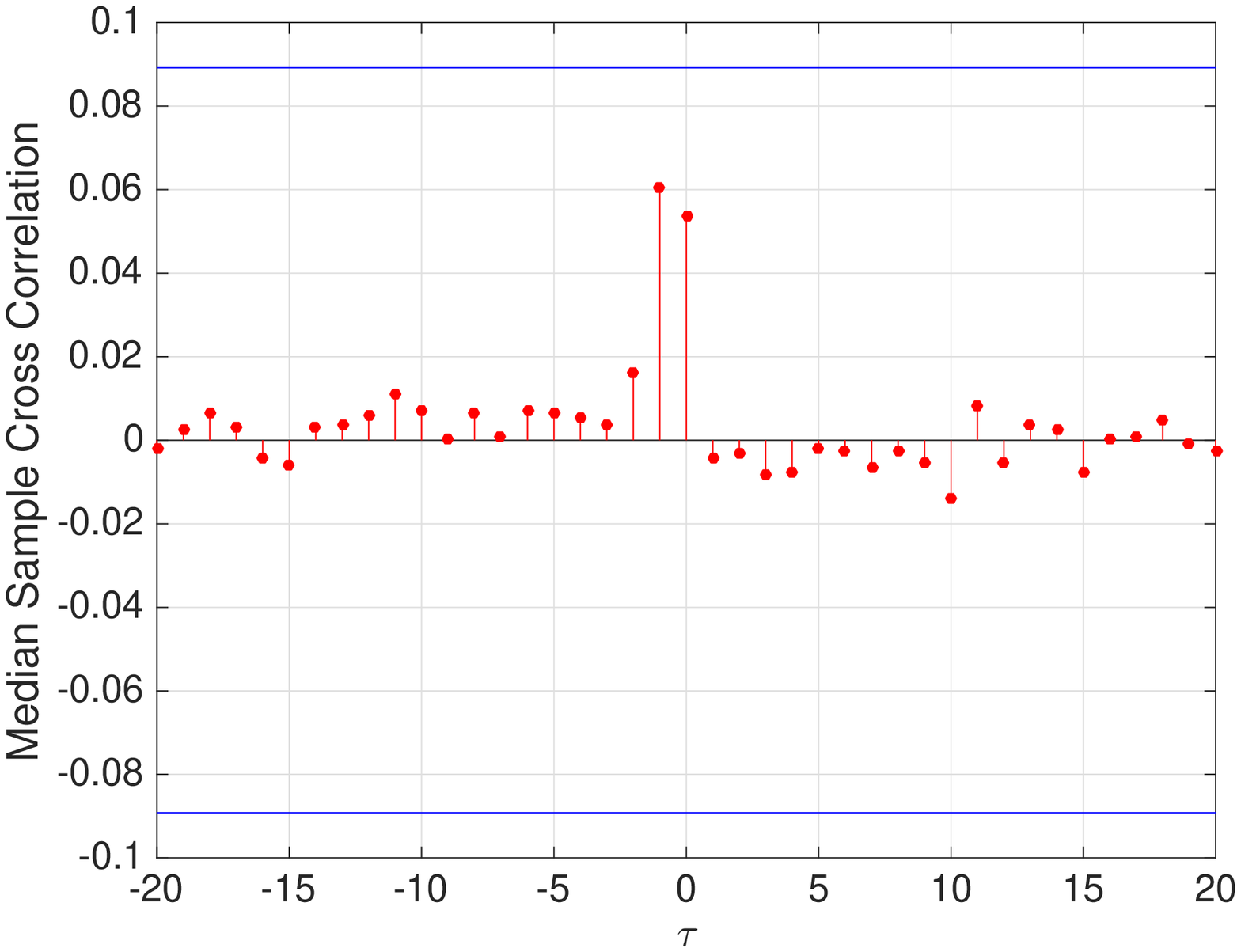}
			\caption{Median cross-correlation}
			\label{fig:Size of transactions - Median LF}
		\end{subfigure}
		\caption{Cross-correlation between volatility and trading volume in
			low-frequency data over 100 simulations} 
		\label{fig:Cross-correlation between volatility and trading volume over 100
			simulations LF}
		\floatfoot{\scriptsize{Note: The blue horizontal lines represent the approximate
				upper and lower confidence bounds $[-0.0892;0.0892]$, assuming volatility and trading
				volume are uncorrelated.}}
	\end{figure}
	
	However, figure \ref{fig:Cross-correlation between volatility and trading volume
		over 100 simulations HF} shows that the cross-correlation between volatility and
	volume for high-frequency data is positive and both volatility and volume show
	the same type of ``long memory'' behaviour \citep{{Lobato2000}, {Cont2007}}.
	The cross-correlation function of the high-frequency data was analysed by
	aggregating and averaging the data over intervals of 30 ticks.
	
	\begin{figure}[!hbpt] \centering
		\begin{subfigure}[H]{0.4\textwidth}
			\includegraphics[width=\textwidth]{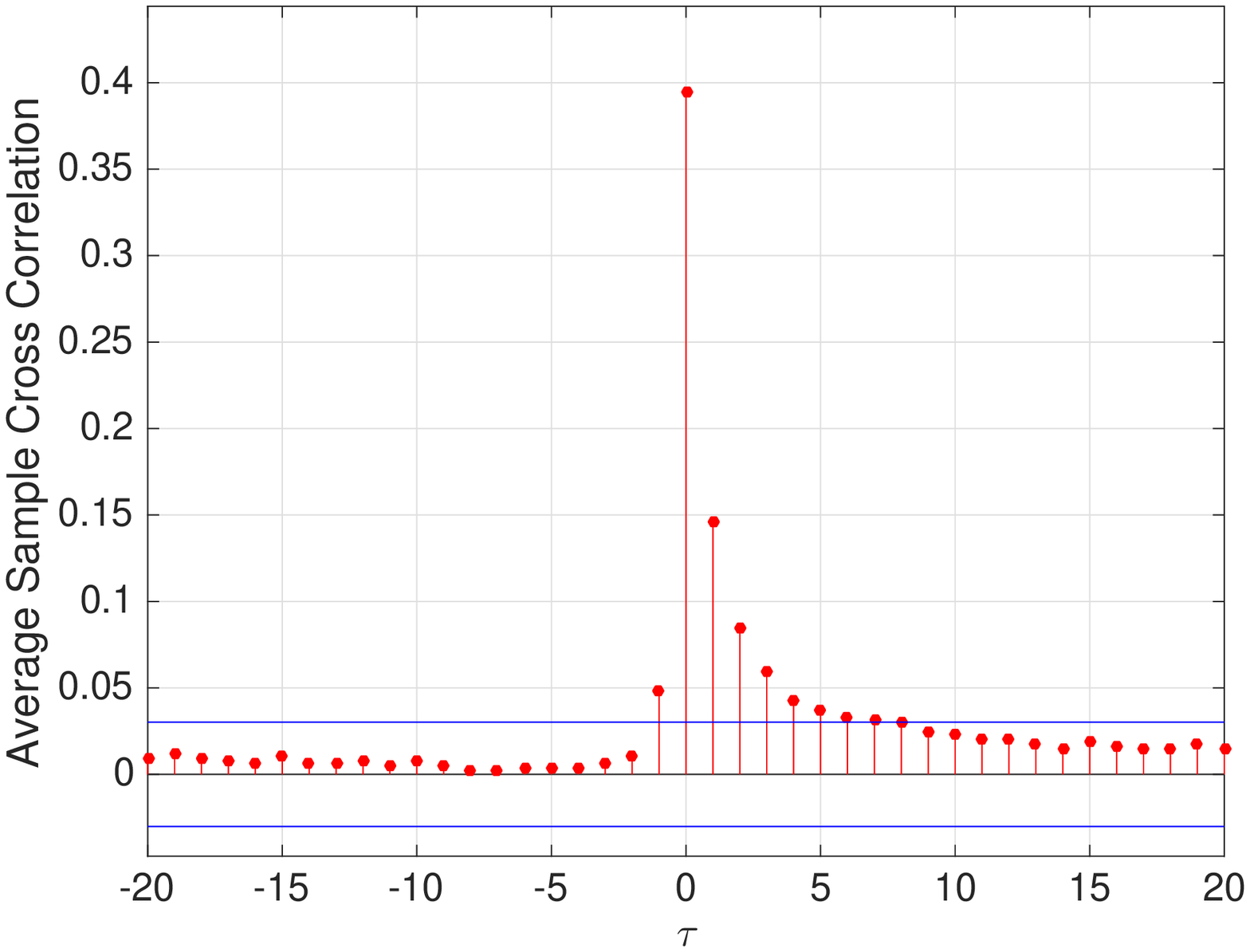}
			\caption{Average cross-correlation}
			\label{fig:Size of transactions - Average HF}
		\end{subfigure}
		\hspace{2cm}
		\begin{subfigure}[H]{0.4\textwidth}
			\includegraphics[width=\textwidth]{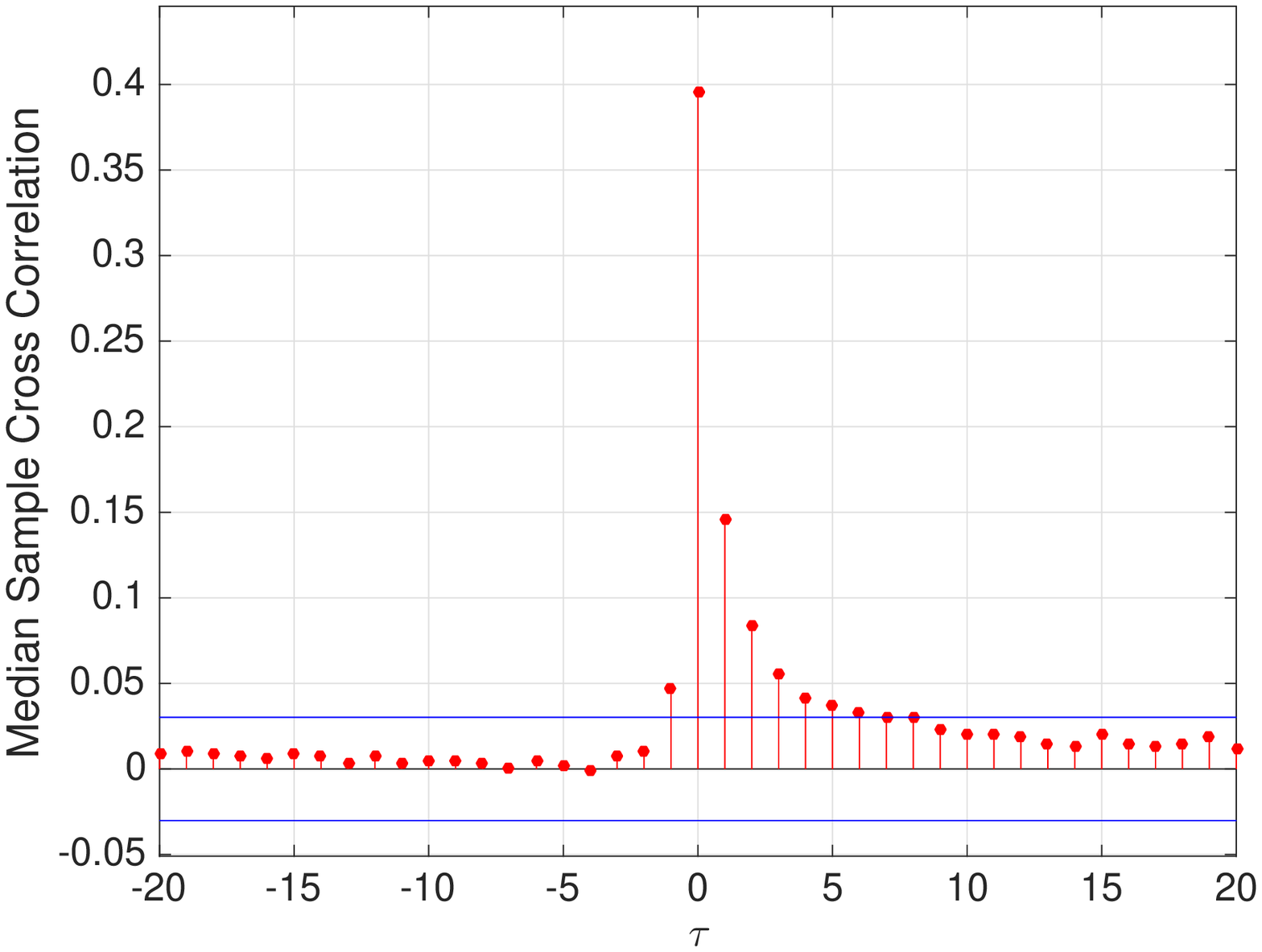}
			\caption{Median cross-correlation}
			\label{fig:Size of transactions - Median HF}
		\end{subfigure}
		\caption{Cross-correlation between volatility and trading volume in
			high-frequency data over 100 simulations}
		\label{fig:Cross-correlation between volatility and trading volume over 100
			simulations HF}
		\floatfoot{\scriptsize{Note: The blue horizontal lines represent the mean of
				approximate upper and lower confidence bounds over 100 simulations $[-0.0345;0.0345]$,
				assuming volatility and trading volume are uncorrelated.}}
	\end{figure}
	
	We also analyse the frequency of trades to identify any relationship
	between volatility and the number of trades. Figure
	\ref{fig:Cross-correlation between volatility and number of trades over 100
		simulations LF} shows that there is no significant cross-correlation for
	low-frequency data.
	
	\begin{figure}[!hbpt] \centering
		\begin{subfigure}[H]{0.4\textwidth}
			\includegraphics[width=\textwidth]{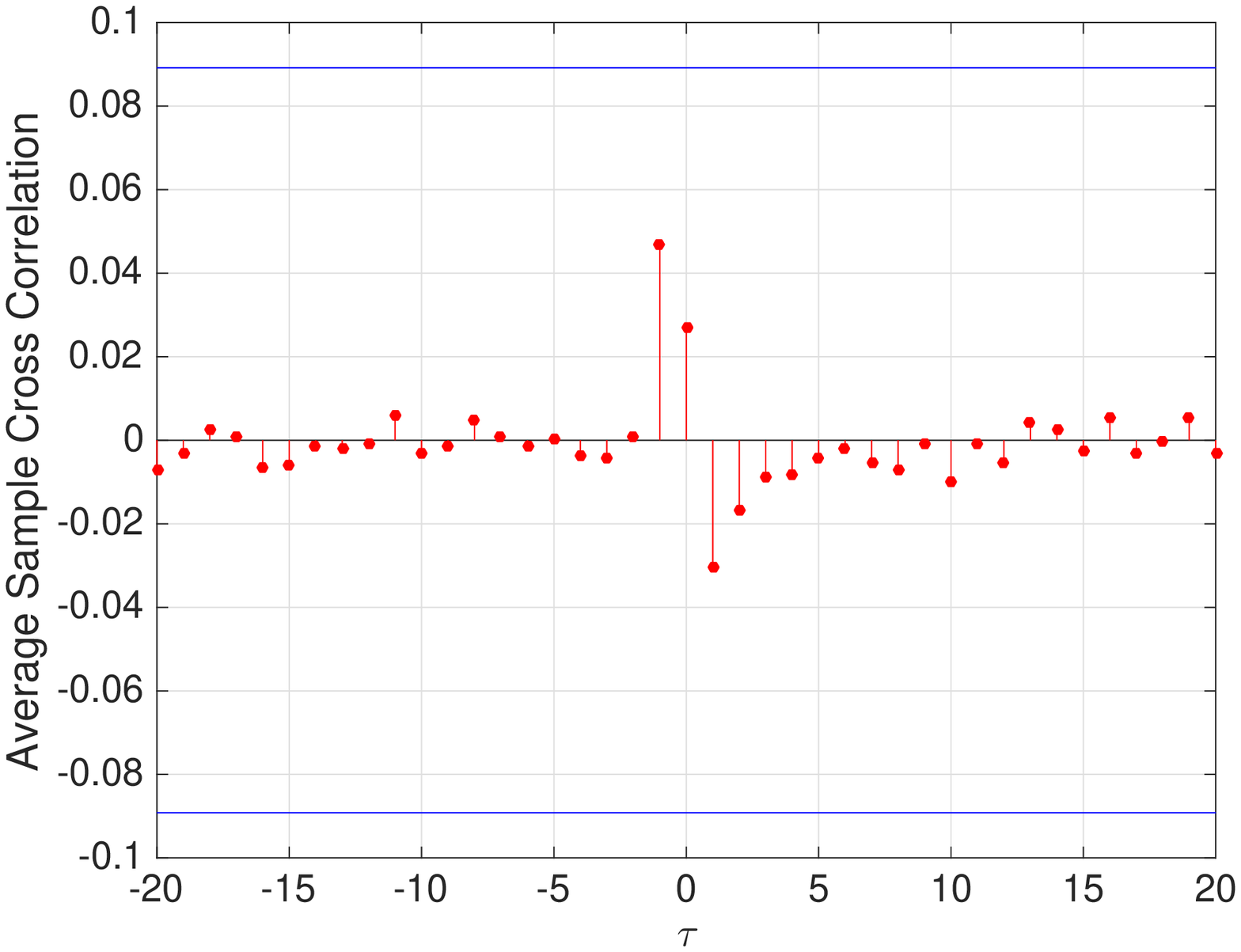}
			\caption{Average cross-correlation}
			\label{fig:Number transactions - Average LF}
		\end{subfigure}
		\hspace{2cm}
		\begin{subfigure}[H]{0.4\textwidth}
			\includegraphics[width=\textwidth]{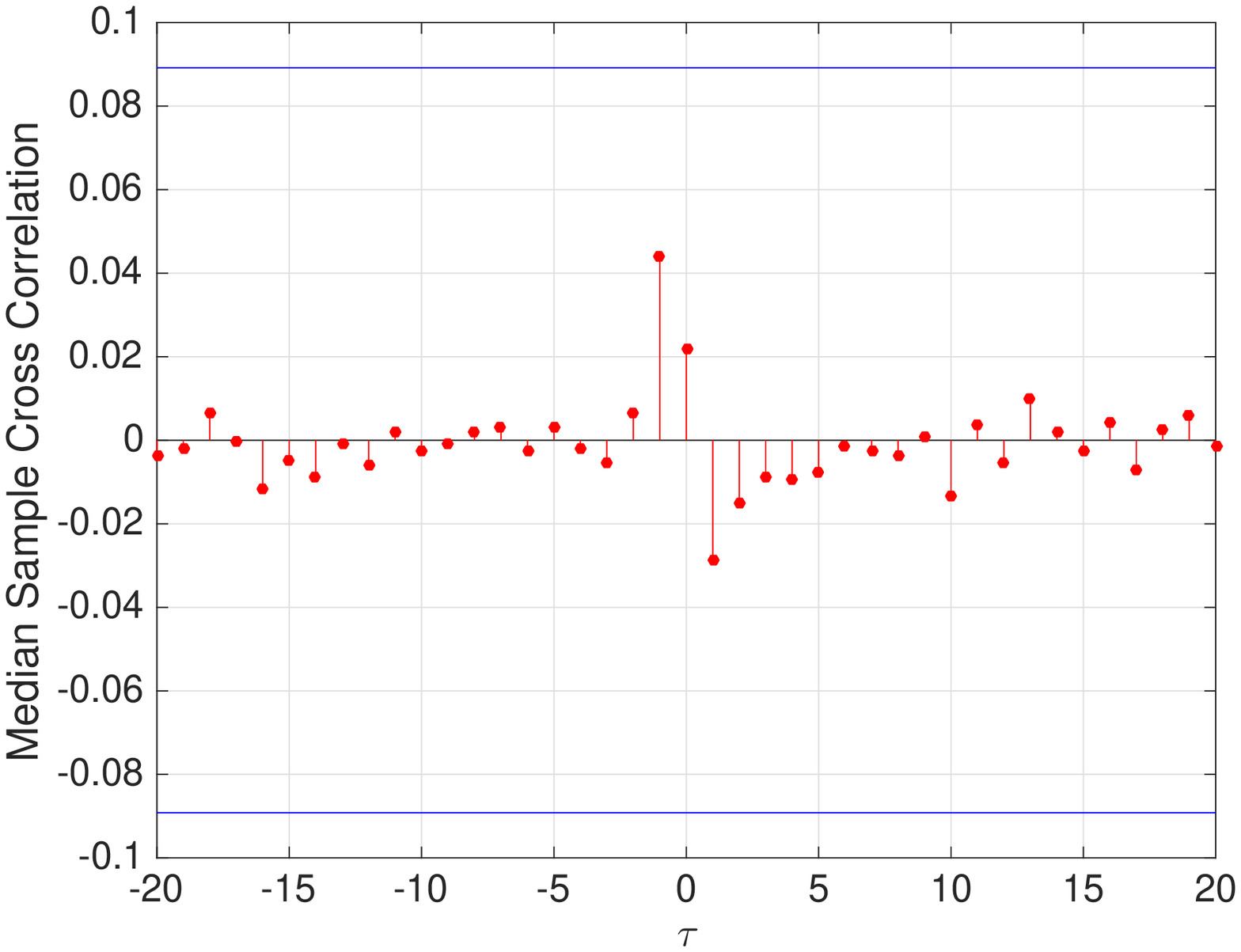}
			\caption{Median cross-correlation}
			\label{fig:Number transactions - Median LF}
		\end{subfigure}
		\caption{Cross-correlation between volatility and number of trades in
			low-frequency data over 100 simulations}
		\label{fig:Cross-correlation between volatility and number of trades over 100
			simulations LF}
		\floatfoot{\scriptsize{Note: The blue horizontal lines represent the approximate
				upper and lower confidence bounds $[-0.0892;0.0892]$, assuming volatility and number of
				trades are uncorrelated.}}
	\end{figure}
	
	In Figure \ref{fig:Cross-correlation between volatility and number of
		trades over 100 simulations HF} we observe that volatility and the number of
	trades only exhibit significant and positive correlation for current number of
	trades in high-frequency data and is only marginally significant for past trades
	in lag -1, as observed in \citep{Westerhoff2003}. \citep{Cont2007} observes that the heterogeneity in agents' time horizons, as in
	our model, may lead to volatility-volume relationships similar to those of
	actual markets.
	\label{subsubsec:Volatility Volume Correlations2}

	\begin{figure}[!hbpt] \centering
		\begin{subfigure}[H]{0.4\textwidth}
			\includegraphics[width=\textwidth]{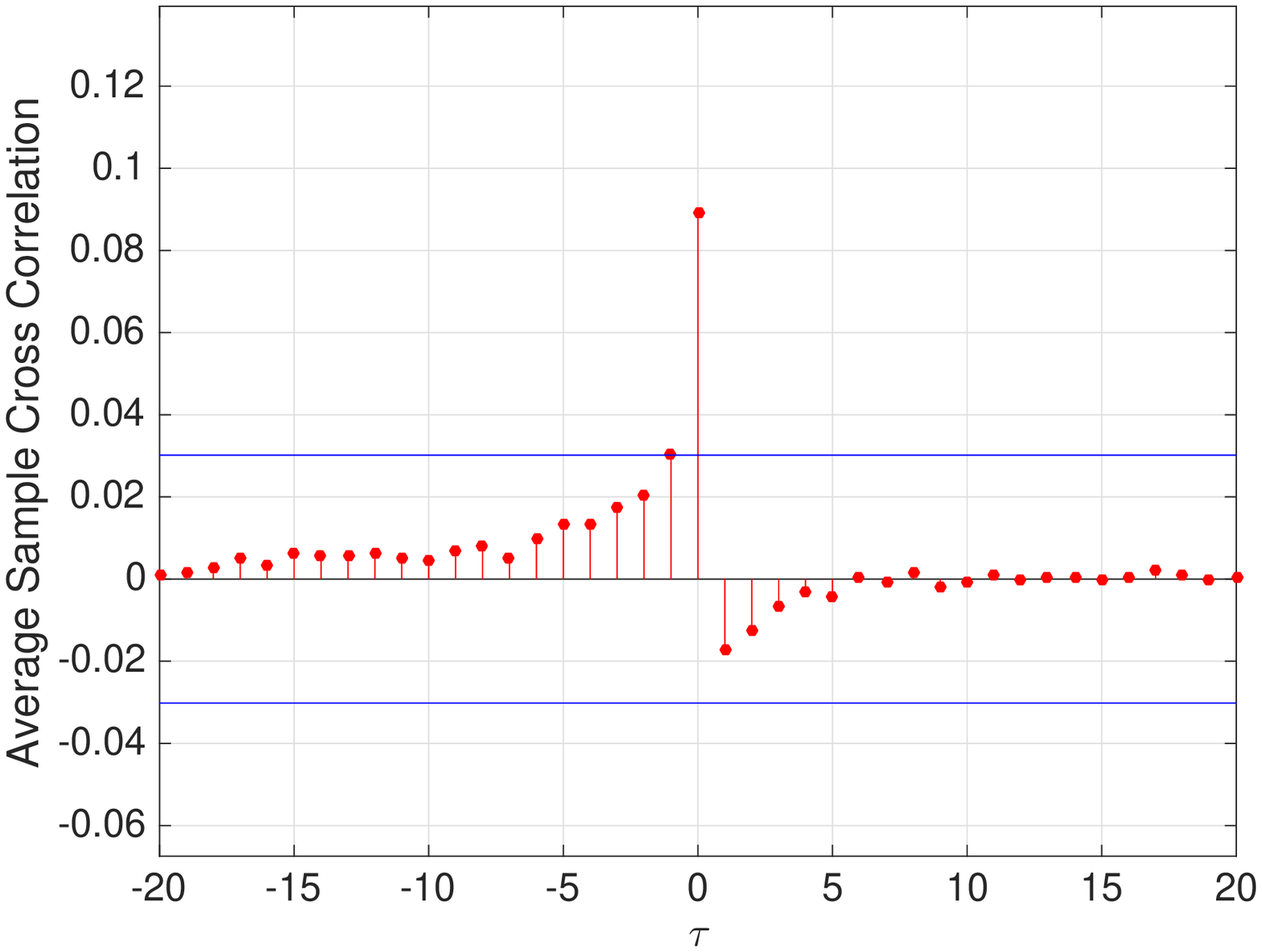}
			\caption{Average cross-correlation}
			\label{fig:Number transactions - Average HF}
		\end{subfigure}
		\hspace{2cm}
		\begin{subfigure}[H]{0.4\textwidth}
			\includegraphics[width=\textwidth]{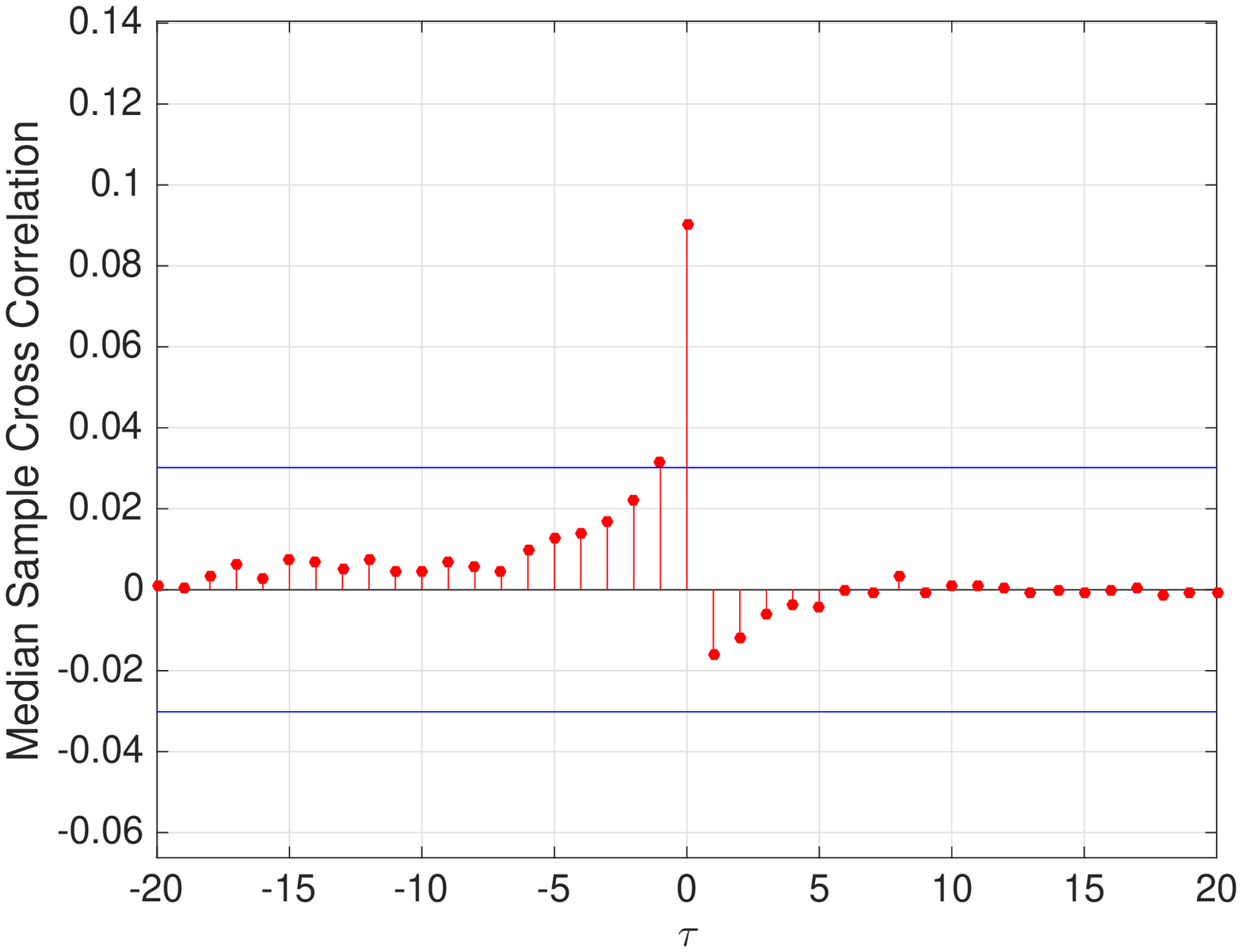}
			\caption{Median cross-correlation}
			\label{fig:Number transactions - Median HF}
		\end{subfigure}
		\caption{Cross-correlation between volatility and number of trades in
			high-frequency data over 100 simulations}
		\label{fig:Cross-correlation between volatility and number of trades over 100
			simulations HF}
		\floatfoot{\scriptsize{Note: The blue horizontal lines represent the mean of
				approximate upper and lower confidence bounds over 100 simulations $[-0.0345;0.0345]$,
				assuming volatility and number of trades are uncorrelated.}}
	\end{figure}
	
	\subsubsection{Unit Roots}
	\label{subsubsec:Unit Roots}

	Table \ref{table:Unit Root Tests} shows the results from the Phillips-Perron
	Unit Root Tests. We can conclude that the time-series of log-return are
	stationary for low- and high-frequency data.
	
	\begin{table}[!hbpt] \centering
		\begin{threeparttable}
			\caption{Phillips-Perron Unit Root Tests}
			\label{table:Unit Root Tests}
			\footnotesize
			\begin{tabular}{c c c c c} 
				\toprule
				{}& {} & \multicolumn{3}{c}{Phillips-Perron}\\
				
				{} & {} & {Test statistics} & {p-value} & {Critical value}\\
				\midrule
				Unregulated & \raisebox{1.5ex}{LF} &
				\raisebox{1.5ex}{[-40.31; -23.24]} & \raisebox{1.5ex}{0.001}
				& \raisebox{1.5ex}{-1.9411}\\
				& HF & {[-422.04; -288.16]} & 0.001 & -1.9416\\
				
				\addlinespace
				\addlinespace
				
				VaR & \raisebox{1.5ex}{LF} &
				\raisebox{1.5ex}{[-39.37; -25.97]} & \raisebox{1.5ex}{0.001}
				& \raisebox{1.5ex}{-1.9411}\\
				
				& HF & {[-727.14; -381.31]} & 0.001 & -1.9416\\
				
				\addlinespace
				\addlinespace
				
				ES & \raisebox{1.5ex}{LF} &
				\raisebox{1.5ex}{[-38.59; -27.43]} & \raisebox{1.5ex}{0.001}
				& \raisebox{1.5ex}{-1.9411}\\
				
				& HF & {[-653.47; -329.11]} & 0.001 & -1.9416\\
				\bottomrule
			\end{tabular}
			\scriptsize
			\begin{tablenotes}
				\item Note: ADF tests present similar results. All unit roots tests indicate
				rejection of the unit-root null in favor of the alternative model, then stationarity. Significance
				level for the hypothesis tests is 0.05. Intervals represent minimum and maximum values of the statistical
				test. The number of lagged difference terms is 0. Model variant is
				autoregressive. Test statistic is standard t statistic using ordinary
				least squares estimates of the coefficients in the alternative model, and
				p-values are left-tail probabilities.
			\end{tablenotes}
		\end{threeparttable}
	\end{table}
	
	In Table \ref{table:Stationarity Tests} we apply the KPSS test which shows that
	all the simulated time-series of returns are stationary. These results are
	consistent with the findings that usually prices are not stationary but returns,
	the differences, are \citep{{deVries1994}, {alexander2008}}.
	
	\begin{table}[!hbpt] \centering
		\begin{threeparttable}
			\caption{Stationarity Tests}
			\label{table:Stationarity Tests}
			\footnotesize
			\begin{tabular}{c c c c c} 
				\toprule
				{} & {} & \multicolumn{3}{c}{KPSS}\\
				
				{} & {} & {Test statistics} & {p-value} & {Critical value}\\
				\midrule
				Unregulated & \raisebox{1.5ex}{LF} & \raisebox{1.5ex}{[0.002; 0.077]} &
				\raisebox{1.5ex}{0.1} & \raisebox{1.5ex}{0.1460}\\
				
				& HF & [$\convert{00006541642591489175}; $\convert{0.00006896426231650868}] &
				0.1 & 0.1460\\
				\addlinespace
				\addlinespace			
				VaR & \raisebox{1.5ex}{LF} &
				\raisebox{1.5ex}{$[0.0027; 0.0787]$} & \raisebox{1.5ex}{0.1} &
				\raisebox{1.5ex}{0.1460}\\
				
				& HF & [$\convert{0.0002624476647748951}$; $\convert{0.00003077004358771957}$] &
				0.1 & 0.1460\\
				\addlinespace
				\addlinespace
				ES & \raisebox{1.5ex}{LF} &
				\raisebox{1.5ex}{$[0.0027;0.0834]$}
				& \raisebox{1.5ex}{0.1} & \raisebox{1.5ex}{0.1460}\\
				\comment{Daily: 19 h=1}
				
				& HF & [$\convert{0.00003398231163871601}$; $\convert{0.0003534097975677140}$] &
				0.1 & 0.1460\\
				\bottomrule
			\end{tabular}
			\scriptsize
			\begin{tablenotes}
				\item Notes: All treatments fail to reject the null hypothesis that returns are
				stationary. Significance level for the hypothesis tests is 0.05. Intervals
				represent minimum and maximum values of the statistical test. The number of autocovariance lags to include in the Newey-West estimator of the long-run variance is 0. The deterministic trend
				term $\delta t$ is included in the model. KPSS tests compute test
				statistics using an ordinary least squares (OLS) regression. Critical values are
				for right-tail probabilities.
			\end{tablenotes}
		\end{threeparttable}
	\end{table}
	
	\subsection{Trading Volume}
	\label{subsection:Trading Volume}
	
	As mentioned in sections \ref{subsection:Trading Volume SF1}--\ref{subsection:Trading Volume SF2}, there is no
	consensus about the distribution of trading volume being L\'{e}vy-stable
	\citep{Vijayaraghavan2011}. The results from our model show scaling parameters
	falling outside the L\'{e}vy-stable interval.
	
	\subsubsection{Power Law Behaviour of Trading Volume}
	\label{subsubsec:Power Law Behaviour of Trading Volume}
	
	Table \ref{table:Power law of trading volume over 100 simulations} exhibits
	extremely high scaling parameters, which indicate the existence of a power law
	behaviour outside the L\'{e}vy-stable interval. Nevertheless, our results show
	that 76 out of 100 simulations exhibit p-values consistent with the existence of
	a power law distribution.
	
	\begin{table}[!hbpt] \centering
		\begin{threeparttable}
			\caption{Power law of trading volume over 100 simulations}
			\label{table:Power law of trading volume over 100 simulations}
			\footnotesize
			\begin{tabular}{c c c c}
				\toprule
				{} & {Mean} & {Median} & {Standard Deviation} \\
				\midrule
				$\hat{\zeta_{V}}$ & 17.50 & 16.86 & (4.32)\\
				$\hat{x}_{min}$ & 11763 & 11801 & (740.67)\\
				p-value & 0.976 & 0.242 & (0.323)\\
				% D statistic & 0.0646 & 0.0666 & & & \\
				Uncertainty $\hat{\zeta_{V}}$ & 4.17 & 3.90 & (1.60)\\
				Uncertainty $\hat{x}_{min}$ & 604.81 & 575.44 & (186.04)\\
				\bottomrule
			\end{tabular}
			\scriptsize
			\begin{tablenotes}
				\item Note: In the low-frequency unregulated treatment, 76 out of 100
				simulations have a p-value greater than 0.05, and therefore are consistent with
				the hypothesis that $x$ is drawn from a distribution of the form of equation \ref{eq: power law behaviour of trading volume}.
			\end{tablenotes}
		\end{threeparttable}
	\end{table}
	
	\subsubsection{Long Memory of Volume}
	\label{subsubsec:Long Memory of Volume}
	
\begin{figure}[!hbpt] \centering
		\begin{subfigure}[H]{0.4\textwidth}
			\includegraphics[width=\textwidth]{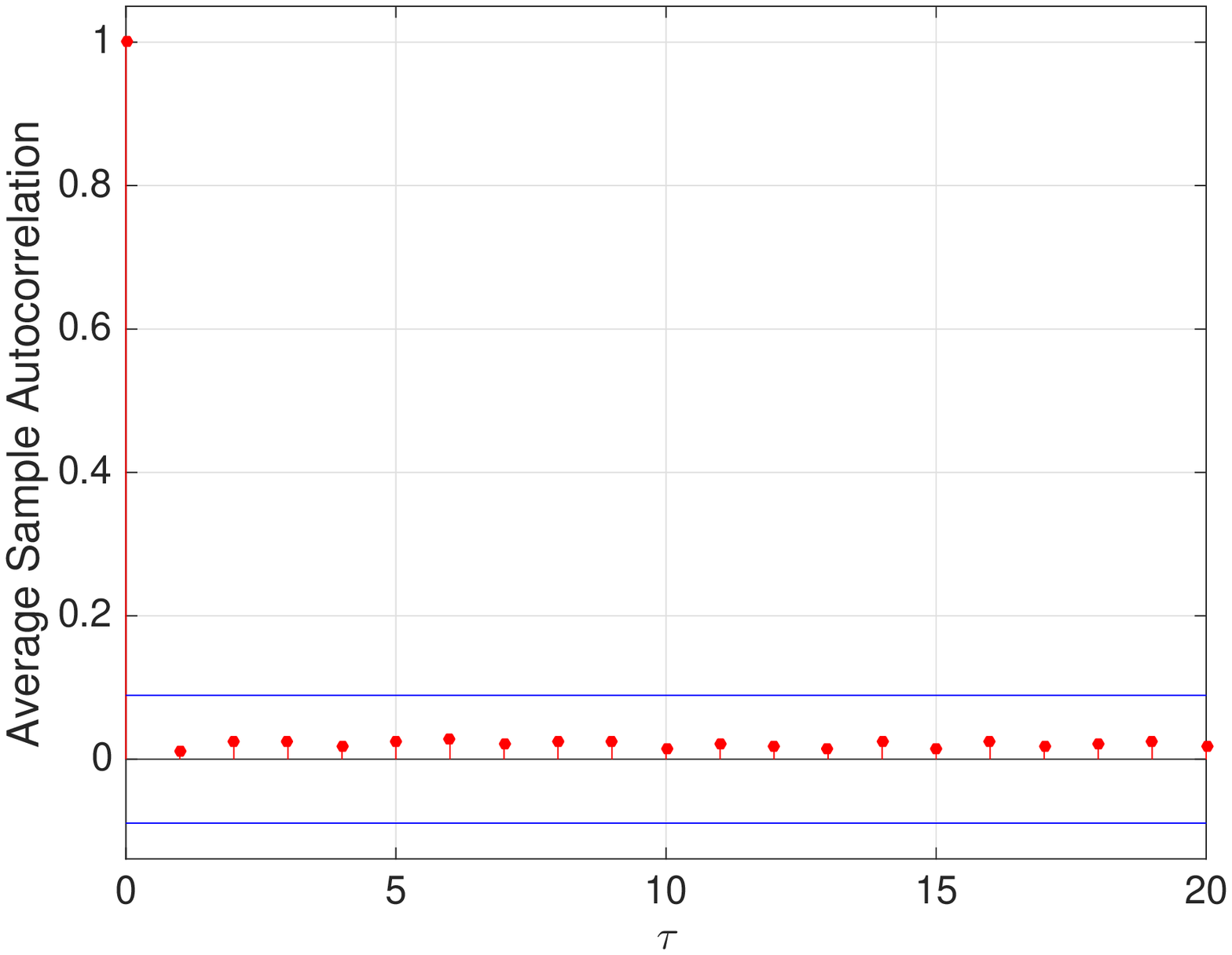}
			\caption{Average autocorrelation}
			\label{fig:Long Memory Volume - Average LF}
		\end{subfigure}
		\hspace{2cm}
		\begin{subfigure}[H]{0.4\textwidth}
			\includegraphics[width=\textwidth]{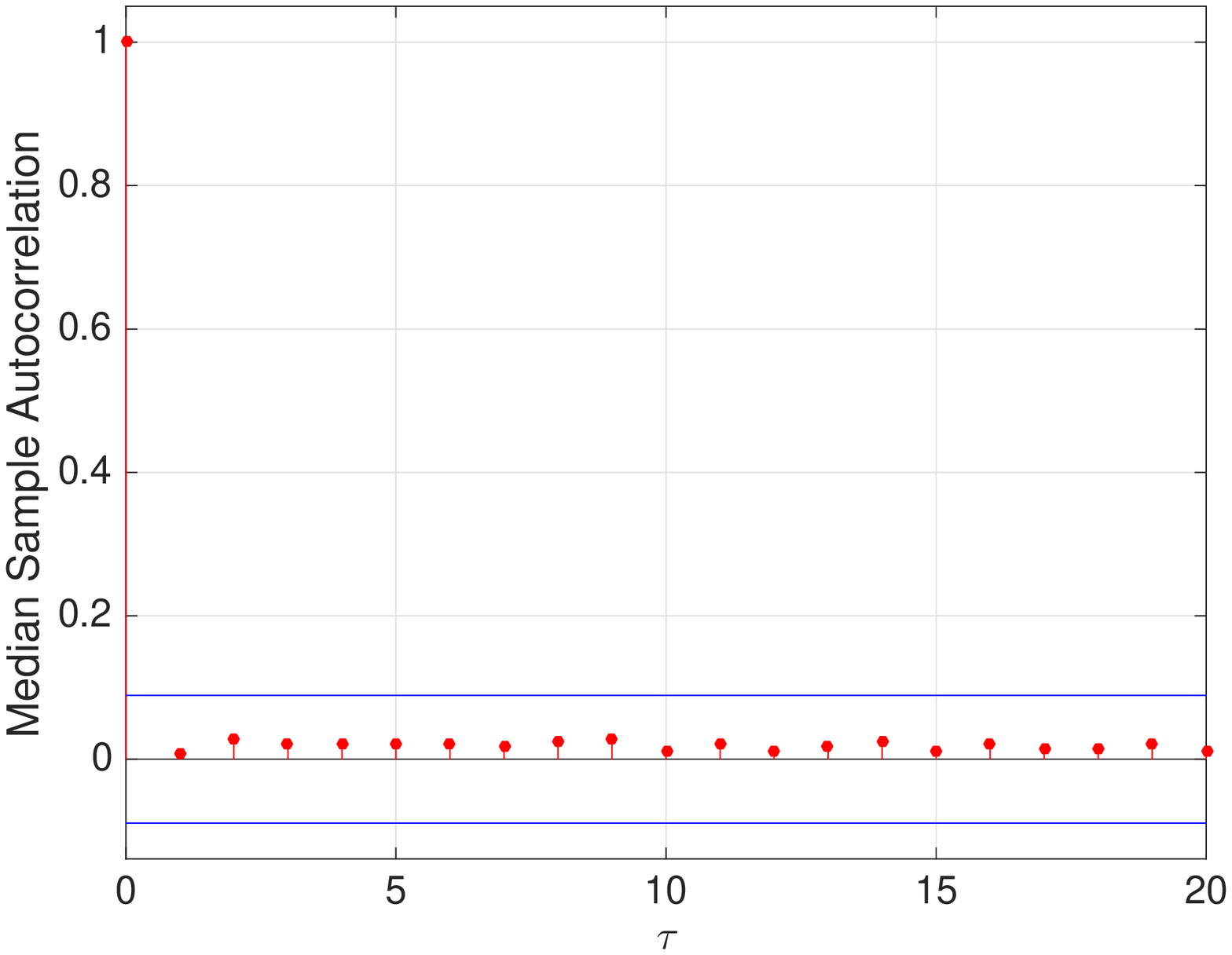}
			\caption{Median autocorrelation}
			\label{fig:Long Memory Volume - Median LF}
		\end{subfigure}
		\caption{Autocorrelation of trading volume in low-frequency data over 100
			simulations} 
		\label{fig:Autocorrelation of trading volume over 100 simulations LF}
		\floatfoot{\scriptsize{Note: The blue horizontal lines represent the approximate
				confidence bounds $[-0.0891;0.0891]$ of the autocorrelation function assuming the time
				serie is a moving average process.}}
	\end{figure}

	Figure \ref{fig:Autocorrelation of trading volume over 100 simulations LF} shows
	that low-frequency trading volume data does not exhibit significant long-memory.
	
	However, Figure \ref{fig:Autocorrelation of trading volume over 100 simulations
		HF} exhibits significant long memory across many transactions for high-frequency
	data, as confirmed by the empirical evidence, e.g. \citep{Russell2010}. These
	results indicate the existence of clustering of volume.
	
	\begin{figure}[!hbpt] \centering
		\begin{subfigure}[H]{0.4\textwidth}
			\includegraphics[width=\textwidth]{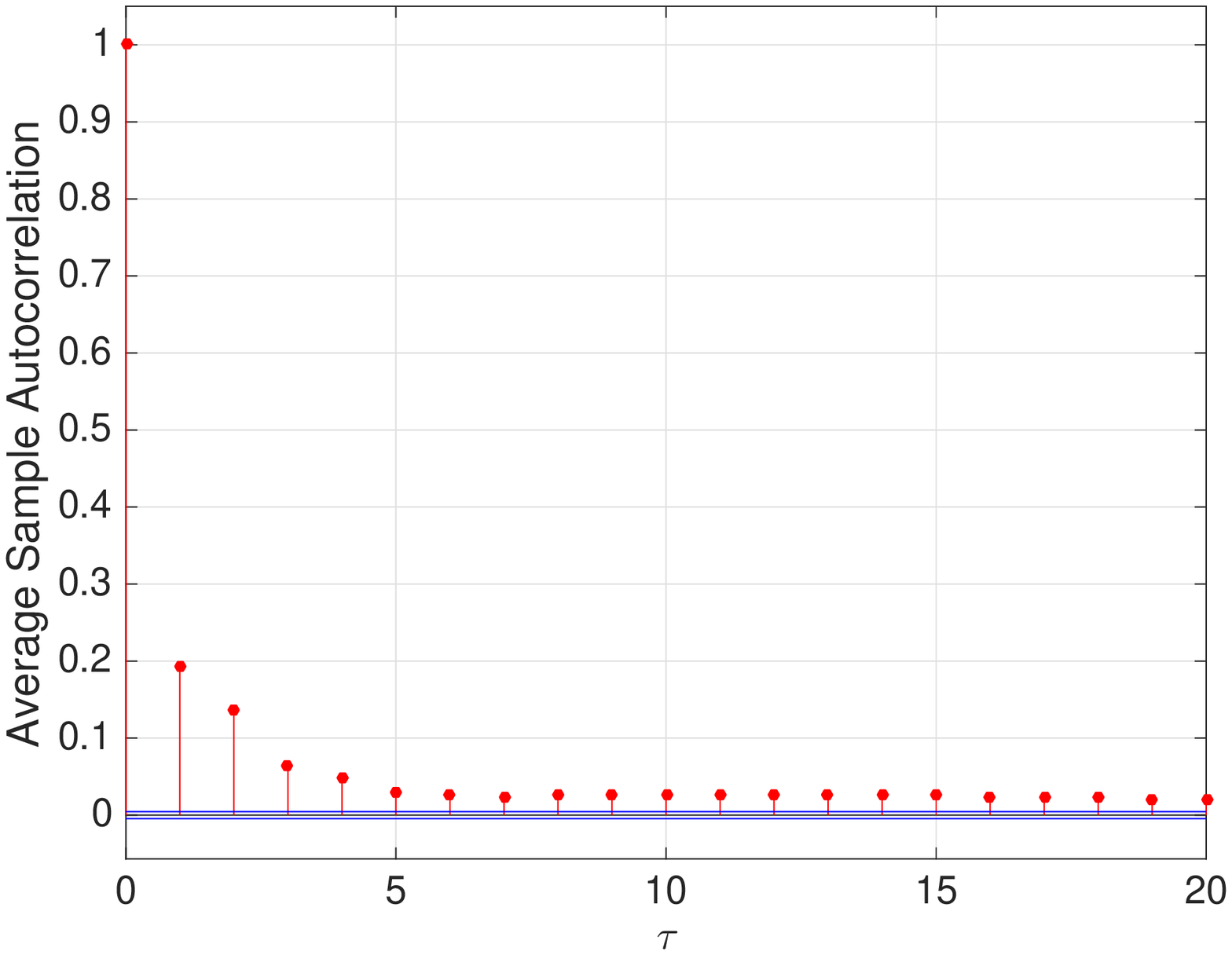}
			\caption{Average autocorrelation}
			\label{fig:Long Memory Volume - Average HF}
		\end{subfigure}
		\hspace{2cm}
		\begin{subfigure}[H]{0.4\textwidth}
			\includegraphics[width=\textwidth]{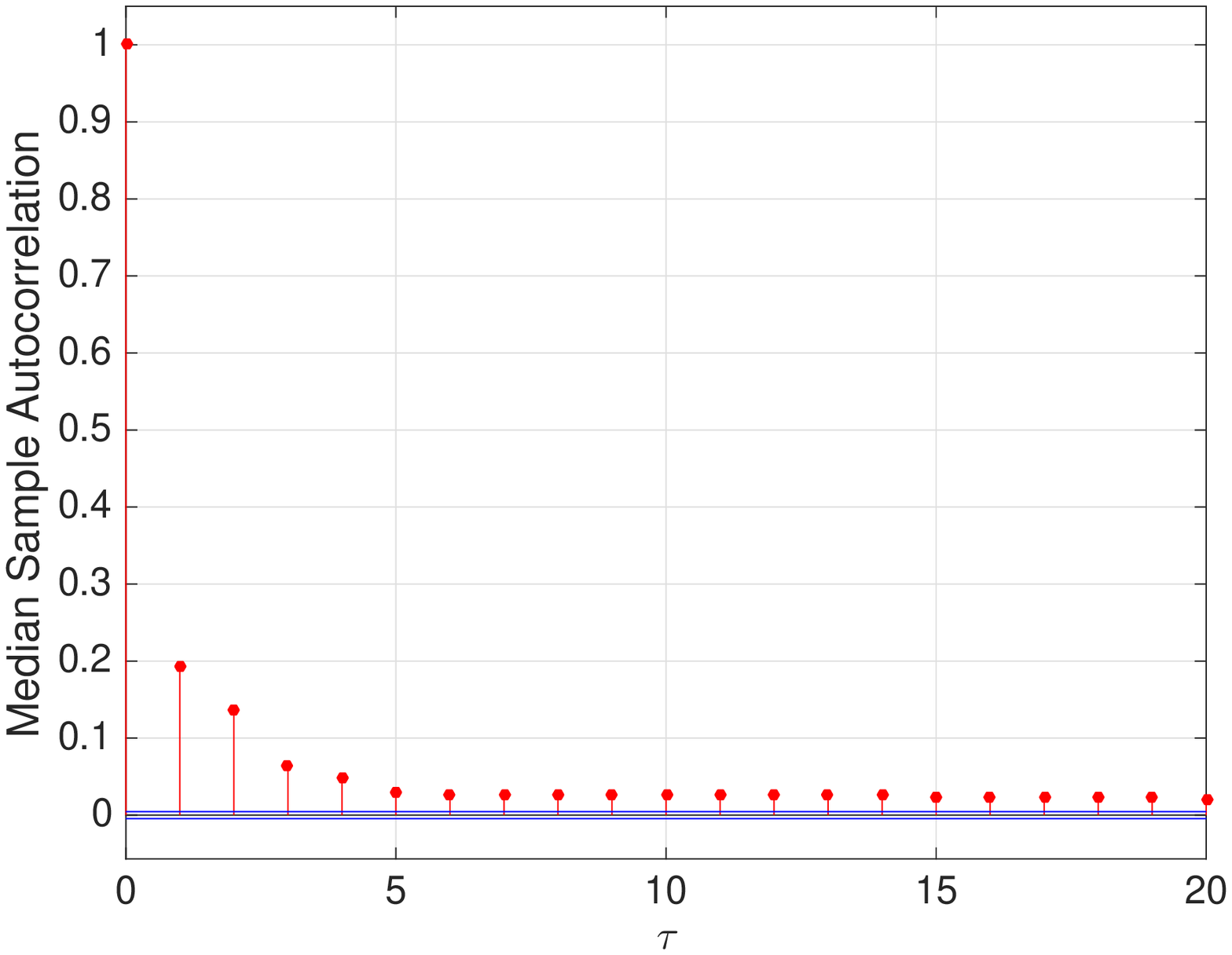}
			\caption{Median autocorrelation}
			\label{fig:Long Memory Volume - Median HF}
		\end{subfigure}
		\caption{Autocorrelation of trading volume in high-frequency data over 100
			simulations} 
		\label{fig:Autocorrelation of trading volume over 100 simulations HF}
		\floatfoot{\scriptsize{Note: The blue horizontal lines represent the approximate
				confidence bounds $[-0.0053;0.0053]$ of the autocorrelation function assuming the time
				serie is a moving average process.}}
	\end{figure}
	
	\subsection{Trading Duration}
	\label{subsection:Trading Duration}
	
	In our baseline treatment across 100 simulations, the minimum time observed
	between events is 1 tick and the maximum duration is 184 ticks. The average
	duration between successive events is 10.24 ticks with a standard deviation of
	9.70 ticks.
	
	\citep{Engle2000} observes that longer durations are associated
	with lower volatilities as predicted by the Easley and O'Hara model \citep{Pacurar2008},
	and is discussed in sections \ref{subsubsec:Volatility Volume Correlations1}--\ref{subsubsec:Volatility Volume Correlations2}.

	\subsubsection{Clustering of Trade Duration}
	\label{subsubsec:Clustering of Trade Duration}
	
	Figure \ref{fig:Autocorrelation of squared trading duration over 100 simulations
		HF} shows the existence of positive autocorrelation of squared trading
	duration starting at low values and exhibiting slow decay of the autocorrelation
	function only in treatments with risk-based capital requirements. The baseline
	treatment does not show any significant autocorrelation of trading duration.
	\citep{Engle1998} observe that these autocorrelations indicate clustering of durations, as identified in empirical
	data.
	
	\begin{figure}[!hbpt] \centering
		\begin{subfigure}[H]{0.3\textwidth}
			\includegraphics[width=\textwidth]{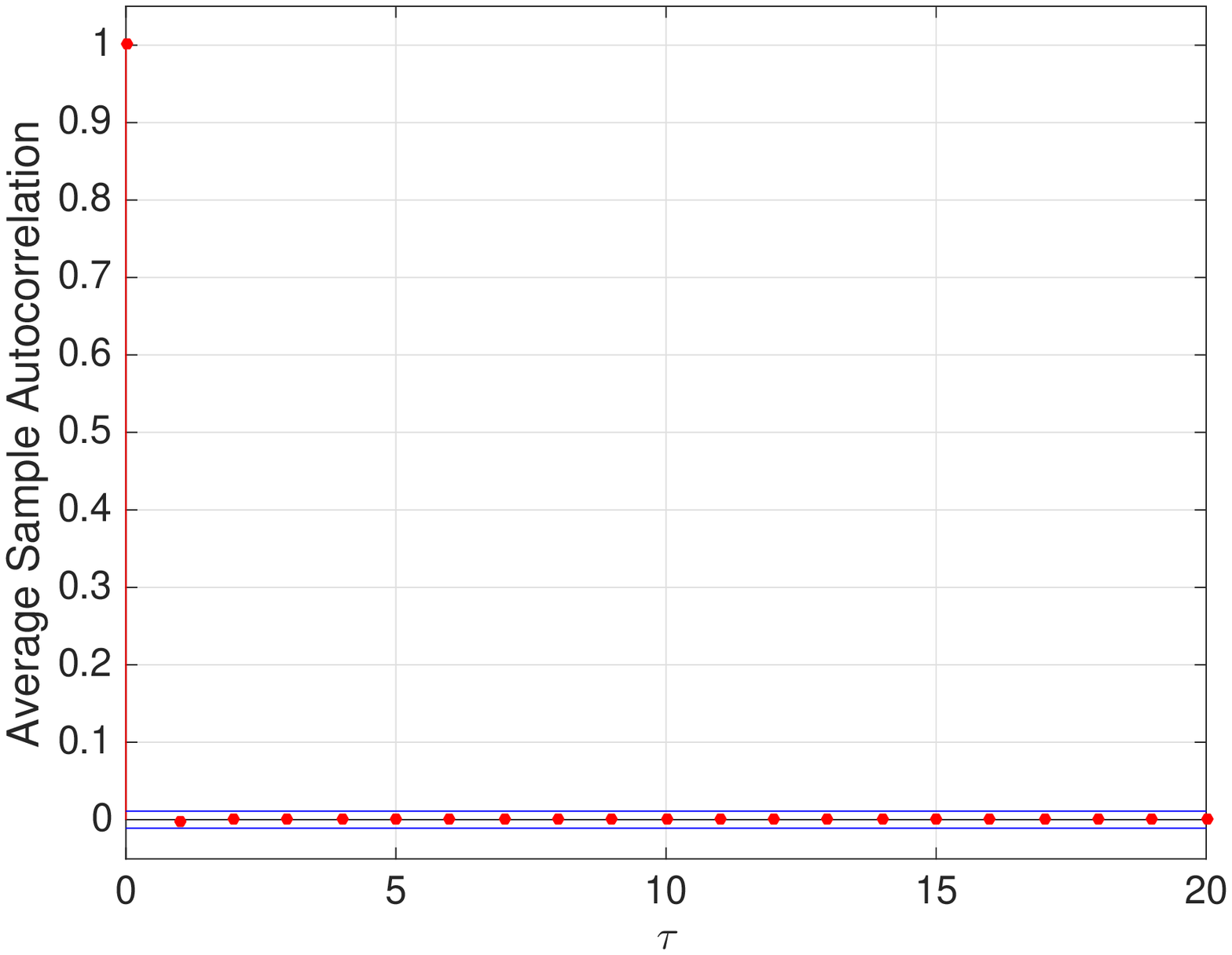}
			\caption{\footnotesize{Baseline (average)}}
			\label{fig:Clustering of Trade Duration - Average HF - Baseline}
		\end{subfigure}
		\hfill
		\begin{subfigure}[H]{0.3\textwidth}
			\includegraphics[width=\textwidth]{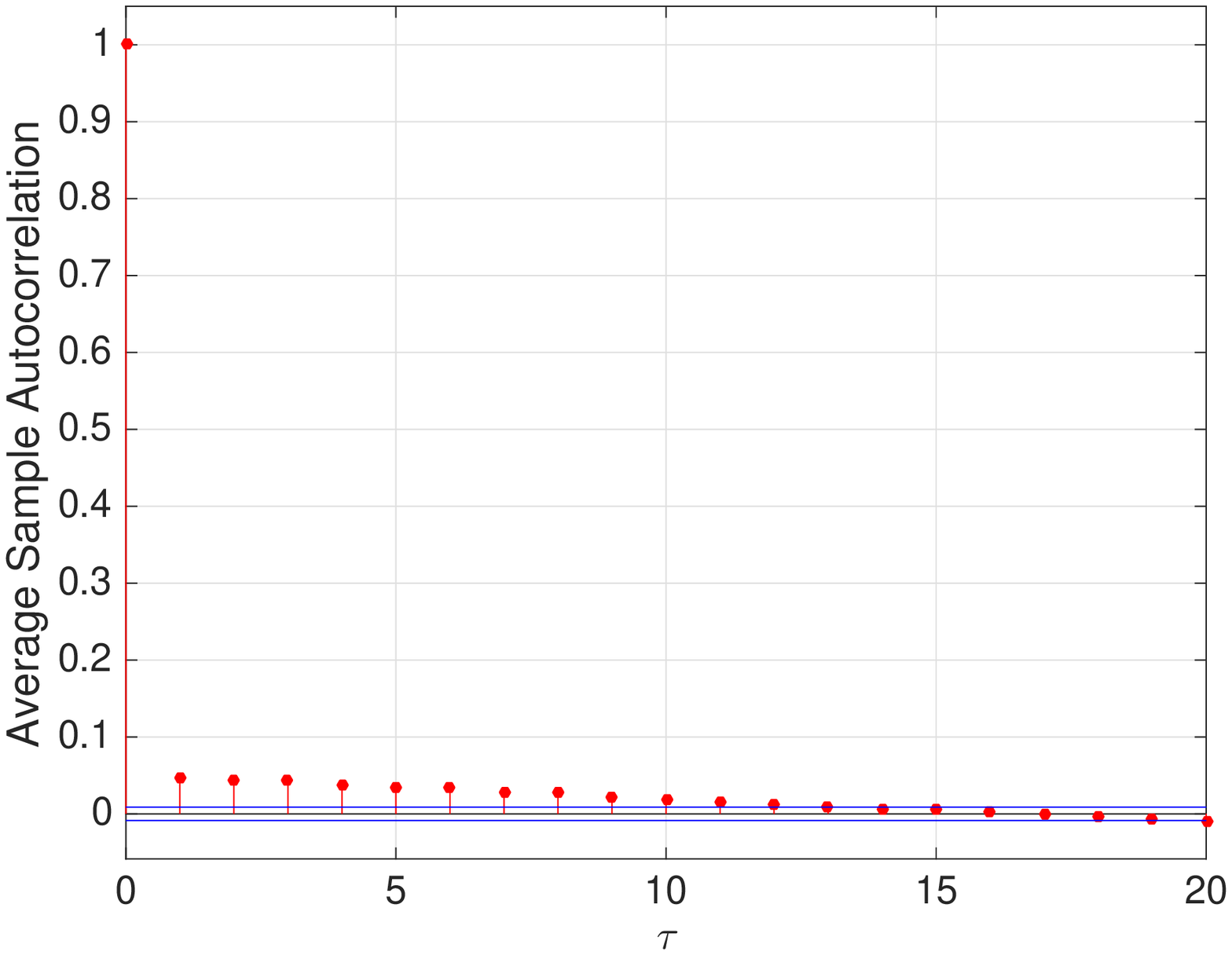}
			\caption{\footnotesize{VaR (average)}}
			\label{fig:Clustering of Trade Duration - Average HF - VaR}
		\end{subfigure}
		\hfill
		\begin{subfigure}[H]{0.3\textwidth}
			\includegraphics[width=\textwidth]{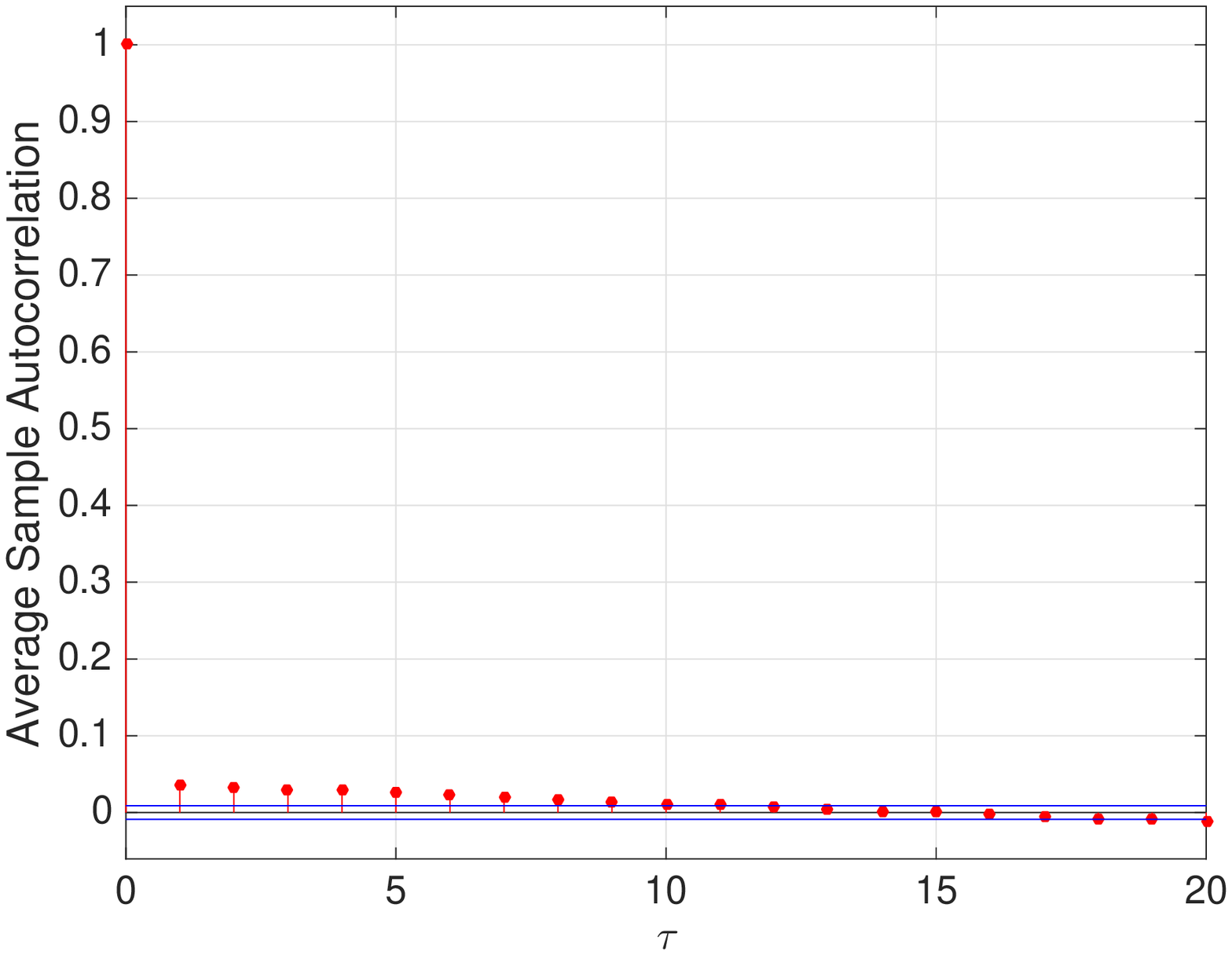}
			\caption{\footnotesize{ES (average)}}
			\label{fig:Clustering of Trade Duration - Average HF - ES}
		\end{subfigure}
		
		\begin{subfigure}[H]{0.3\textwidth}
			\includegraphics[width=\textwidth]{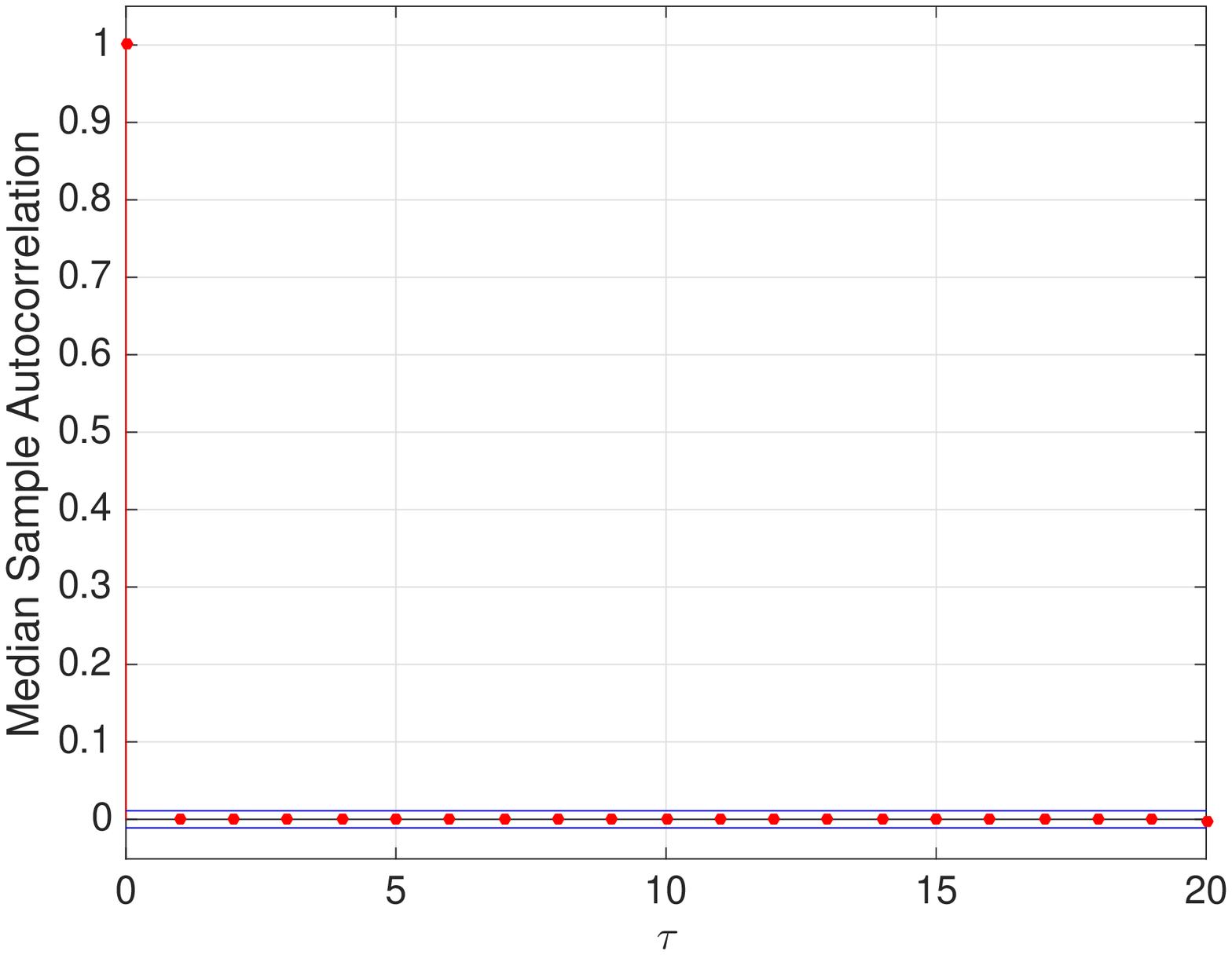}
			\caption{\footnotesize{Baseline (median)}}
			\label{fig:Clustering of Trade Duration - Median HF - baseline}
		\end{subfigure}
		\hfill
		\begin{subfigure}[H]{0.3\textwidth}
			\includegraphics[width=\textwidth]{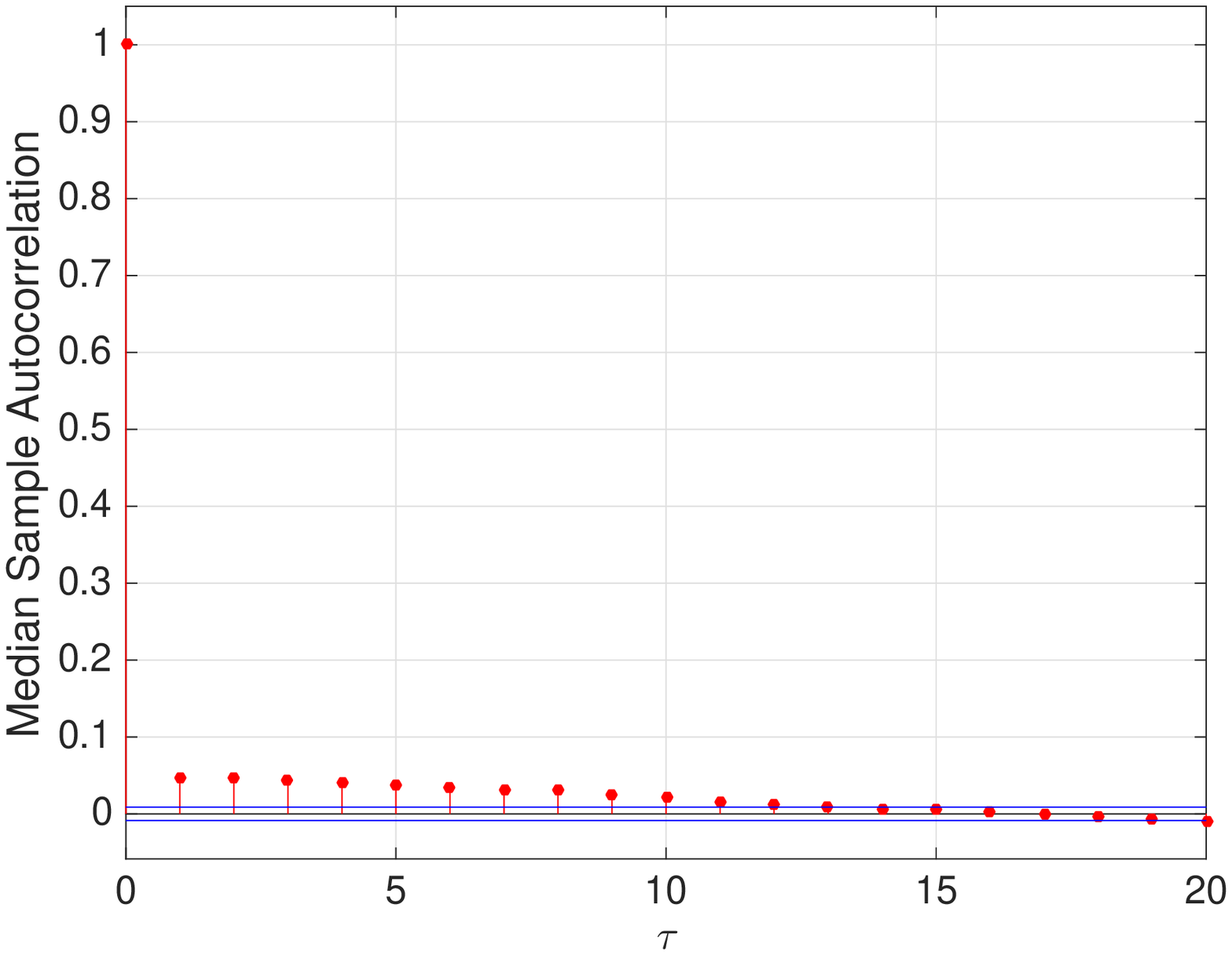}
			\caption{\footnotesize{VaR (median)}}
			\label{fig:Clustering of Trade Duration - Median HF - VaR}
		\end{subfigure}
		\hfill
		\begin{subfigure}[H]{0.3\textwidth}
			\includegraphics[width=\textwidth]{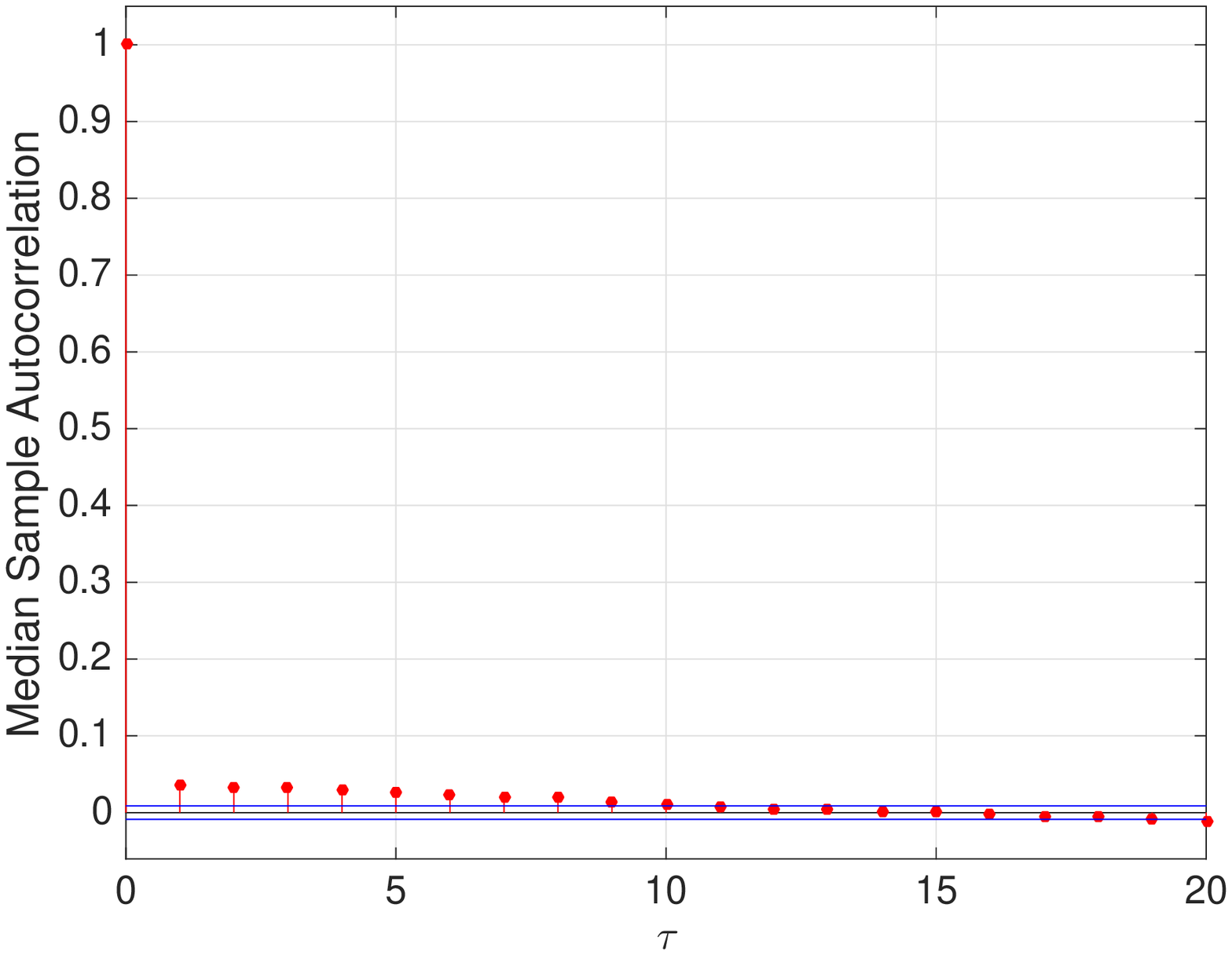}
			\caption{\footnotesize{ES (median)}}
			\label{fig:Clustering of Trade Duration - Median HF - ES}
		\end{subfigure}
		
		\caption{Autocorrelation of squared trading duration in high-frequency data over
			100 simulations}
		\label{fig:Autocorrelation of squared trading duration over 100 simulations HF}
		\floatfoot{\scriptsize{Note: The blue horizontal lines represent the average
				confidence bounds for, respectively, baseline, VaR and ES treatments
				($[-0.0126;0.0126]$, $[-0.0104;0.0104]$, $[-0.011;0.011]$) of the
				autocorrelation function assuming the time-series is a moving average process.}}
	\end{figure}
	
	\subsubsection{Long Memory of Trade Duration}
	\label{subsubsec:Long Memory of Trade Duration}
	
	As for clustering of trade durations, Figure \ref{fig:Autocorrelation of trading
		duration over 100 simulations HF} shows the existence of positive
	autocorrelations of trading duration starting at low values and exhibiting slow
	decay of the autocorrelation function, but only for treatments with
	risk-based capital requirements.
	
	\begin{figure}[!hbpt] \centering
		\begin{subfigure}[H]{0.3\textwidth}
			\includegraphics[width=\textwidth]{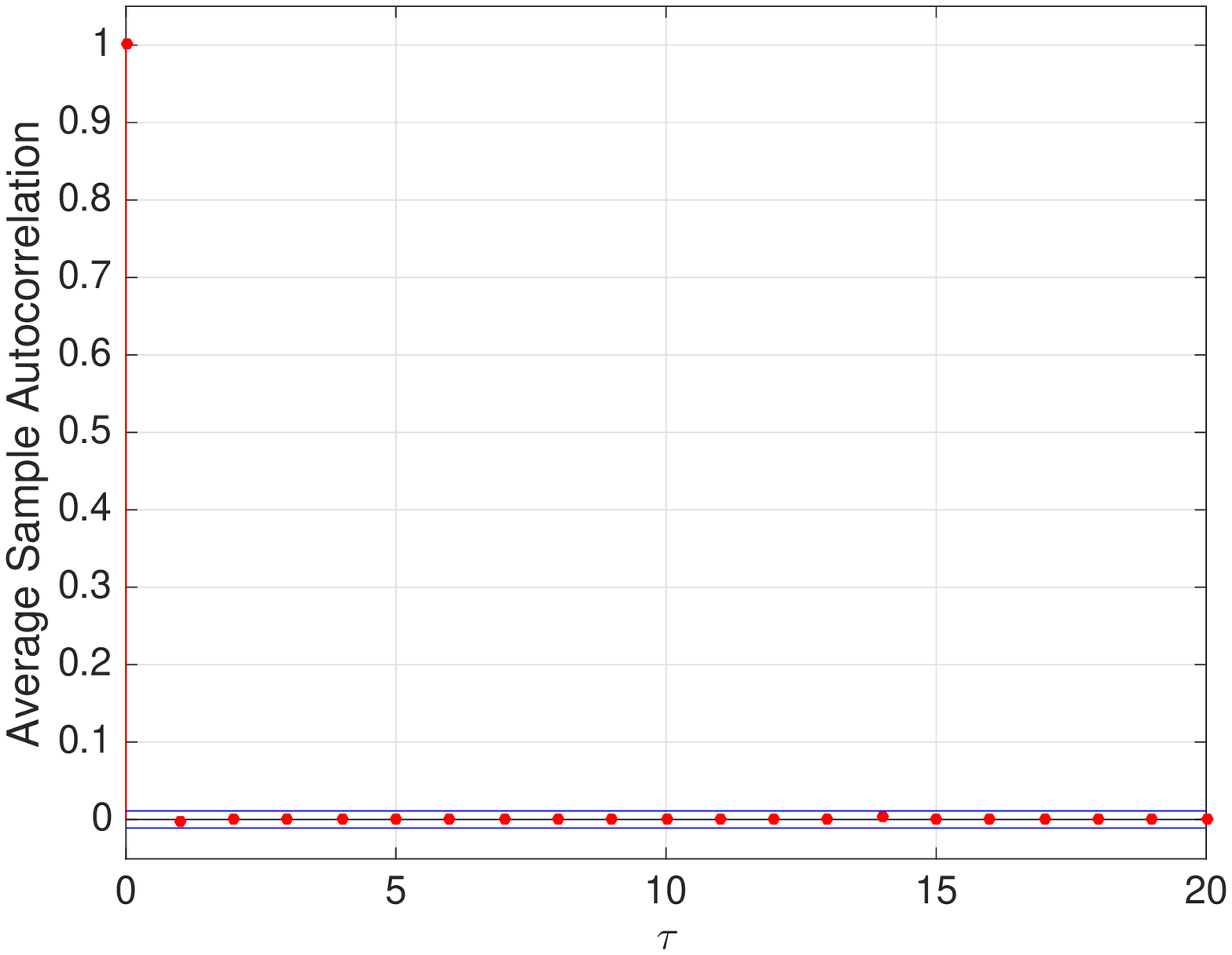}
			\caption{\footnotesize{Baseline (average)}}
			\label{fig:Long Memory of Trade Duration - Average HF - baseline}
		\end{subfigure}
		\hfill
		\begin{subfigure}[H]{0.3\textwidth}
			\includegraphics[width=\textwidth]{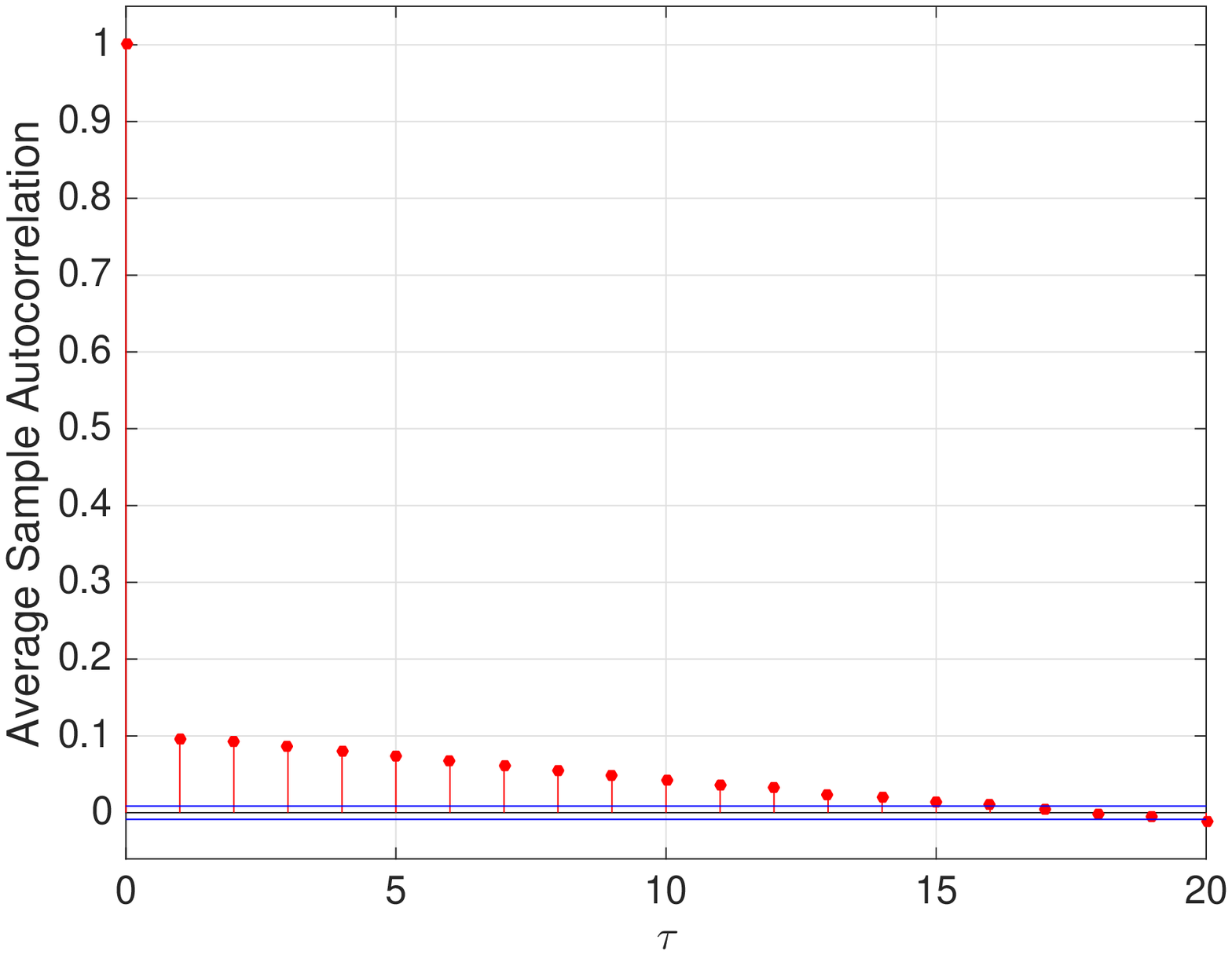}
			\caption{\footnotesize{VaR (average)}}
			\label{fig:Long Memory of Trade Duration - Average HF - VaR}
		\end{subfigure}
		\hfill
		\begin{subfigure}[H]{0.3\textwidth}
			\includegraphics[width=\textwidth]{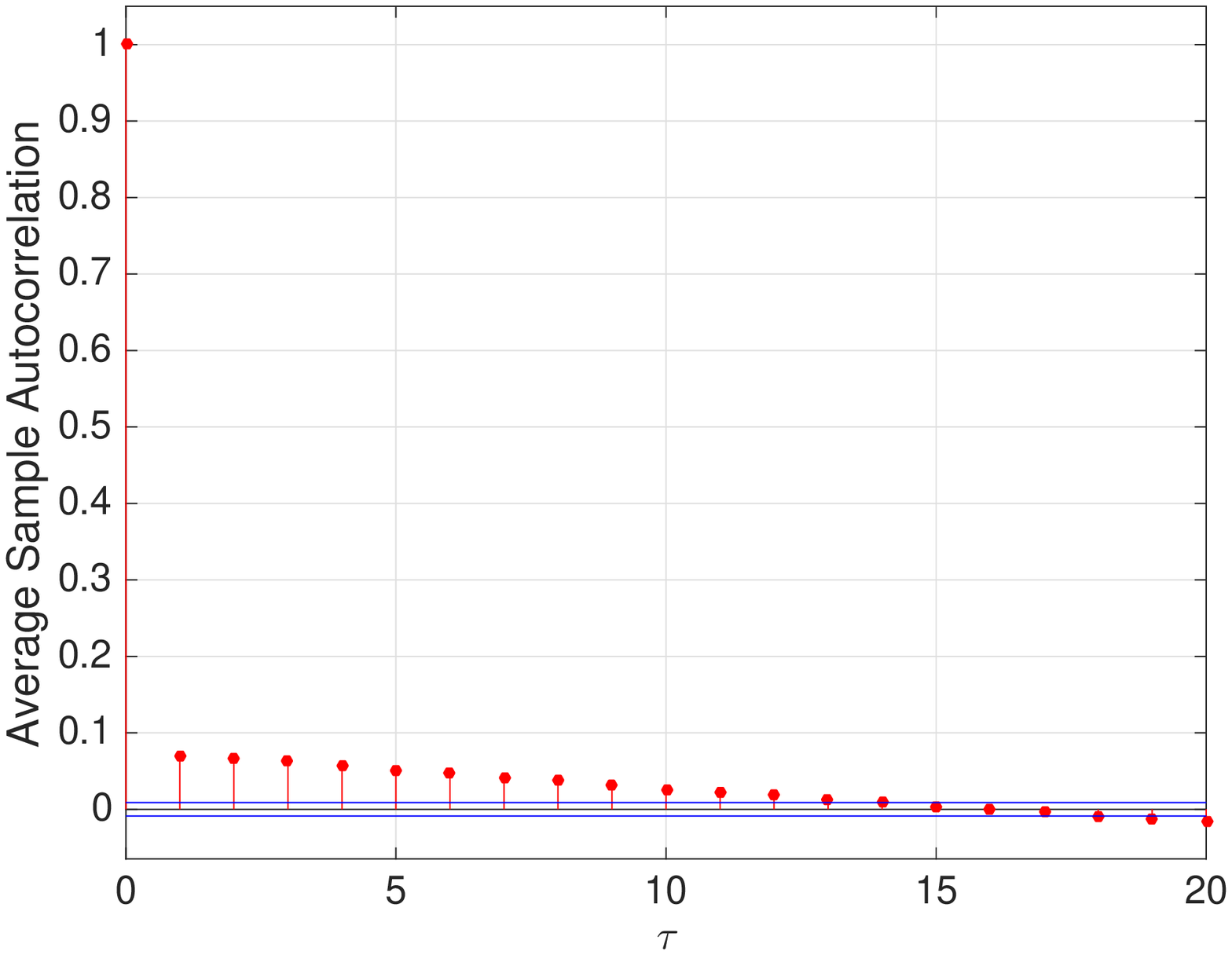}
			\caption{\footnotesize{ES (average)}}
			\label{fig:Long Memory of Trade Duration - Average HF - ES}
		\end{subfigure}
		
		\begin{subfigure}[H]{0.3\textwidth}
			\includegraphics[width=\textwidth]{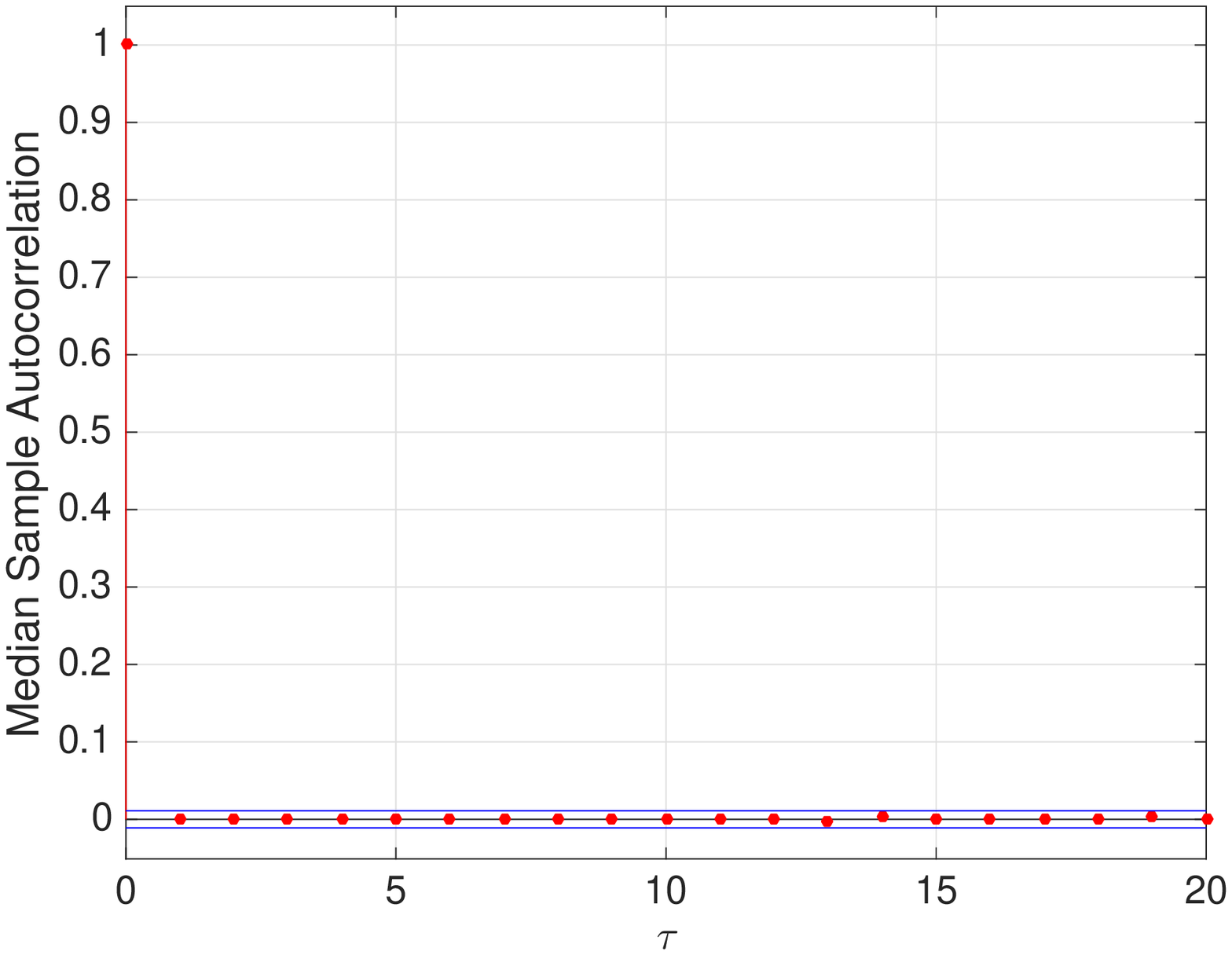}
			\caption{\footnotesize{Baseline (median)}}
			\label{fig:Long Memory of Trade Duration - Median HF - baseline}
		\end{subfigure}
		\hfill
		\begin{subfigure}[H]{0.3\textwidth}
			\includegraphics[width=\textwidth]{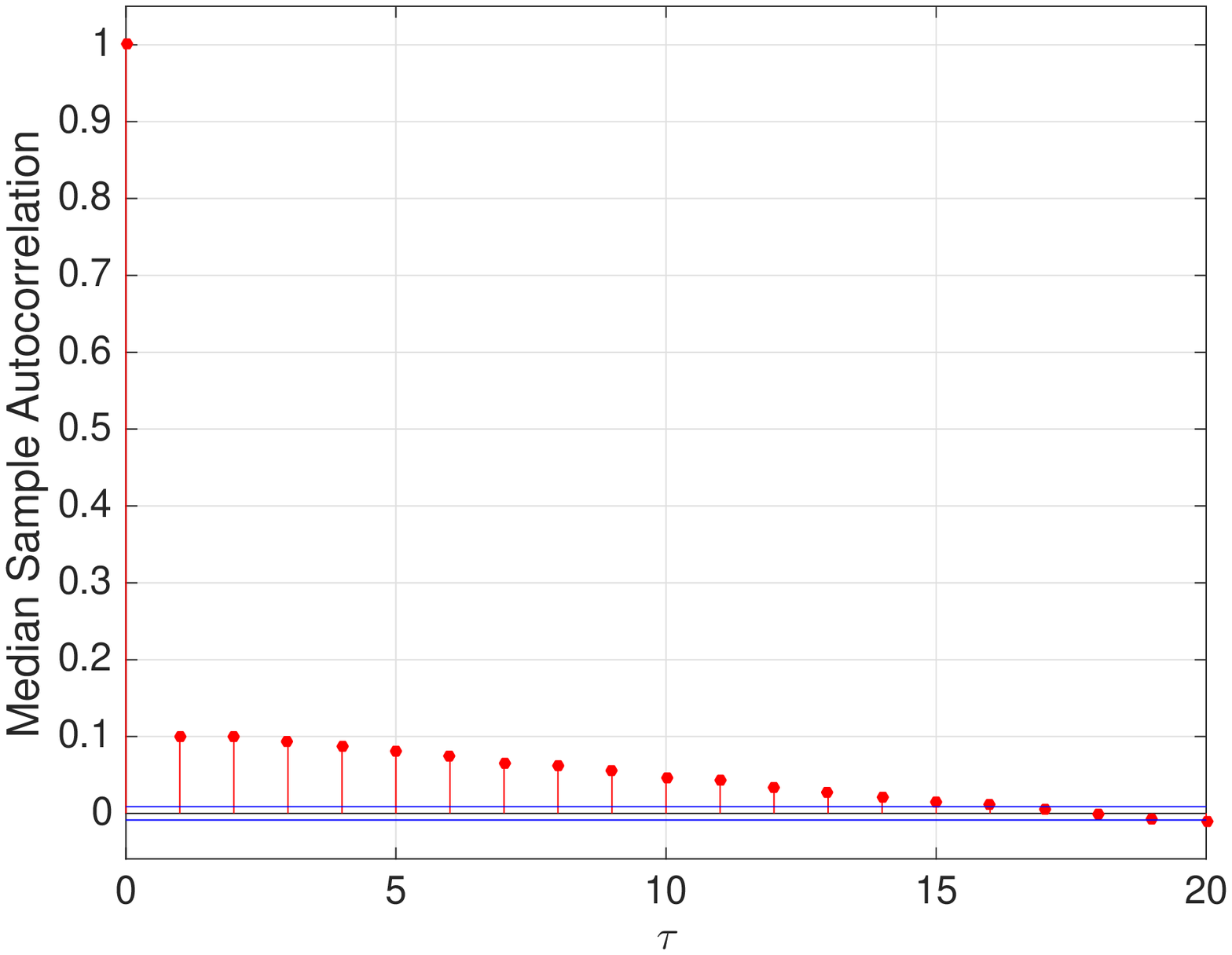}
			\caption{\footnotesize{VaR (median)}}
			\label{fig:Long Memory of Trade Duration - Median HF - VaR}
		\end{subfigure}
		\hfill
		\begin{subfigure}[H]{0.3\textwidth}
			\includegraphics[width=\textwidth]{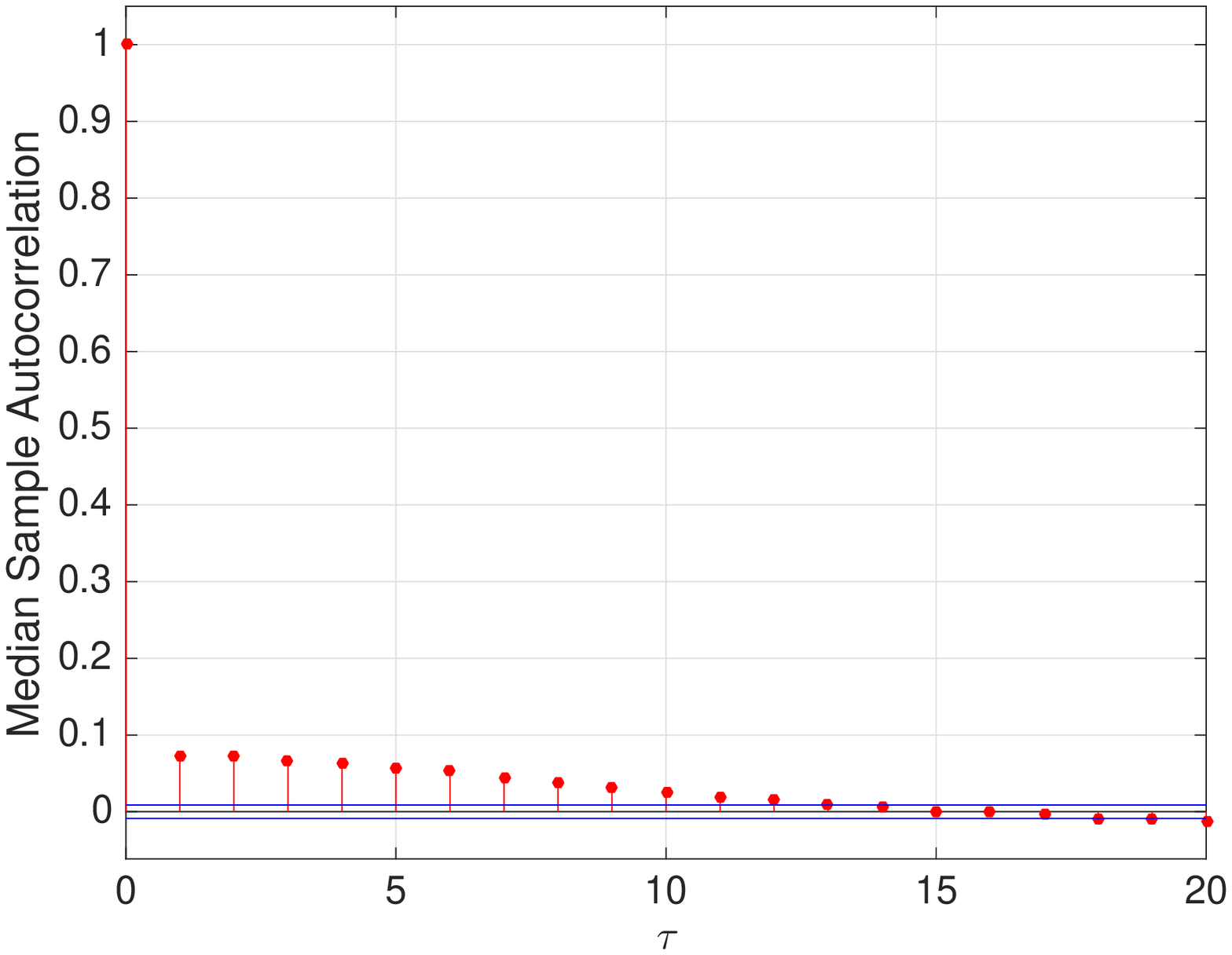}
			\caption{\footnotesize{ES (median)}}
			\label{fig:Long Memory of Trade Duration - Median HF - ES}
		\end{subfigure}
		
		\caption{Autocorrelation of trading duration in high-frequency data over 100
			simulations}
		\label{fig:Autocorrelation of trading duration over 100 simulations HF}
		\floatfoot{\scriptsize{Note: The blue horizontal lines represent the average
				confidence bounds ($[-0.0126;0.0126]$, $[-0.0104;0.0104]$, $[-0.011;0.011]$) of
				the autocorrelation function assuming the time-series is a moving average
				process.}}
	\end{figure}
	
	\subsubsection{Overdispersion}
	\label{subsubsec:Overdispersed}
	
\begin{figure}[!hbpt]\centering
		\begin{subfigure}[h]{0.4\textwidth}
			\includegraphics[width=\textwidth]{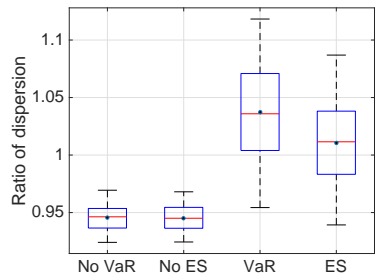}
		\end{subfigure}
		\caption{Boxplot of overdispersation of trade durations}
		\label{fig:Boxplot of overdispersation of trade durations}
	\end{figure}

	Figure \ref{fig:Boxplot of overdispersation of trade durations} shows that the
	mean of trade durations in treatments with risk-based capital requirements are
	characterised by overdispersion. The mean of ratio of standard deviation to mean
	of the duration series is greater than one in VaR (1.038) and ES (1.010)
	treatments, while the baseline treatments both reveal the same level of
	underdispersion (0.954).
	
	We conclude from these results that the stylised facts of trading duration are
	replicated by our model, but only in treatments with risk-based capital
	requirements.
	
	\subsection{Transaction Size}
	\label{subsection:Transaction Size}
	
	In this section we investigate if distributions of trading volume are consistent
	with a L\'{e}vy-stable distribution.
	\label{subsection:Transaction Size1}
	
	\subsubsection{Power Law Behaviour of Trades}
	\label{subsection:Power Law Behaviour of Trades}
	
	Table \ref{table:Power law of number of trades over 100 simulations} exhibits a
	different behaviour relative to the previous analysed power laws. 
	
	\begin{table}[!hbpt] \centering
		\begin{threeparttable}
			\caption{Power law of number of trades over 100 simulations}
			\label{table:Power law of number of trades over 100 simulations}
			\footnotesize
			\begin{tabular}{c c c c}
				\toprule
				{} & {Mean} & {Median} & {Standard Deviation} \\
				\midrule
				$\hat{\zeta_{N}}$ & 3.9 & 3.9 & (0.0)\\
				$\hat{x}_{min}$ & 227.04 & 226 & (34.08)\\
				p-value & 0.0 & 0.0 & (0.0)\\
				% D statistic & 0.0646 & 0.0666 & & & \\
				Uncertainty $\hat{\zeta_{N}}$ & 0.0 & 0.0 & (0.0)\\
				Uncertainty $\hat{x}_{min}$ & 4.10 & 3.94 & (1.08)\\
				\bottomrule
			\end{tabular}
			\scriptsize
			\begin{tablenotes}
				\item Note: In the low-frequency unregulated treatment, all the simulations
				have a p-value of 0, which rules out the hypothesis that $x$ is drawn from a
				distribution of the form of equation \ref{eq: power law behaviour of number of
					trades}.
			\end{tablenotes}
		\end{threeparttable}
	\end{table}
	
	As in \citep{Plerou2001}, we find a mean value $\zeta_{N}=3.9$, which is greater
	than 2 and therefore outside the L\'{e}vy-stable distribution.
	The fact that all p-values are 0 confirms the evidence in
	\citep{Vijayaraghavan2011} supporting the existence of a non-invariant
	distribution.
	\label{subsection:Transaction Size2}
	
	\subsection{Bid-ask spread}
	\label{subsubsection:Bid-ask spread}
	
	\subsubsection{Spread Correlated with Price Change}
	\label{subsubsec:Spread Correlated with Price Change}
	
	The existence of bid-ask spread, although small in magnitude, has several
	important consequences in time-series properties of asset returns. The bid-ask
	spread introduces a negative lag 1 serial correlation in the series of observed
	price changes, as observed in figure \ref{fig:Absence of HF Autocorrelation}.
	This serial correlation in an asset return is referred to as the bid-ask bounce
	in the finance literature \citep{tsay2002}. The same result was also reported
	in \citep{Stephan1990} where the authors computed serial correlations of price
	changes to test for the bid-ask spread effect, which produced a negative average
	first-order serial correlation meaningfully different from zero.
	
	\subsubsection{Thinness and Large Spread}
	\label{subsubsec:Thinness and Large Spread}
	
	% We use a proportional bid-ask spread as in, for example, \cite{McInish1992} and
	% \cite{Abhyankar1997}:
	% 
	% \begin{equation}
	% \label{eq:proportional bid-ask spread}
	% BAS_{t}=\frac{ask_{t}-bid_{t}}{\frac{ask_{t}+bid_{t}}{2}}
	% \end{equation}
	
	We calculate the bid-ask spread as in equation \ref{eq:bid-ask spread} and
	investigate the impact of market thinness in both the bid and ask sides of the
	order-book. In figure \ref{fig:Average cross-correlation between spread and
		volume in high-frequency data over 100 simulations} we confirm the statistically
	significant negative correlation between spread and bid volume. However, there
	is no significant dependence between spread and ask volume.
	
	\begin{figure}[!hbpt] \centering
		\begin{subfigure}[H]{0.4\textwidth}
			\includegraphics[width=\textwidth]{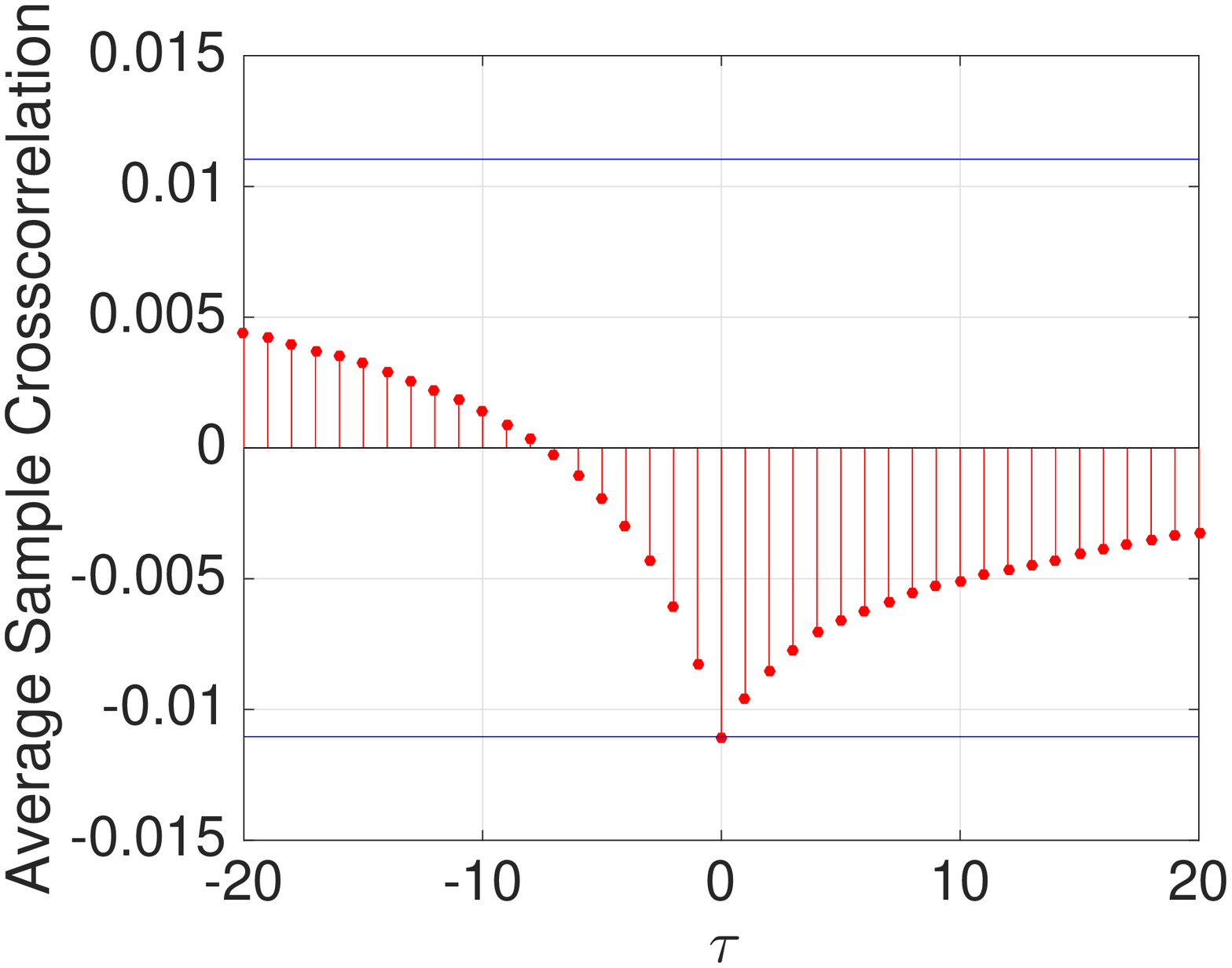}
			\caption{Ask}
			\label{fig:Cross-correlation between spread and volume - Ask}
		\end{subfigure}
		\hspace{2cm}
		\begin{subfigure}[H]{0.4\textwidth}
			\includegraphics[width=\textwidth]{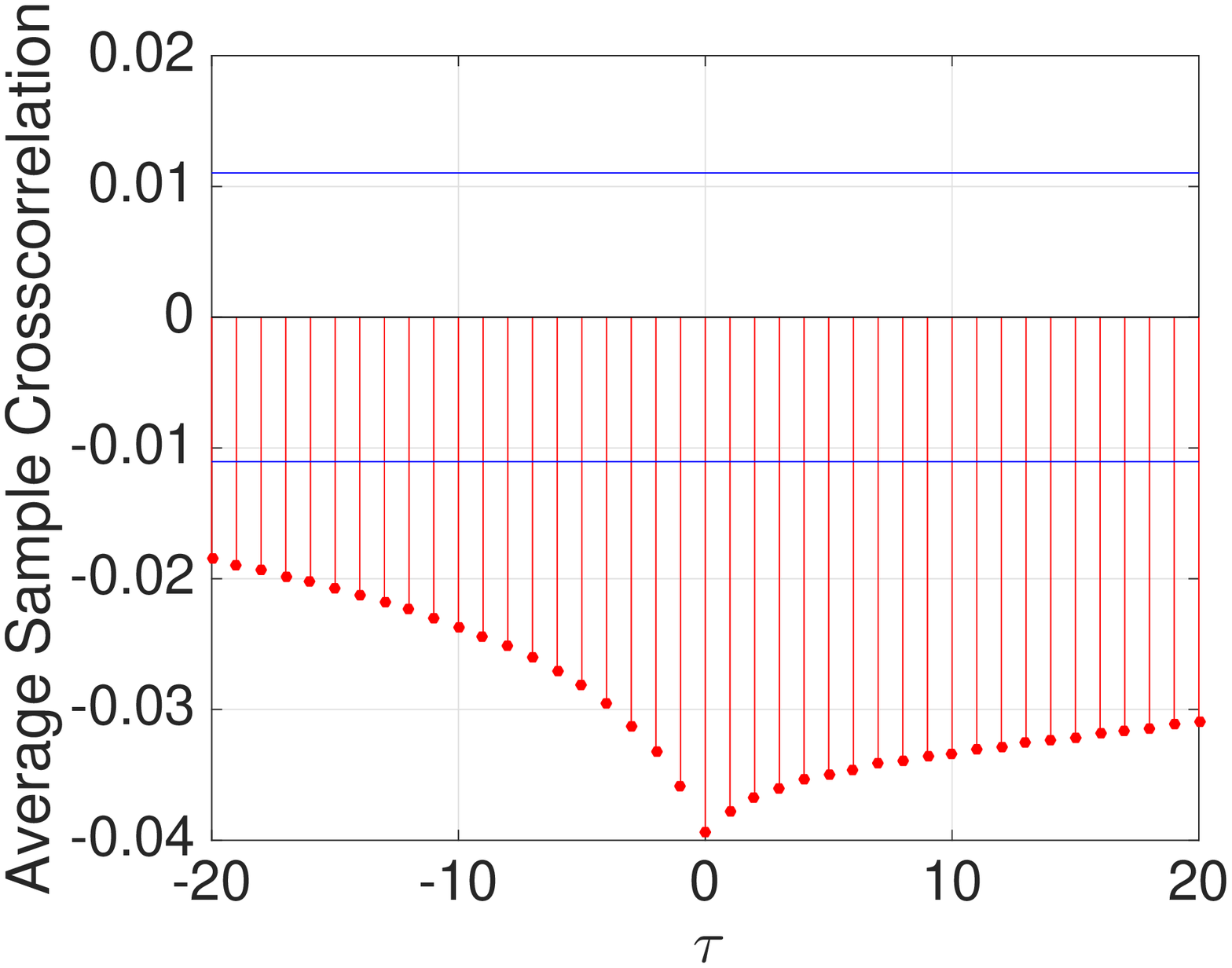}
			\caption{Bid}
			\label{fig:Cross-correlation between spread and volume - Bid}
		\end{subfigure}
		\caption{Cross-correlation between spread and volume in high-frequency data over 100
			simulations}
		\label{fig:Average cross-correlation between spread and volume in high-frequency
			data over 100 simulations}
		\floatfoot{\scriptsize{Note: The blue horizontal lines represent the mean of
				approximate upper and lower confidence bounds over 100 simulations
				($[-0.0126,0.0126]$), assuming spread and volume are uncorrelated.}}
	\end{figure}
	
	\section{Conclusion}
	\label{sec:Conclusion chapter 4}
	
	In this paper we demonstrate that our model is able to robustly and
	consistently replicate most of the well-known stylised facts of financial
	time-series data, either with the simplest baseline treatment or treatments with
	risk-based capital requirements. To the best of our knowledge, this is the first
	agent-based model which is able to replicate a vast number of stylised facts of
	financial time-series of returns, trading volume, trading duration, transaction
	size and bid-ask spread. The ability of our agent-based model to correctly
	reproduce statistical properties of financial time-series, including the
	qualitative properties of financial time-series (e.g.
	heavy tails), their correct quantitative properties (e.g. tail exponent or
	moments of distribution), and the distribution of particular behaviours (e.g.
	power laws), demonstrates the success of our model in replicating stylised facts
	as defined by the literature \citep{Panayi2013}.
	
	Only factual evidence, the statistical properties of financial time-series, can
	show whether a model has a meaningful empirical counterpart. By using the
	benchmark of the empirical validation of agent-based models, our model shows
	that it can be taken as an adequate representation of reality and, hence, be
	accepted as valid. As a valid abstraction from reality, our model aims to give
	valid and meaningful explanations of certain phenomena, and by replicating most
	of the stylised facts it adds additional robustness to the conclusions
	concerning the hypothesis designed to explain particular features of reality.
	
	\citep{Cont2001} concludes that the statistical properties of financial data work
	as constraints that a stochastic process has to verify in order to reproduce
	these statistical properties accurately. However, some authors (e.g.
	\citep{{Cont2001}, {Chiarella2006}}) observe that most of the ABM with
	heterogeneous agents, a characteristic of our model, have difficulty in
	replicating realistic time-series and most currently existing models fail to
	reproduce all these statistical features at once, confirming how constraining
	those properties are. \citep{Cont2001} concludes that these stylised facts,
	albeit qualitative in some cases, are so constraining that it is consequently
	not easy to exhibit a stochastic process able to reproduce them within a single
	model.
	
	Through our attempt to study a comprehensive list of the statistical properties
	of financial data, and contrary to the conclusions found in the ABM literature, our
	model demonstrates that it is well suited to replicate a vast number of stylised
	facts of the financial time-series, simultaneously, and not only the most common
	statistical properties of real financial markets. This empirical validation
	allows us to confidently use our model to investigate real phenomena that we
	aim to better understand and explain. The results produced by our model provide
	a response to a key challenge faced by ABM as they demonstrate that the model
	generated financial time-series can be consistent with most of the known
	empirical facts.
	We can therefore be confident that our model is empirically adequate and offers
	some empirical validity as a basis for future modelling.
	
	Further research using the CPT framework, namely exploring other experimental
	treatments as in the M-V framework, would lead to a better understanding
	of financial stylised facts under different market conditions, i.e. leverage and shortselling.
	
	\section*{References}
	
	\bibliography{mybibfile}
	
\end{document}